\documentclass[aps,twocolumn,superscriptaddress,floatfix]{revtex4-1}
\usepackage{graphicx,amsmath,amssymb,verbatim,color}
\usepackage{booktabs}
\usepackage[utf8]{inputenc}
\usepackage{fancyhdr, fancybox, empheq} 
\usepackage{color}
\usepackage{url}

\usepackage[colorlinks=true,citecolor=blue,linkcolor=blue,urlcolor=blue, backref=false,pdfborder={0 0 0}]{hyperref}

\newcommand{\beq}{\begin{equation}}
\newcommand{\eeq}{\end{equation}}
\newcommand{\barr}{\begin{eqnarray}}
\newcommand{\earr}{\end{eqnarray}}
\usepackage[normalem]{ulem}

\newcommand{\aap}{A\&A}
\newcommand{\aapr}{A\&ARv}
\newcommand{\aaps}{A\&A Suppl.}

\newcommand{\mnras}{MNRAS}
\newcommand{\apjl}{ApJL}

\newcommand{\bs}{\boldsymbol}
\newcommand{\ho}{\hat{\Omega}}
\newcommand{\hp}{\hat{p}}
\newcommand{\hq}{\hat{q}}
\newcommand{\bsm}[1]{{\bs{\mathcal{#1}}}}
\newcommand{\bsmhat}[1]{\skew{6}\widehat{\bs{\mathcal{#1}}}}

\usepackage{array}
\newcolumntype{P}[1]{>{\centering\arraybackslash}p{#1}}

\begin{document}

\title{Insights into searches for anisotropies in the nanohertz gravitational-wave background} 

\author{Yacine Ali-Ha\"imoud}
\affiliation{Center for Cosmology and Particle Physics, Department of Physics, New York University, New York, NY 10003, USA}

\author{Tristan L.~Smith}
\affiliation{Department of Physics and Astronomy, Swarthmore College, 500 College Ave., Swarthmore, PA 19081, United States}

\author{Chiara M.~F.~Mingarelli}
\affiliation{Department of Physics, University of Connecticut, 196 Auditorium Road, U-3046, Storrs, CT 06269-3046, USA}
\affiliation{Center for Computational Astrophysics, Flatiron Institute, 162 Fifth Ave, New York, NY 10010, USA}

\date{\today}

\begin{abstract}
Within the next several years pulsar timing arrays (PTAs) are positioned to detect the stochastic gravitational-wave background (GWB) likely produced by the collection of inspiralling super-massive black holes binaries, and potentially constrain some exotic physics. So far most of the pulsar timing data analysis has focused on the monopole of the GWB, assuming it is perfectly isotropic. The natural next step is to search for anisotropies in the GWB. In this paper, we use the recently developed PTA Fisher matrix to gain insights into optimal search strategies for GWB anisotropies. For concreteness, we apply our results to EPTA data, using realistic noise characteristics of its pulsars. We project the detectability of a GWB whose angular dependence is assumed to be a linear combination of predetermined maps, such as spherical harmonics or coarse pixels. We find that the GWB monopole is always statistically correlated with these maps, implying a loss of sensitivity to the monopole when searching simultaneously for anisotropies. We then derive the angular distributions of the GWB intensity to which a PTA is most sensitive, and illustrate how one may use these ``principal maps" to approximately reconstruct the angular dependence of the GWB. Since the principal maps are neither perfectly anisotropic nor uncorrelated with the monopole, we also develop a frequentist criterion to specifically search for anisotropies in the GWB without any prior knowledge about their angular distribution. Lastly, we show how to recover existing EPTA results with our Fisher formalism, and clarify their meaning. The tools presented here will be valuable in guiding and optimizing the computationally demanding analyses of pulsar timing data.
\end{abstract}

\maketitle

\section{Introduction}

Pulsar timing arrays (PTAs; see e.g.~\cite{Burke-Spolaor:2018bvk, Mingarelli2019} for recent reviews) are poised to detect the nanohertz stochastic gravitational wave background (GWB) produced by the collection of inspiralling super-massive black hole binaries (SMBHB) within the next few years \cite{Taylor+:2016, Kelley:2017lek}. A detection of the GWB will provide a wealth of information about empirical scaling relations between galaxies and SMBH masses, and shed light on the dynamical processes that affect the evolution of SMBHBs after galaxy mergers (e.g., Refs.~\cite{Sesana13, Simon:2016ibt,2017NatAs...1..886M}). In addition, PTAs are sensitive to a range of low-frequency GWBs produced by more exotic but no less interesting processes, such as cosmic string networks, non-standard inflation, or phase transitions in the early Universe (see, e.g., Refs.~\cite{siemensetal2019, Burke-Spolaor:2018bvk, lms+16}).

Three major PTAs have been operating for over a decade: the North American Nanohertz Observatory for Gravitational Waves (NANOGrav \footnote{\url{http://nanograv.org}}), the European Pulsar Timing Array (EPTA \footnote{\url{http://www.epta.eu.org}}), and the Parkes Pulsar Timing Array (PPTA \footnote{\url{https://www.atnf.csiro.au/research/pulsar/ppta/}}). They moreover join effort within the International Pulsar Timing Array (IPTA \footnote{\url{http://ipta4gw.org}}). These PTA collaborations have been setting increasingly stringent limits to the GWB amplitude, and refining their analysis methods as systematics are better understood. In the near future, the Square Kilometer Array (SKA) \footnote{\url{https://www.skatelescope.org}} is expected to detect hundreds of new low-noise millisecond pulsars \cite{Keane_15}, which will allow the construction a highly sensitive PTA \cite{Janssen_15}. With such large datasets, it is critical to have simple but robust and versatile forecasting tools, in order to understand the sensitivity of a PTA without having to run computationally-demanding simulations. In a previous paper \cite{Ali-Haimoud:2020ozu} (hereafter, Paper I), we laid down the bases of a Fisher formalism to assess the sensitivity of a PTA to a GWB with generic frequency, angular, and polarization dependence. In the present work, we explore in detail the angular dependence aspect, using realistic PTA noise properties.

Most PTA data analyses assume that the GWB is isotropic \cite{Arzoumanian:2020vkk,Arzoumanian_18, Lentati_15, Shannon_15, Verbiest_16}. Although this is expected to be a good approximation, some amount of anisotropy should arise from the finite number of loud, nearby SMBHBs, as well as the non-uniform distribution of cosmological sources tracing large-scale structure \cite{Mingarelli:2013dsa, 2017NatAs...1..886M}. Other sources of nanohertz GWs such as cosmic string networks \cite{Blanco-Pillado:2017oxo, Olmez:2011cg, Jenkins_18} or domain walls \cite{Liu_20} may also produce an anisotropic GWB. A handful of studies have looked into characterizing the sensitivity of PTAs to GWB anisotropies. Refs.~\cite{Mingarelli:2013dsa,Gair:2014rwa, Hotinli_19} computed the response of pulsar timing residual correlations to an anisotropic GWB, parametrized by its spherical-harmonic expansion. Refs.~\cite{Taylor:2013esa, Taylor_20} developed a Bayesian analysis pipeline and tested it against simulated GWBs. Lastly, Ref.~\cite{Taylor:2015udp} (hereafter TMG15) conducted a search for GWB anisotropies by performing Markov Chain Monte Carlo (MCMC) analyses of EPTA data. In order to keep computational costs manageable, they limited their analysis to a subset of 6 EPTA pulsars determined to be the most sensitive for \emph{continuous} gravitational-wave searches \cite{Babak:2015lua}. TMG15 found this dataset was not informative enough to constrain anisotropies, as their upper limits saturate the physical prior which imposes a positive GWB intensity. 

The main thrust of this paper is to build on the Fisher formalism derived in Paper I, in order to assess the information content of a real PTA, and in particular its sensitivity to the angular distribution of the GWB intensity. We start by laying out the basic formalism in Section \ref{sec:formalism}, where we derive the weak-signal-limit Fisher matrix for a GWB with a fixed frequency dependence, but arbitrary angular dependence. To construct this Fisher matrix for a specific PTA requires an accurate characterization of the noise properties of each pulsar. To provide a specific application of our approach as well as make contact with the analysis in TMG15, we focus on the EPTA in this paper. In Section \ref{sec:dataset} we describe how we infer the characteristic noise strains of the EPTA pulsars using publicly available timing data, and the public code \texttt{Hasasia} \cite{Hazboun:2019nqt,Hazboun:2019vhv}. As a first application of the Fisher formalism, we show how one can rank-order pulsar pairs -- rather than individual pulsars -- by their contribution to the signal-to-noise ratio (SNR) of an isotropic GWB. In particular, in Fig.~\ref{fig:SNR2_cumul} we identify the best 44 EPTA pulsar pairs (out of 861 available) that contribute 90\% of the SNR$^2$.

Furthermore we show how to use the Fisher matrix to forecast the detectability of GWB anisotropies using standard bases in Section \ref{sec:known-dependence}. Specifically, we first consider a toy model of a stochastic ``hot spot" in a known direction on top of an isotropic background, then forecast the sensitivity of the EPTA to the GWB amplitude in a finite number of coarse pixels, as well as to its spherical-harmonic coefficients. Importantly, we show that a PTA with $N_{\rm pair}$ pulsar pairs can constrain the projections of the GWB on at most $N_{\rm pair}$ independent basis functions. For generic choices of bases (e.g.~coarse pixels or spherical harmonics), these projections are, in general, statistically correlated with one another and with the amplitude of the GWB monopole. This implies that the sensitivity to each projection and to the monopole is systematically degraded as one expands the set of basis functions -- e.g.~by increasing the cutoff $\ell_{\max}$ in a spherical-harmonic expansion. In other words, the sensitivity to various GWB anisotropies strongly depends on the priors.

In Section \ref{sec:PCA}, we explicitly derive the $N_{\rm pair}$ statistically independent functions of direction that a PTA is most sensitive to. These ``principal maps" are simply the eigenmaps of the Fisher matrix, and provide a model-independent and PTA-specific basis on which to decompose the GWB. They allow one to {\it search under the lamppost} provided by a given PTA. As an example application, we discuss how one may use principal maps to attempt to reconstruct the angular dependence of the GWB. We find, however, that a useful reconstruction likely requires a very large amplitude of the GWB, at least for the EPTA data we used. In Section \ref{sec:frequentist}, we propose an alternative approach to search for unknown anisotropies, in a model-independent fashion, using a frequentist criterion. 

The hurried reader may want to skip ahead to Section \ref{sec:Taylor}, where we apply our formalism to the subset of 6 EPTA pulsars used in TMG15. We derive sensitivity estimates for the low-order spherical-harmonic amplitudes in good agreement with their upper limits. We also clarify the meaning of their pixel-by-pixel upper-limit map, as well as their limits on high-order spherical harmonic amplitudes, for which the number of independent parameters is larger than $N_{\rm pair}$. We prove that the parameters they constrain are not the pixel-by-pixel or spherical-harmonic amplitudes of the true GWB, but rather of its projection on the $N_{\rm pair}$-dimensional space of sky maps observable by the PTA. As a consequence, the derived limits get worse with more pulsars, and cannot be interpreted as true upper limits. We sharply demonstrate our point by closely reproducing their results using the Fisher matrix and its principal maps. This comparison with existing results ought to provide a clear demonstration of the usefulness of our Fisher formalism to forecasting the sensitivity of real PTAs to anisotropies. 

We discuss our results and possible extensions in Section \ref{sec:discussion}. While we apply our formalism to the EPTA in this paper, we emphasize that our approach can be easily applied to any other PTA given the pulsar timing and parameter files. To this end, we will make a version of the \texttt{Python} script used in this paper public. 

\section{Basic conventions and formalism} \label{sec:formalism}

\subsection{Map notation and dot product in map space}

Throughout this paper, we denote by $\ho$ a direction in the sky (i.e.~a unit-length vector). We refer to functions $\bs{M}: \ho \mapsto M(\ho)$ as ``maps" and denote them with bold symbols when referring to the function itself. We denote by $\bs{1}: \ho \mapsto 1$ the monopole map. Lastly, we define the dot product of two maps $\bs{M}_1, \bs{M}_2$ as follows:
\beq
\bs{M}_1 \cdot \bs{M}_2 \equiv \int \frac{d^2 \ho}{4 \pi} M_1(\ho) M_2(\ho).
\eeq
With this convention, the monopole map has unit norm, and the average of a map $\bs{M}$ over directions is just 
$\bs{1} \cdot \bs{M}$.

\subsection{Stochastic gravitational-wave background}

We denote by $h_{ab}(f, \ho)$ the GW strain at frequency $f$ and with propagation direction $\ho$. We assume that it is a stationary Gaussian random field (as would be the case if it is generated by a large number of uncorrelated sources). It is therefore entirely determined by its rank-4 power spectrum, for which we give a complete geometric description in Paper I. Here we focus on the \emph{total intensity} map $\bsm{I}(f)$ of the GWB, i.e.~we assume that the strain power spectrum takes the form
\barr
\langle h_{ab}(f, \ho) h_{cd}^*(f', \ho') \rangle &=& \frac{\delta_{\rm D}(\ho', \ho)}{4 \pi} \frac{\delta_{\rm D}(f'-f)}{2} \nonumber\\
&& \times \mathcal{I}(f, \ho) \mathfrak{I}_{abcd}(\ho),  \label{eq:h-pow-spec}
\earr
where the purely geometric (frequency-independent) rank-4 tensor $ \mathfrak{I}_{abcd}(\ho)$ is given by
\barr
\mathfrak{I}_{abcd}(\ho) &\equiv& \delta_{ac}^{\bot \ho}\delta_{bd}^{\bot \ho} +   \delta_{ad}^{\bot \ho}\delta_{bc}^{\bot \ho} - \delta_{ab}^{\bot \ho}\delta_{cd}^{\bot \ho}, \\
\delta_{ab}^{\bot \ho} &\equiv& \delta_{ab} - \ho_a \ho_b.
\earr
We refer the reader to Paper I for the correspondence of our frame-independent expressions with the frame-dependent notation usually adopted in the literature, as well as the generalization of Eq.~\eqref{eq:h-pow-spec} to a linearly and circularly polarized GWB. 

The GWB intensity is related to the characteristic strain $h_c(f)$ through
\beq
\bs{1} \cdot \bsm{I} = \int \frac{d^2 \ho}{4 \pi} \mathcal{I}(f, \ho) = \frac12 \frac{h_c^2(f)}{f}. \label{eq:I-hc}
\eeq
Equivalently, we may write 
\beq
\mathcal{I}(f, \ho) = \frac12 \frac{h_c^2(f)}{f} P(f, \ho), \label{eq:I-hc-P}
\eeq
where $\bs{P}(f)$ is the (frequency-dependent) angular distribution of GWB intensity, with unit angle average.

\subsection{Pulsar timing residuals correlations}

Consider a pulsar $p$, in the direction $\hp$. The times of arrival (TOAs) of its pulses are a combination of a deterministic term due to intrinsic processes (such as the spin-down of the pulsar or variations in the position of the pulsar on the sky), a pulsar-specific intrinsic noise term $N_p$, uncorrelated between different pulsars (due to measurement noise, variations of the dispersion measure, as well as any intrinsic stochastic processes, such as pulsar `jitter'), and additional stochastic contributions correlated among pulsars. In this paper, we assume that the latter are exclusively generated by the GWB, and neglect additional sources of correlated noise, such as global clock and ephemeris errors \cite{Tiburzi16, Arzoumanian_18, Vallisneri20}. After fitting for a deterministic timing model, the pulsar time residual $R_p$ is thus  
\begin{equation}
    R_p = N_p + R_p^{\rm GW}. 
\end{equation}

We define the (one-sided) cross-power spectrum $\mathcal{R}_{pq}(f)$ of the total time residual at pulsars $p, q$ as follows:
\beq
\langle R_p(f) R_q^*(f') \rangle = \frac{\delta_{\rm D}(f'-f)}{2} \mathcal{R}_{p q}(f), \label{eq:R_pq-def}
\eeq
and similarly define $\mathcal{R}_{p q}^{\rm GW}(f)$ as the cross-power spectrum of the GW-induced timing residuals $R_p^{\rm GW}$. The cross-power spectrum of the intrinsic noise is defined as 
\beq
\langle N_p(f) N_q^*(f') \rangle = \frac{\delta_{\rm D}(f'-f)}{2} \delta_{pq} \sigma_p^2(f).
\eeq
In Paper I we derived the following simple compact form for the linear relation between $\mathcal{R}_{pq}^{\rm GW}(f)$ and $\bsm{I}(f)$:
\beq
\mathcal{R}_{pq}^{\rm GW}(f) = \frac{1 + \delta_{pq}}{(4 \pi f)^2}~ \bs{\gamma}_{\hp \hq}\cdot \bsm{I}(f),\label{eq:I-to-R}
\eeq
where the \emph{pairwise timing response function}, $\bs{\gamma}_{\hp \hq}$, is the map with values
\barr
\gamma_{\hp \hq}(\ho) &\equiv& 2\frac{\left( \hat p \cdot \hat q - (\hat p \cdot \ho) (\hat q \cdot \ho)\right)^2}{(1+ \hat p \cdot \ho)(1 + \hat q \cdot \ho)} \nonumber\\
&&  -  (1 - \hat p\cdot \ho)(1 - \hat q\cdot \ho). ~~ 
\earr
Using Eq.~\eqref{eq:I-hc-P}, we may rewrite Eq.~\eqref{eq:I-to-R} as
\barr
\mathcal{R}_{pq}^{\rm GW}(f) = \Gamma_{pq}(f) \frac{h_c^2(f)}{12 \pi^2 f^3},\label{eq:ORF} 
\earr
where
\beq
\Gamma_{pq}(f) \equiv \frac38 (1 + \delta_{pq}) \bs{\gamma}_{\hp \hq} \cdot \bs{P}(f)
\eeq
is the frequency-dependent ``overlap reduction function". 

If the GWB is a pure monopole, the timing residual cross correlation is is proportional to the Hellings and Downs function \cite{Hellings:1983fr}, and only only depends on the dot product $\hp \cdot \hq$ through the function
\beq
\mathcal{H}(\hp \cdot \hq) \equiv \bs{\gamma}_{\hp \hq} \cdot \bs{1}. \label{eq:HD-def}
\eeq
Specifically, this function is given by, for $\mu \equiv \hp \cdot \hq$,
\beq
\mathcal{H}(\mu) \equiv \frac{3 + \mu}{3} + 2 (1 - \mu) \ln\left(\frac{1 - \mu}{2} \right). \label{eq:HD}
\eeq
Note that we do not impose any normalization on the Hellings and Downs function $\mathcal{H}(\mu)$: it is simply defined as the response to an isotropic GWB intensity with unit amplitude, through Eq.~\eqref{eq:HD-def}. In particular, $\mathcal{H}(1) = 4/3$.

\subsection{Gaussian likelihood for the GWB intensity}

In what follows we use capital indices to label unique pairs of distinct pulsars, e.g.~$I \equiv (p, q) = (q, p)$ for $p \neq q$. We then define $\mathcal{R}_I \equiv \mathcal{R}_{pq}$, $\gamma_{I} \equiv \gamma_{\hp \hq}$, and $\mathcal{H}_I \equiv \mathcal{H}(\hp \cdot \hq)$.

Let us define $\widehat{\mathcal{R}}_{I}(f)$ to be unbiased estimators of the timing residual cross-power spectra at a given frequency $f$. Assuming that the only source of correlated timing residuals for distinct pulsars is the GWB, these estimators have mean value $\langle \widehat{\mathcal{R}}_I \rangle = \mathcal{R}_I^{\rm GW}$. In practice, one cannot directly construct finite-variance estimators of the cross-power spectra, but only of finite bandpowers, from which one may then extract $\widehat{\mathcal{R}}_{I}(f)$ for sufficiently narrow bands, see Paper I for details. Mathematically, the covariance matrix of the estimators $\mathcal{C}_{IJ} \equiv \textrm{cov}\left(\widehat{\mathcal{R}}_{I}(f), \widehat{\mathcal{R}}_{J}(f)\right)$ is inversely proportional to the bandwidth $\Delta f$ used to estimate the bandpowers: we denote it as 
\beq
\mathcal{C}_{IJ} = \frac1{\Delta f} \mathfrak{C}_{IJ},
\eeq
where both $\mathcal{C}_{IJ}$ and $\mathfrak{C}_{IJ}$ depend on frequency. 

Estimators of $\mathcal{R}_I$ at frequencies separated by more than $\sim 1/T$ are uncorrelated, where $T$ is the characteristic observation time. If the estimators are constructed from a sufficiently large number of uncorrelated data samples, we may approximate their probability distribution $\mathcal{P}$ as the product of uncorrelated multivariate Gaussians:
\barr
&&\mathcal{P}\left(\left\{\widehat{\mathcal{R}}_I \right\} \Big{|} \bsm{I} \right) \propto \prod_{f} \exp\Bigg{[}- \frac12 \Delta f \sum_{IJ} \left(\widehat{\mathcal{R}}_I(f) - \mathcal{R}_I^{\rm GW}(f) \right) \nonumber\\
&& ~~~~~~~~~~~~~~~~~~~~~~~~~~~~~~\mathfrak{C}^{-1}_{IJ}  \left(\widehat{\mathcal{R}}_J(f) - \mathcal{R}_J^{\rm GW}(f) \right) \Bigg{]}.\nonumber\\
&&\approx \exp\Bigg{[}- \frac12 \int df  \sum_{IJ} \left(\widehat{\mathcal{R}}_I(f) - \mathcal{R}_I^{\rm GW}(f) \right) \nonumber\\
&&~~~~~~~~~~~~~~~~~~~~\mathfrak{C}^{-1}_{IJ}  \left(\widehat{\mathcal{R}}_J(f) - \mathcal{R}_J^{\rm GW}(f) \right) \Bigg{]},
\earr
where the second approximation holds provided the frequency bands are narrow, $\Delta f \ll f$. 

As is standard in Bayesian data analysis, one may interpret this probability distribution as the \emph{likelihood} $\mathcal{L}$ of the signal $\bsm{I}(f)$ given the data (assuming flat priors):
\beq
\mathcal{L}\left(\bsm{I} \right) \propto \mathcal{P}\left(\left\{\widehat{\mathcal{R}}_I \right\} \Big{|} \bsm{I} \right). \label{eq:Likelihood}
\eeq
It is useful to write this probability explicitly as a Gaussian distribution for $\bsm{I}$. To do so, we first define the maps $\bs{\gamma}^*_I$, as the unique linear combinations of the $N_{\rm pair}$ pairwise timing response functions that satisfy $\bs{\gamma}^*_I \cdot \bs{\gamma}_J = \delta_{IJ}$. These are the \emph{dual maps} (not to be confused with complex conjugates) of the pairwise timing response functions. We then define the following estimator of the intensity map: 
\beq
\widehat{\bsm{I}}(f) \equiv (4 \pi f)^2 \sum_I \widehat{\mathcal{R}}_I(f) \bs{\gamma}^*_I. \label{eq:hat-I}
\eeq
We may then rewrite the likelihood \eqref{eq:Likelihood} in the form
\barr
&&\mathcal{L}\left(\bsm{I} \right)  \propto \nonumber\\
&&\exp \left[ - \frac12 \int df \left(\bsm{I}(f) - \widehat{\bsm{I}}(f)\right)\cdot \bsm{G} \cdot \left(\bsm{I}(f) - \widehat{\bsm{I}}(f)\right)\right],~~~ \label{eq:L(I)}
\earr
where $\bsm{G}(f, \ho, \ho')$ is a continuous inverse covariance ``matrix", given by
\beq
\bsm{G}(f, \ho, \ho') = \frac1{(4 \pi f)^4} \sum_{I,J} \gamma_I(\ho) \mathfrak{C}^{-1}_{IJ} \gamma_J(\ho').
\eeq

\subsection{Weak-signal limit}

In Paper I we derived the weak-signal limit of $\mathfrak{C}_{IJ}$:
\beq
\mathfrak{C}_{IJ} \approx \frac1{2 T_{pq}}\frac{\sigma_p^2(f) \sigma_q^2(f)}{\mathcal{T}_p(f) \mathcal{T}_q(f)} \delta_{IJ}, \ \ \ \ \ I \equiv (p,q),
\eeq
where $\mathcal{T}_p(f)$ is the transmission function, which accounts for the loss of information when fitting a deterministic timing model \cite{Hazboun_19}, and $T_{pq} \equiv \min(T_p, T_q)$ is the minimum observation time of the pulsar pair $(p, q)$ (or more generally, the time of overlap between the two pulsar observations). This limit holds if the GWB-induced timing residual is subdominant to the intrinsic timing noise for each pulsar, and at every frequency, i.e.~if
\beq
\mathcal{I}(f) \ll (4 \pi f)^2 \sigma_p^2(f),  \ \ \ \forall p, \ \ \  \forall f.
\eeq
In what follows we adopt the compact notation of Ref.~\cite{Hazboun_19} and define the characteristic noise strain for pulsar $p$ as
\beq
h_{c, p}^2(f) \equiv \frac34 \frac{(4 \pi f)^2 f \sigma_p^2(f)}{\mathcal{T}_p(f)}.
\eeq
The factor of 3/4 arises from the Hellings and Downs normalization convention in Ref.~\cite{Hazboun_19}; we chose to use exactly the same definition for $h_{c, p}$ as in that paper to avoid any confusion. With this notation, we find the following expression for the weak-limit intensity inverse-covariance:
\barr
\bsm{G}(f, \ho, \ho') \approx  \sum_I \mathcal{G}_I(f) \gamma_I(\ho) \gamma_I(\ho'),
\earr
where, for $I = (p, q)$, we have defined
\beq
\mathcal{G}_I(f) \equiv \frac{9}{8} \frac{f^2  T_{pq}}{h_{c,p}^2(f) h_{c, q}^2(f)}.
\eeq

\subsection{Application to a GWB with a power-law frequency dependence}

It is of course easier to search for signals with a specific frequency dependence. In the remainder of this paper, we will specialize to a GWB whose frequency and direction dependence factorize, and with a power-law frequency dependence, of the form
\beq
\mathcal{I}(f, \ho) = \frac{\mathcal{A}(\ho)}{2f} \left(f/f_{\rm yr}\right)^{-2\alpha}, \ \ \ \ \alpha  > 0,  \label{eq:I-A}
\eeq
where $f_{\rm yr} \equiv (1~\textrm{year})^{-1}$. With this convention, and as can be seen from Eq.~\eqref{eq:I-hc} the characteristic strain takes the form $h_c(f) = A_h (f/f_{\rm yr})^{-\alpha}$, where $A_h^2 \equiv \bs{1} \cdot \bsm{A}$ is the angle average of $\bsm{A}$.

While we keep our expressions general, for numerical applications we will specialize to a power law $\alpha = 2/3$. Indeed, this is what is expected from a population of SMBHBs on circular orbits, shrinking exclusively due to GW emission \cite{Phinney:2001di}. The true frequency dependence may differ from this simple form, especially at low-frequencies if the binaries are eccentric or if environmental effects such as stellar hardening and/or gas torques are driving the binaries to merge instead of GWs~\cite{Sesana13, Chen_17, Burke-Spolaor:2018bvk}.
Note that our formalism can readily be applied to an arbitrary frequency dependence.

We now want to rewrite Eq.~\eqref{eq:L(I)} as a likelihood for $\bsm{A}$. To do so, we define the dimensionless Fisher ``matrix"
\barr
\bsm{F}(\ho, \ho') &\equiv& \sum_I \mathcal{F}_I ~\gamma_I(\ho) \gamma_I(\ho'), \label{eq:Fisher-A}\\
\mathcal{F}_I &\equiv& \int df ~ \frac{(f/f_{\rm yr})^{-4\alpha}}{4 f^2}~\mathcal{G}_I(f),
\earr
as well as the inverse-variance-weighted estimator amplitude map
\barr
\bsmhat{A} &\equiv& \sum_I \widehat{\mathcal{A}}_I ~\bs{\gamma}_I^*, \label{eq:hat-A}\\
\widehat{\mathcal{A}}_I &\equiv& \frac1{\mathcal{F}_I} \int df  ~ \frac{(f/f_{\rm yr})^{-2\alpha}}{2 f}~\mathcal{G}_I(f) (4 \pi f)^2 \widehat{\mathcal{R}}_I(f).~ \label{eq:hat-AI}
\earr
The likelihood of the GWB amplitude $\bsm{A}$ is then proportional to
\beq
\mathcal{L}(\bsm{A}) \propto \exp\left[- \frac12 (\bsm{A} - \bsmhat{A}) \cdot \bsm{F} \cdot(\bsm{A} - {\bsmhat{A}})\right]. \label{eq:P(A)}
\eeq
Equations \eqref{eq:Fisher-A} and \eqref{eq:P(A)} are fundamental to all the forthcoming calculations in this work.

The total SNR$^2$ for an arbitrary GWB with amplitude $\bsm{A}$ is given by 
\beq
\textrm{SNR}^2 = \bsm{A} \cdot \bsm{F} \cdot \bsm{A} = \sum_I \mathcal{F}_I \left[ \bs{\gamma}_I \cdot \bsm{A}\right]^2.\label{eq:SNR2}
\eeq
We see that for each pulsar pair, the coefficient $\mathcal{F}_I$ characterizes its overall sensitivity to a power-law GWB, and the pairwise-timing response function $\bs{\gamma}_I$ characterizes its angular response. 

Before moving on, let us remark that, given $N_{\rm pair}$ distinct pulsar pairs, there are at most $N_{\rm pair}$ independent detectable components of the GWB. In particular, a PTA is completely blind to the projection of the GWB perpendicular to all the pairwise timing response functions. Specifically, given an amplitude $\bsm{A}$, we may always decompose it as a piece $\bsm{A}_{||}$ which is a linear combination of the $\bs{\gamma}_{I}$'s, and a piece $\bsm{A}_{\bot}$ orthogonal to all of them; the latter is completely undetectable by the PTA. The estimator $\bsmhat{A}$ generalizes the ``optimal statistic" $\widehat{A}_{\rm GWB}^2$ \cite{Anholm:2008wy, Arzoumanian_18} to an anisotropic GWB. It is an unbiased estimator of the \emph{observable} piece of the signal $\bsm{A}_{||}$, but of course not of the full map $\bsm{A}$. We can therefore equally write $\bsmhat{A} \equiv \bsmhat{A}_{||}$.

\subsection{Contribution of anisotropies to timing residual autocorrelations} \label{sec:auto-corr}

The GWB also contributes to each pulsar's timing residual autocorrelation (or power spectrum). These autocorrelations differ from the cross correlations in an essential aspect: one \emph{does} expect a non-zero autocorrelation even in the absence of a GWB, due to intrinsic noise. In the analysis of real data, one must therefore search simultaneously for an appropriately parametrized intrinsic noise spectrum, for each pulsar, jointly with a GWB-induced noise common to all pulsars (a ``common red noise"). Because of the unknown intrinsic noise, one therefore cannot hope to ever achieve a \emph{detection} of the GWB with autocorrelations alone \cite{Siemens:2013zla}. Still, autocorrelations do contain information about the GWB: at the very least, they can be used to set upper limits on its amplitude.

Here we expand on a point that we made in Paper I: a pulsar's autocorrelation term is sensitive to the GWB anisotropy. In other words, GWB anisotropies contribute a common red noise. While this fact can be readily inferred from the expressions for the overlap reduction functions given in Ref.~\citep{Mingarelli:2013dsa}, it is unclear whether this important point has been accounted for in previous searches for anisotropic GWB. Moreover, here we show that only the GWB monopole, dipole and quadrupole contribute to pulsars' autocorrelations.

From Eqs.~\eqref{eq:I-to-R} and \eqref{eq:I-A}, the GWB contribution to the single-pulsar timing residual power spectrum is proportional to $\bs{\gamma}_{\hp \hp} \cdot \bsm{A}$. As we saw in Paper I, the single-pulsar timing response function $\gamma_{\hp \hp}(\ho)$ is a linear combination of a monopole, dipole projected along $\hp$, and quadrupole twice projected onto $\hp$:
\beq
\gamma_{\hp \hp}(\ho) = \frac43- 2 \hp \cdot \ho + \hp^i \hp^j  \left(\ho^i \ho^j - \frac13 \delta^{ij} \right).
\eeq
Let us define the following geometric decomposition of the GWB dimensionless amplitude:
\barr
&&\mathcal{A}(\ho) = \mathcal{A}_0  + 3\sum_{i=1}^3 \mathcal{A}_1^i \ho^i \nonumber\\
&& ~~~~ + \frac{15}2 \sum_{i,j = 1}^3 \mathcal{A}_2^{ij} \left(\ho^i \ho^j - \frac13 \delta^{ij} \right) + \Delta \mathcal{A}(\ho), ~~
\earr
where the tensor $\mathcal{A}_2^{ij}$ is symmetric and trace-free (thus has 5 independent components), and $\Delta \bsm{A}$ only contains contributions that are octupolar and higher order. The components $\mathcal{A}_0, \mathcal{A}_1^i$ and $\mathcal{A}_2^{ij}$ are uniquely defined through
\barr
\mathcal{A}_0 \equiv \bsm{A} \cdot \bs{1}, \ \ \ \ \  \mathcal{A}_1^i \equiv \int \frac{d^2 \ho}{4 \pi} \mathcal{A}(\ho) \ho^i, \\
\mathcal{A}_2^{ij} \equiv \int \frac{d^2 \ho}{4 \pi} \mathcal{A}(\ho) \left(\ho^i \ho^j - \frac13 \delta^{ij} \right).
\earr
They are, respectively, the monopole, dipole and quadrupole components of the GWB. With this decomposition, we obtain
\barr
\bs{\gamma}_{\hp \hp} \cdot \bsm{A} =  \frac43 \mathcal{A}_0  - 2 \sum_{i = 1}^3 \hp^i \mathcal{A}_1^i + \sum_{i,j = 1}^3 \hp^i \hp^j \mathcal{A}_2^{ij}.
\earr
Thus, as noted in Paper I, the auto-correlated power spectra constrain specific combinations of the GWB monopole, dipole and quadrupole. Importantly, when allowing for anisotropies, one must self-consistently propagate their impact on the single-pulsar noises, i.e.~one must account for ``common red processes" which arise from the dipole and quadrupole pieces of the GWB, in addition to its monopole. If autocorrelations are used to set upper limits on a common red process, their dependence on dipole and quadrupole GWB anisotropies systematically degrades the inferred upper limits on the monopole.

One of our main objectives here is to explore the ability of PTAs to \emph{detect} the GWB. While autocorrelations do contain information, they cannot be used to claim a detection \cite{Siemens:2013zla}. As a consequence we focus on the correlations between different pulsars in the remainder of this paper.

\section{Dataset used in this paper} \label{sec:dataset}

\begin{figure}
    \centering
    \includegraphics[width = \columnwidth]{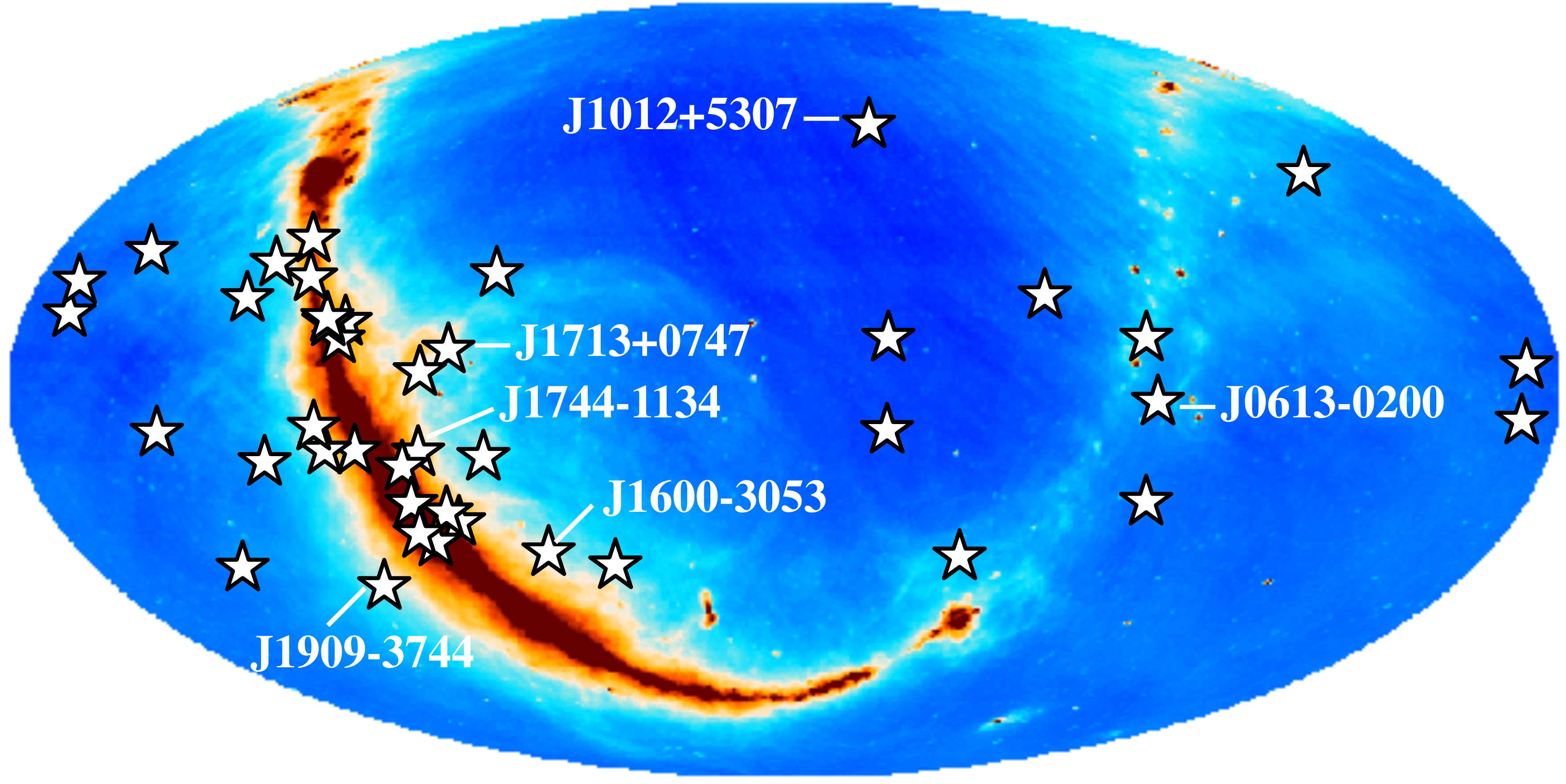}
    \caption{Locations of the EPTA pulsars, in equatorial coordinates. We use the full EPTA dataset in Sections \ref{sec:dataset}-\ref{sec:frequentist}. The subset of 6 highlighted pulsars is used when compare our results to those of Ref.~\cite{Taylor:2015udp} in Section \ref{sec:Taylor}. The background map is the Stockert and Villa-Elisa 1.4 GHz continuum map \cite{Reich_86, Reich_01}, which situates the positions of the pulsars relative to the Galactic plane.} 
    \label{fig:epta_loc}
\end{figure}

Throughout most of this paper, to provide a concrete example, we will apply our formalism to the full EPTA dataset described in the last data release \cite{Desvignes_16}. The EPTA is composed of the high-precision timing of 42 millisecond pulsars for up to 17 years with an overlap of 21 pulsars with the NANOGrav Nine-Year dataset \cite{2015ApJ...813...65N}. We show the locations of the EPTA pulsars relative to the Galactic plane in Fig.~\ref{fig:epta_loc}. 

Out of the 42 EPTA pulsars, six were determined to contribute 90\% of the SNR$^2$ in simulated \emph{continuous} GW searches \cite{Babak:2015lua}, and were used for searches of anisotropies in the GWB in Ref.~\cite{Taylor:2015udp} (hereafter, TMG15). Specifically, these six pulsars are J0613-0200, J1012+5307, J1600-3053, J1713+0747, J1744-1134, and J1909-3744. Their locations in the sky are shown in Fig.~\ref{fig:epta_loc}, and their characteristic noise strains in Fig.~\ref{fig:hc-epta6}. We will use this subset of the EPTA in Section \ref{sec:Taylor} to compare our sensitivity estimates to those of TMG15.

\subsection{Evaluation of the characteristic noise strains}

As described in the previous section, the primary quantity characterizing pulsar noise properties is the characteristic noise strain. Here, we explain how we estimate the characteristic noise strains of all 42 EPTA pulsars. 

We model the intrinsic white noise in the standard way:
\barr
\left[\langle \delta R_{p}(t_i) \delta R_{p}(t_j) \rangle\right]_{\rm white} &=& \Big{(} {\rm EFAC}_{p,b}~ \sigma_{p,i}^2 \nonumber\\
&& + {\rm EQUAD}_{p,b}^2 \Big{)} \delta_{ij},
\earr
where $t_i$ is the $i^{\rm th}$ time-of-arrival of pulsar $p$, $\sigma_{p,i}$ is the uncertainty in the $i^{\rm th}$ timing residual of pulsar $p$, and the white noise parameters, EFAC$_{p,b}$ and EQUAD$_{p,b}$, are included for each observing system $b$ (i.e.~different telescope and/or backends). EFAC is dimensionless and EQUAD has dimensions of time (we use units of seconds for our analysis). Our priors on these parameters were a flat-linear prior on EFAC $\in$ [0.01,10] and a flat-log prior on EQUAD $\in$ $[10^{-8.5},10^{-3}]$ sec. The effects of variations in the dispersion measure (DM) and intrinsic red-noise (RN) were modeled with a power-law spectral density of the form 
\barr
    \sigma^2_{p,\rm X}(f) =  \frac{A_{\rm X}^2}{12 \pi^2}\left(f/f_{\rm yr}\right)^{-\gamma_{\rm X}} \textrm{yr}^3, \ \ \textrm{X = RN or DM},~~~
\earr
where $A_{\rm RN}$ is independent of radio frequency $\nu$ whereas $ A_{\rm DM} \propto \nu^{-2}$. We use a flat-linear prior on the power-law indices, $\gamma_{\rm X}$ $\in$ $[0,7]$ and a flat-log prior on the dimensionless amplitudes $A_{\rm X}$ $\in$ $[10^{-20},10^{-11}]$, for both RN and DM. The prior on the power-law indices have been chosen so that they range from white noise ($\gamma_{\rm X} = 0$) to the steepest power-law for which the fit to the timing model removes any dependence on the functional form for these spectral densities at low frequencies ($f<1/T_{\rm obs}$) \cite{vanHaasteren:2012hj}. This range also covers the expected variation in the power-law index due to random walks in phase, period and period derivatives (which give $\gamma_{\rm RN} = 2, 4, 6$, respectively) \cite{Arzoumanian_18}. Note that we \emph{do not} include a ``common red noise" process in our analysis. This is consistent with our weak-signal assumption, and allows us to fit for each pulsar noise properties independently.

In order to extract values for these noise parameters we used the Parallel Tempering Markov-Chain Monte Carlo sampler \texttt{PTMCMCSampler} \cite{PTMCMCSampler}, the PTA software \texttt{Enterprise} \cite{enterprise}, and the TOA and timing model parameters used in the EPTA data release \cite{Desvignes:2016yex} from the EPTA repository \footnote{\url{http://www.epta.eu.org/aom.html}}. We set the noise parameters to equal their median values from MCMC chains that contain $10^6$ samples. We have confirmed that our noise parameters match well with the noise analysis described in \cite{Caballero:2015srj}. 

\begin{figure}
    \centering
    \includegraphics[width = \columnwidth]{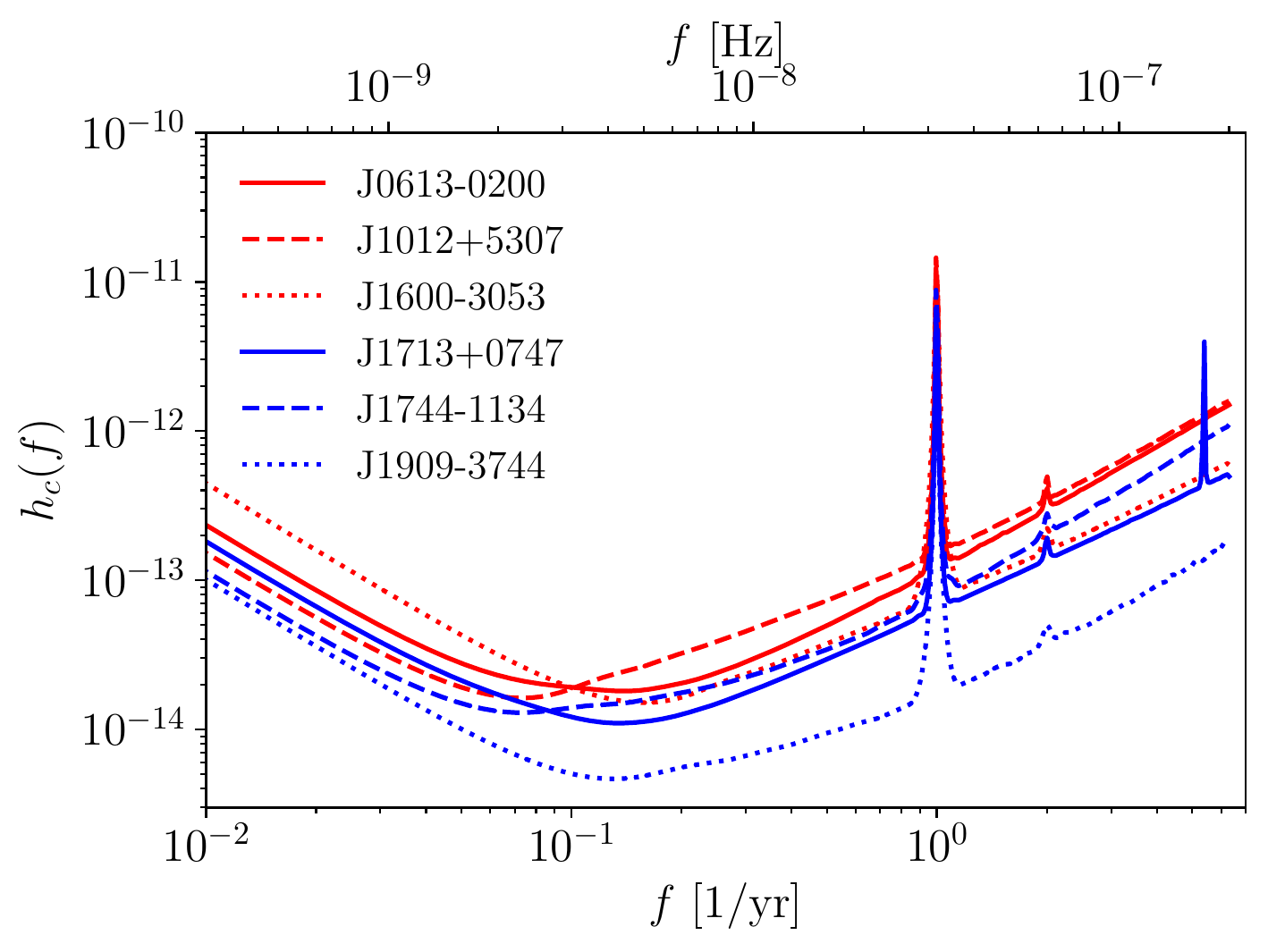}
    \caption{Characteristic noise strains for 6 of the EPTA pulsars, produced using publicly available EPTA data \cite{Desvignes:2016yex} and \texttt{Hasasia} \cite{Hazboun:2019nqt,Hazboun:2019vhv}, see main text for details. The spikes at 1/yr and 2/yr arise from degeneracies with the Earth's orbital motion when fitting the pulsar position and distance. For J1713-0747, the spike at $f \approx 5.4/$yr is at the 67.8-day period of this binary system.}
    \label{fig:hc-epta6}
\end{figure}

After we have estimated the noise parameters for each pulsar we use \texttt{Hasasia} \cite{Hazboun:2019nqt,Hazboun:2019vhv} to compute the total noise power spectrum $\sigma_p^2(f)$, transmission function $\mathcal{T}_p(f)$, and characteristic noise strain $h_{c, p}(f)$, for each one of the EPTA pulsars. As an illustration, we show the characteristic noise strains of 6 of the EPTA pulsars in Fig.~\ref{fig:hc-epta6}. We will use these specific 6 pulsars when comparing of our results with those of TMG15 in Section \ref{sec:Taylor}.

\subsection{Best pulsar pairs for isotropic GWB searches} \label{sec:best-pulsars}

In what follows we determine which pulsar pairs contribute most to the total SNR$^2$ from a cross-correlation analysis. We will see that only a small number of pulsar pairs are needed to get most of the SNR$^2$. A similar analysis was performed in Ref.~\cite{Yardley_11} for the Parkes PTA; here we consider the EPTA and provide a much simpler, analytic calculation.

As can be seen from Eq.~\eqref{eq:SNR2}, for a general GWB, each pair $I$ contributes $\mathcal{F}_I (\bs{\gamma}_I \cdot \bsm{A})^2$ to the SNR$^2$. The relative ranking of pulsar pairs thus depends on the angular dependence of the GWB, and there is no universal ranking that holds for an arbitrary GWB. Since we expect the GWB to be predominantly isotropic, it is sensible to rank pulsar pairs under this assumption.

For an isotropic GWB $\mathcal{A}(\ho) = A_h^2$, from Eq.~\eqref{eq:SNR2} we see that the SNR$^2$ is given by
\beq
\textrm{SNR}^2 = A_h^4 \sum_{I} \mathcal{F}_I ~\mathcal{H}_I^2. \label{eq:SNR2-mono}
\eeq
First, to check the soundness of our analysis, we estimate the 95\% sensitivity to a monopole, $A_h^{95\%}$, such that SNR[$A_h^{95\%}$] = 2. Using Eq.~\eqref{eq:SNR2-mono}, we find 
\beq
A_h^{95\%} = \sqrt{2}\left(\sum_{I} \mathcal{F}_I~ \mathcal{H}_I^2\right)^{-1/4}.
\eeq
Using the characteristic noise strains we computed for the EPTA dataset, we find $A_h^{95\%} \approx 2.5 \times 10^{-15}$. If we restrict ourselves to the 6 EPTA pulsars used in Refs.~\cite{Lentati_15, Taylor:2015udp}, we find instead $A_h^{95\%} \approx 3.4 \times 10^{-15}$. This lies between the 95\% upper-limit values found in these references, of $3.0 \times 10^{-15}$ \cite{Lentati_15} and $3.9 \times 10^{-15}$ \cite{Taylor:2015udp}. This gives us confidence that our characteristic noise strains provide a realistic description of the data.

\begin{figure*}[ht]
     \includegraphics[width = 2\columnwidth]{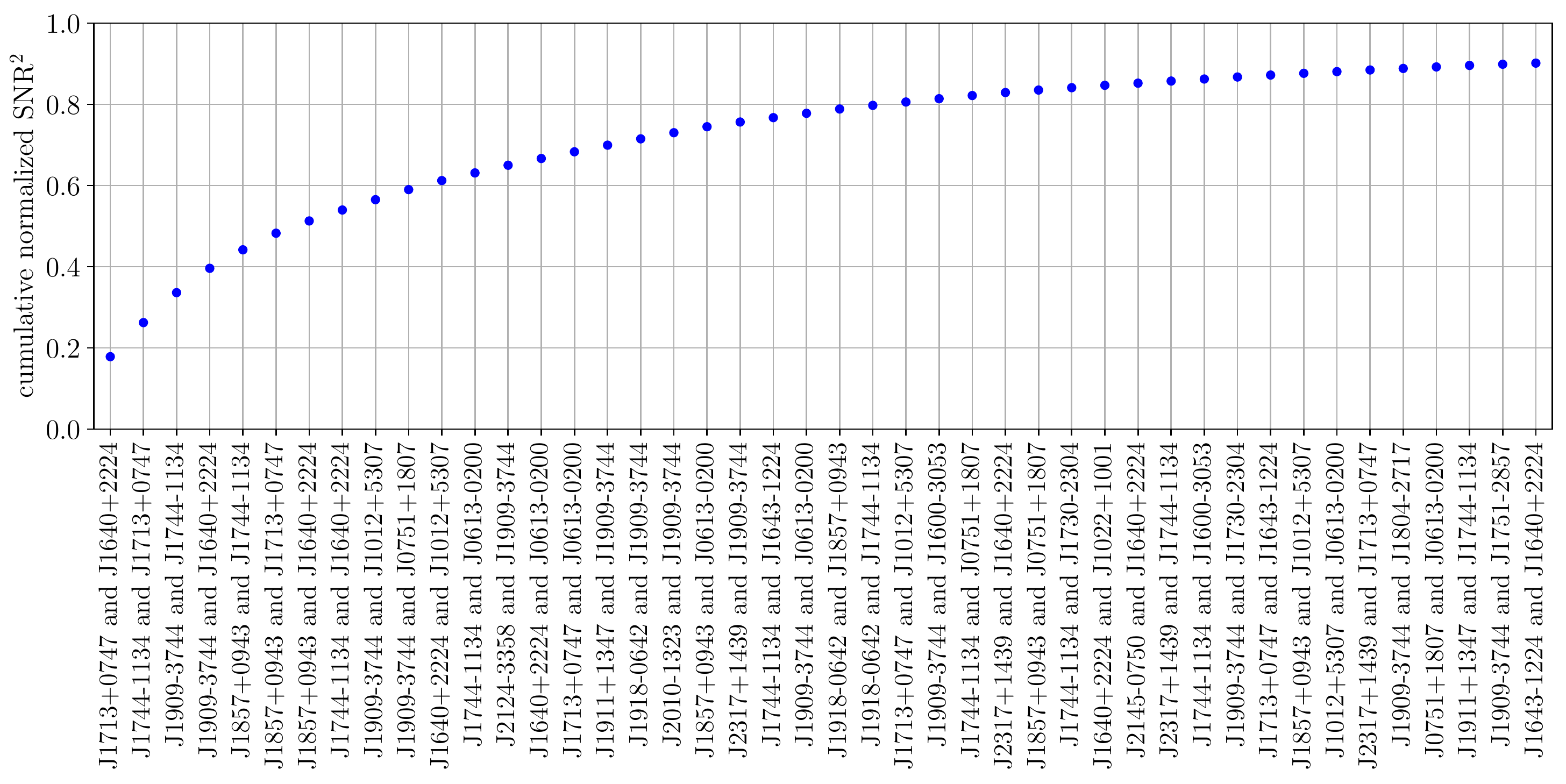}
    \caption{Normalized cumulative SNR$^2$ for an isotropic GWB with characteristic strain $h_c(f) \propto f^{-2/3}$. The total SNR$^2$ is computed with the full EPTA array of 42 pulsars \cite{Desvignes_16}, corresponding to 861 distinct pairs. This figure only shows the contributions of the best 44 pairs, labeling the $x$-axis, required to reach 90\% of the SNR$^2$.}
    \label{fig:SNR2_cumul}
\end{figure*}

From Eq.~\eqref{eq:SNR2-mono}, we see that for an isotropic GWB, each pair $I = (p, q)$ contributes $A_h^4 \mathcal{F}_I \mathcal{H}_I^2$ to the SNR$^2$. We have ranked the 861 pulsar pairs of the full EPTA dataset in terms of their contributions to the SNR$^2$. In Fig.~\ref{fig:SNR2_cumul} we show the normalized cumulative contributions of the best 44 pulsar pairs, which contributed 90\% of the SNR$^2$. While this is three times as many pairs as what can be constructed with the 6 pulsars used in Refs.~\cite{Lentati_15, Taylor:2015udp}, these 44 pairs represent 5\% of the total number of pairs in the full dataset, and should still constitute a manageable collection of data. Moreover, we find that the 6 pulsars used in Refs.~\cite{Lentati_15, Taylor:2015udp} only contribute 26\% of the total SNR$^2$ for an isotropic GWB, while the best 15 pairs would amount to 68\% of the SNR$^2$, as can be seen in Fig.~\ref{fig:SNR2_cumul}. 

This result is of significant importance to speed up future analyses of real data. It is well known that accounting for all the correlations between pulsar pairs in a full Bayesian analysis of timing residuals is computationally challenging. Here we see that it suffices to include a small, manageable number of pulsar pairs to recover most of the SNR$^2$. Our simple Fisher formalism allows us to efficiently determine which pairs to use for any given dataset. This can be done not only for an isotropic GWB, but also for any assumed angular dependence, if desired.

\section{Sensitivity to GWB anisotropies using standard bases} \label{sec:known-dependence}

\subsection{General approach} \label{sec:known-general}

Suppose that we model the GWB amplitude to be a linear combination of known maps $\bs{M}_n$: 
\beq
\bsm{A} = \sum_n \mathcal{A}_n \bs{M}_n, \label{eq:A-Mn}
\eeq
where $\mathcal{A}_n$ are scalar amplitudes. The maps $\bs{M}_n$ could be, for instance, spherical harmonics, or Dirac functions centered at specific directions in the sky. They can be, in general, any set of linearly independent maps, and need not even be orthogonal nor normalized. We can always define the dual maps $\bs{M}_n^*$ (not to be confused with complex conjugates), such that $\bs{M}_n \cdot \bs{M}_m^* = \delta_{mn}$. Given the estimator map ${\bsmhat{A}}$, we may then define $\widehat{\mathcal{A}}_n \equiv \bs{M}_n^* \cdot {\bsmhat{A}}$. The probability distribution $\mathcal{L}(\bsm{A})$ given in Eq.~\eqref{eq:P(A)} can then be interpreted as a joint probability distribution for the set of coefficients $\{\mathcal{A}_n \}$, and be recast in the form 
\barr
\mathcal{P}(\{ \mathcal{A}_n\}) &\propto& \exp\left[- \frac12 \sum_{mn} (\mathcal{A}_m - \widehat{\mathcal{A}}_m)  F_{mn} (\mathcal{A}_n - \widehat{\mathcal{A}}_n)\right],~~\\
F_{mn} &\equiv& \bs{M}_m \cdot \bsm{F} \cdot \bs{M}_n.
\earr
In other words, the coefficients $\mathcal{A}_n$ are Gaussian-distributed, with mean $\widehat{\mathcal{A}}_n$, and covariance 
\beq
\textrm{cov}(\mathcal{A}_n, \mathcal{A}_m) = (F^{-1})_{mn}. \label{eq:cov-params}
\eeq
In particular, upon marginalizing over all other parameters, the probability distribution of any single parameter $\mathcal{A}_n$ is a Gaussian with mean $\widehat{\mathcal{A}}_n$ and variance $(F^{-1})_{nn}$. This means that, to be able to claim a 95\%-confidence detection of a non-zero coefficient $\mathcal{A}_n$, the underlying true parameter needs to satisfy $\mathcal{A}_n \geq \mathcal{A}_n^{95\%}$, where
\beq
\mathcal{A}_n^{95\%} \equiv 2\sqrt{(F^{-1})_{nn}}. 
\eeq
This represents a threshold for \emph{detection}, i.e.~allows us to forecast the sensitivity of a given PTA. Note that $(F^{-1})_{nn} \geq 1/F_{nn}$, unless the map $\bs{M}_n$ is statistically uncorrelated with all other maps. Therefore the sensitivity to any given map amplitude $\mathcal{A}_n$ is always degraded if one simultaneously searches for other amplitudes. 

It is important to note that at most $N_{\rm pair}$ map amplitudes can be simultaneously searched for. Indeed, the Fisher matrix $\bsm{F}$ only has rank $N_{\rm pair}$, as can be seen by writing Eq.~\eqref{eq:Fisher-A} formally as a sum of tensor products $\bsm{F} = \sum_I \mathcal{F}_I~ \bs{\gamma}_I \otimes \bs{\gamma}_I$. Therefore, the matrix $F_{mn}$ is guaranteed to be singular if it has more than $N_{\rm pair}$ rows and columns. In other words, at most $N_{\rm pair}$ parameters can have finite variance. In practice, one can still set upper limits even with more than $N_{\rm pair}$ parameters, if additional priors are included (e.g.~physical priors, or a prior based on the auto-correlation functions of individual pulsars). But in order to be able to claim a detection, at most $N_{\rm pair}$ parameters can be simultaneously searched for. 

\subsection{A toy problem: monopole and hot spot with a known direction} \label{sec:hot-spot}

As a first application of this formalism, let us consider a simple toy problem, with only two basis maps. Suppose that external considerations lead us to expect a GWB ``hot spot" in known direction $\ho_*$, in a addition to a background monopole. Note that we are still considering a \emph{stochastic} ``hot spot", i.e.~this cannot be generated by a single binary, but rather must originate from a concentration of many binaries radiating incoherently. Both the hot spot and monopole have an unknown amplitude. 

Mathematically, we assume that the GWB amplitude takes the form
\beq
\mathcal{A}(\ho) = \mathcal{A}_0 + 4 \pi \mathcal{A}_1 \delta_{\rm D}(\ho; \ho_*). \label{eq:hot-spot}
\eeq
Note that this is a different split between monopole and hot-spot than adopted in Ref.~\cite{Hotinli_19} and Paper I; with the convention \eqref{eq:hot-spot}, the amplitude of the hot spot is unbounded from above. One can interpret $\mathcal{A}_0$ as the amplitude of a background monopole, on top of which the hot spot sits. The total monopole amplitude is then $\mathcal{A}_0 + \mathcal{A}_1$. 

In the notation of Section \ref{sec:known-general}, the maps $\bs{M}_0, \bs{M}_1$ are
\beq
\bs{M}_0 = \bs{1}, \ \ \ \  \ \bs{M}_1: \ho \mapsto 4 \pi \delta_{\rm D}(\ho; \ho_*).
\eeq
The two maps $\bs{M}_0, \bs{M}_1$ are not orthogonal, and more importantly, are in general not statistically independent, in the sense that $\bs{M}_0 \cdot \bsm{F} \cdot \bs{M}_1 \neq 0$.

\begin{figure}[ht]
    \centering
    \includegraphics[width = \columnwidth]{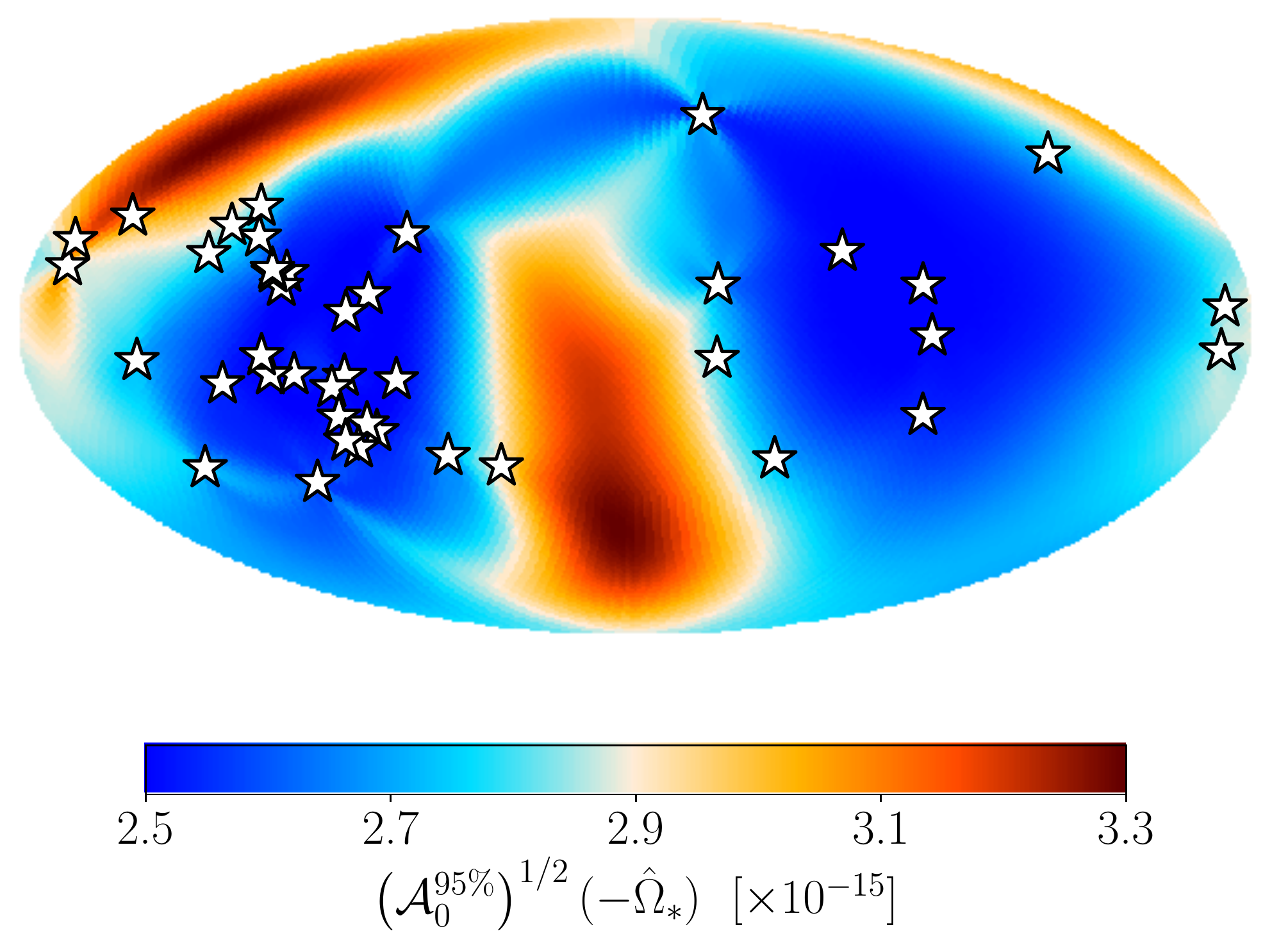}
    \includegraphics[width = \columnwidth]{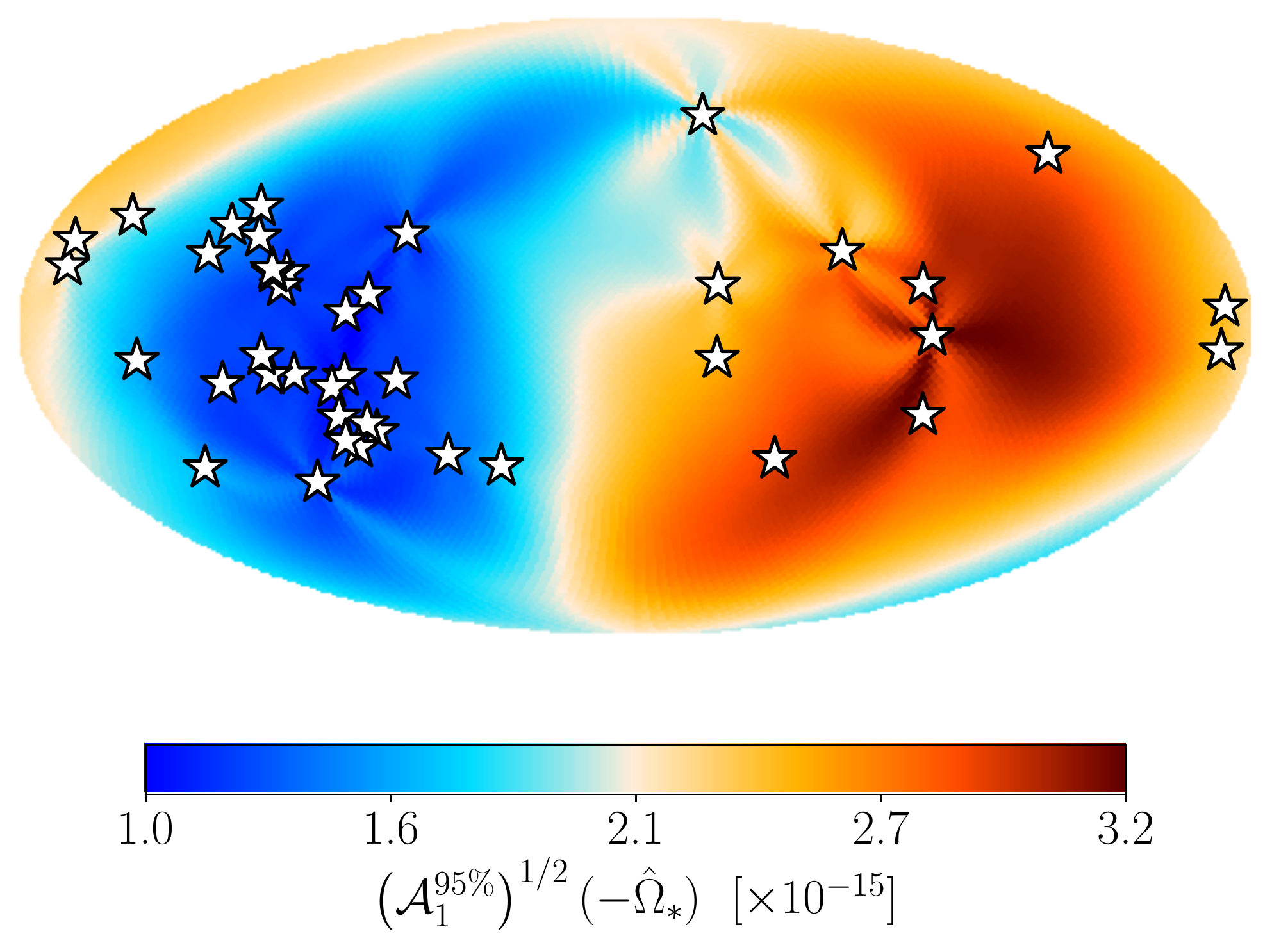}
    \caption{Minimum amplitudes of a background monopole (top) and hot spot (bottom) detectable at the 95\% confidence level by the EPTA, as a function of the hot spot direction $-\ho_*$. This assumes the GWB consists of a monopole and a single hot spot originating from a known direction.}
    \label{fig:hot_spot}
\end{figure}

Applying the formalism of Section \ref{sec:known-general}, we must first compute the 2 by 2 inverse-covariance matrix $F_{mn}$. For any given $\ho_*$, its components are
\barr
F_{00} &=& \sum_{ I } \mathcal{F}_I \mathcal{H}_I^2,\\
F_{01} &=& F_{10} = \sum_{ I} \mathcal{F}_I \mathcal{H}_I \gamma_{I}(\ho_*),\\
F_{11} &=& \sum_{ I } \mathcal{F}_I [\gamma_I(\ho_*)]^2.
\earr
The covariance matrix of the amplitudes $(\mathcal{A}_0, \mathcal{A}_1)$ is then the inverse of $F$. In Fig.~\ref{fig:hot_spot}, we show the minimum background monopole and hot-spot amplitudes detectable at the 95\% confidence level, as a function of the direction of the hot spot. We see that the sensitivity to the background monopole amplitude $\sqrt{\mathcal{A}_0^{95\%}}$ varies from $2.5 \times 10^{-15}$ to 3.3 $\times 10^{-15}$, depending on the assumed location of the hot spot. This is to be compared with the estimated EPTA sensitivity to a monopole-only GWB, $A_h^{95\%} \approx 2.5 \times 10^{-15}$. We thus see that searching for a monopole simultaneously with another map, in this case a hot spot, systematically degrades the sensitivity to the monopole relative to the case where one assumes the GWB is a pure monopole. This is due to the statistical correlation (i.e.~non-zero matrix element $F_{01}$) between the two maps.

It is important to emphasize that this analysis applies to a hot spot in a \emph{known direction}. Without prior knowledge of the hot spot direction, it is not possible to set any constraint on its amplitude, as the number of free parameters would then be infinite, thus exceeding $N_{\rm pair}$. As mentioned, this is mostly a toy problem, as in practice it is highly unlikely that one would expect only one single spot at a time. Next we consider a more realistic setup, where we simultaneously constrain the amplitudes of the GWB in multiple coarse pixels.  

\subsection{Detectability of coarsely pixelized anisotropies} \label{sec:pixelized}

As explained earlier, we may simultaneously search for at most $N_{\rm pair}$ independent amplitudes of the GWB. It is therefore not possible to simultaneously constrain the GWB in all pixels in the sky. We can, however, meaningfully analyze the amplitudes of GWB in $N_{\rm pix} \leq N_{\rm pair}$ coarse pixels. For this analysis, we will use the HEALPix pixelization \cite{Healpix} into $N_{\rm pix} = 12 N_{\rm side}^2$ pixels of equal area $4\pi/N_{\rm pix}$. We define the $N_{\rm pix}$ coarse-pixel basis functions $\bs{P}_n$ such that $P_n(\ho) = 1$ if $\ho$ is within the $n$-th pixel, and 0 otherwise, and search for coarse-grained anisotropies of the form
\beq
\bsm{A} = \sum_{n = 1}^{N_{\rm pix}} \mathcal{A}_n \bs{P}_n. \label{eq:A-pixels}
\eeq
To compute the discrete Fisher matrix coefficients $F_{nm}$, we need to first computed the dot products $\bs{\gamma}_{\hp \hq} \cdot \bs{P}_n = \frac1{N_{\rm pix}} \langle \bs{\gamma}_{\hp \hq}\rangle_n$, where $\langle \bs{M} \rangle_n$ is the average of map $\bs{M}$ in the $n$-th pixel. The latter is obtained by ``degrading" the map to a resolution of $N_{\rm pix}$.

\begin{figure*}
    \centering
    \includegraphics[width = 0.66\columnwidth]{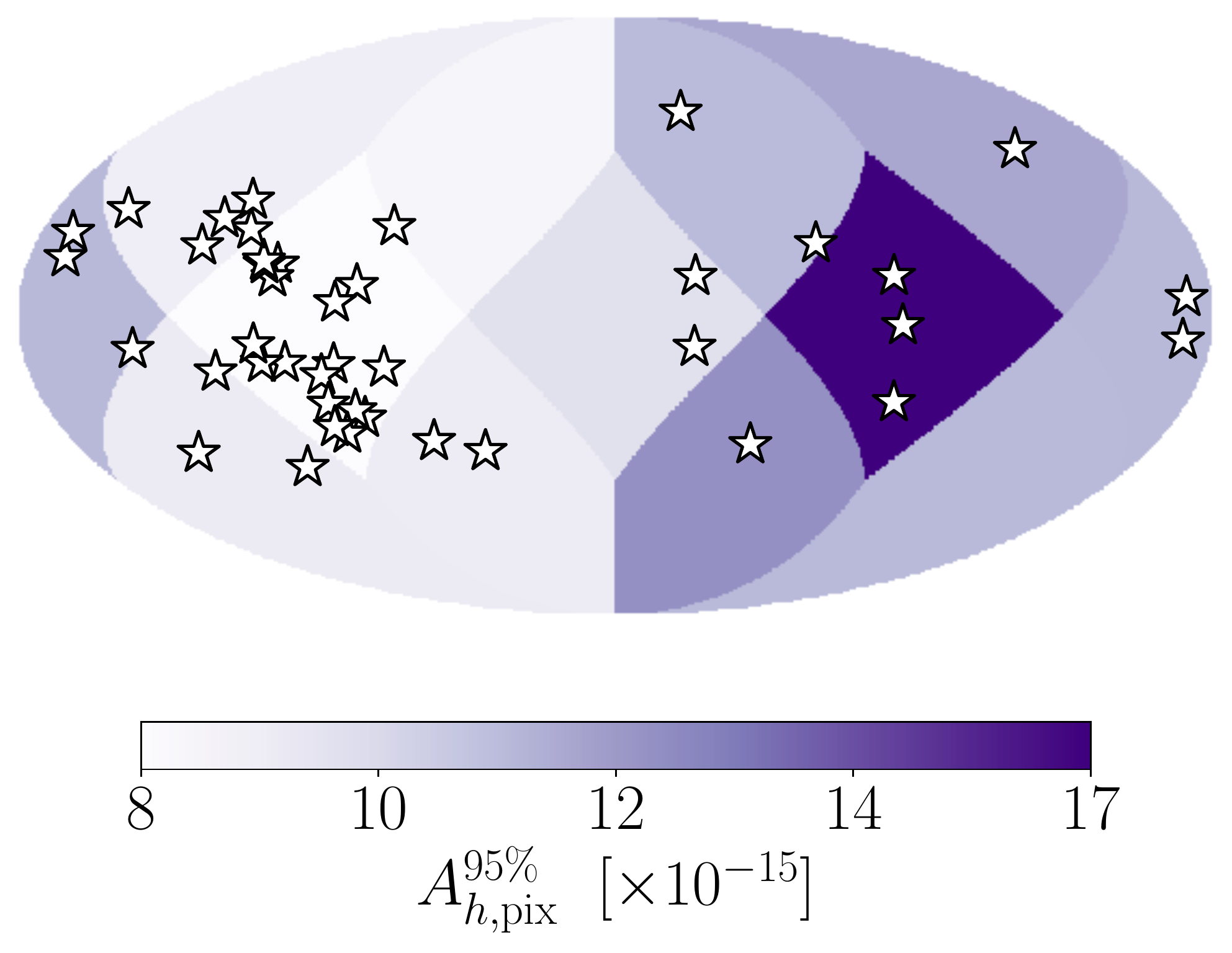}
    \includegraphics[width = 0.66\columnwidth]{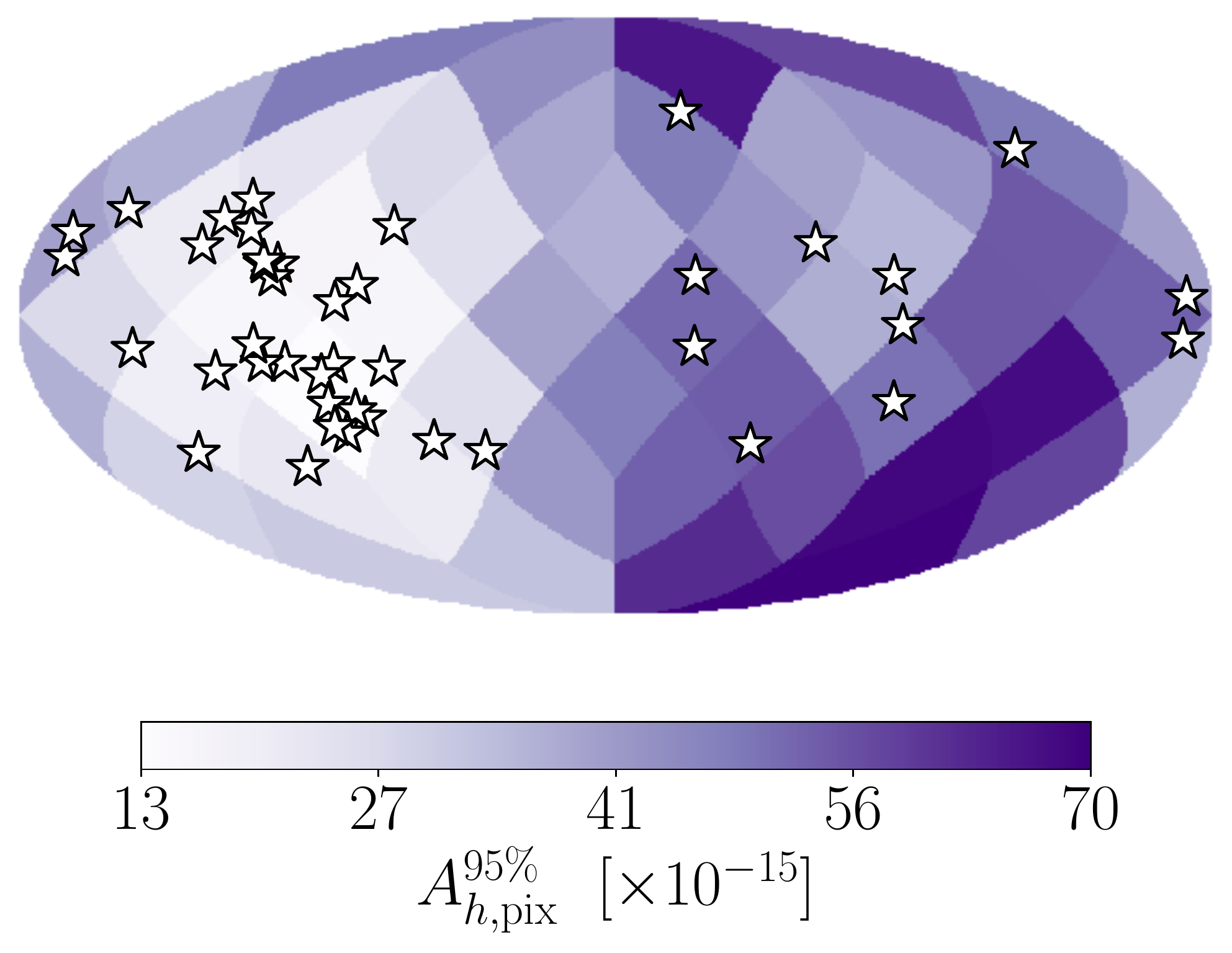}
    \includegraphics[width = 0.66\columnwidth]{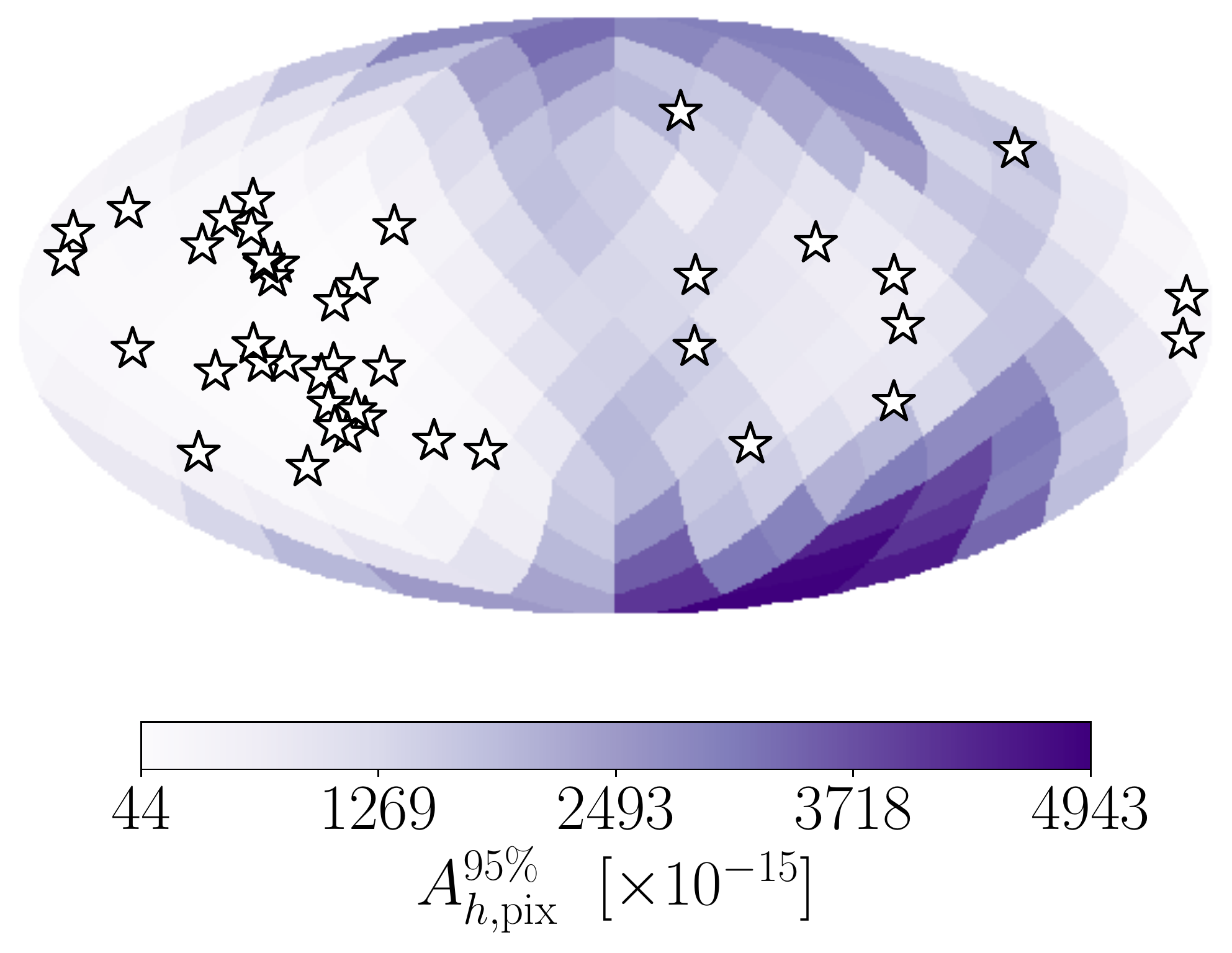}
    \caption{Sensitivity of the EPTA to a coarsely pixelized GWB, for $N_{\rm pix} = 12, 48$ and 192 HEALPix pixels. Specifically, this shows the sensitivity in each coarse pixel after marginalizing over all other pixels.}
    \label{fig:coarse-pixels}
\end{figure*}

The 861 distinct pulsar pairs afforded by the EPTA allow us to use $N_{\rm side}$ as large as 8. We show the 95\% sensitivity $A_{h, \rm pix}^{95\%} \equiv \sqrt{\mathcal{A}_n^{95\%}}$ in each coarse pixel in Fig.~\ref{fig:coarse-pixels}, for $N_{\rm side} = 1, 2, 4$, corresponding to $N_{\rm pix} = 12, 48$ and 192, respectively. For $N_{\rm side} = 8$, we found that the $768\times 768$ coarse-pixel covariance matrix is numerically ill-conditioned, thus do not consider this case. We see that the pixel-by-pixel sensitivity degrades dramatically as $N_{\rm side}$ is increased (especially in the less well-constrained parts of the sky). Note also that the lower $N_{\rm pix}$ limits cannot be simply recovered by degrading the higher $N_{\rm pix}$ limits, as the order of map-degrading and inverting the Fisher matrix cannot be interchanged. 

The average characteristic strain squared is given by
\beq
A_h^2 = \frac1{N_{\rm pix}} \sum_{n = 1}^{N_{\rm pix}} \mathcal{A}_n, \label{eq:A00-coarse}
\eeq
which has variance  
\beq
\textrm{var}\left(A_h^2\right) = \frac1{N_{\rm pix}^2} \sum_{n,m} \textrm{cov}(\mathcal{A}_n, \mathcal{A}_m),
\eeq
translating to a 95\% sensitivity $\left(A_h^{95\%}\right)^2 = 2 \sqrt{\textrm{var}\left(A_h^2\right)}$. For the EPTA, we find $A_h^{95\%} = (4.6, 5.0, 7.8) \times 10^{-15}$ for $N_{\rm side} = 1, 2, 4$, respectively. This is to be contrasted with $A_h^{95\%} = 2.5 \times 10^{-15}$ if the GWB were known to be strictly isotropic. This highlights once again that the general form assumed for the GWB strongly affects the sensitivity to the monopole.

\subsection{Detectability of spherical-harmonic amplitudes}
\label{sec:sph_harm}

\subsubsection{Setup and physical prior}

Following common practice \cite{Mingarelli:2013dsa, Taylor:2015udp}, we may decompose GWB anisotropies  on the basis of \emph{real} spherical harmonics $\mathcal{Y}_{\ell m}(\ho)$, which we normalize such that $\bs{\mathcal{Y}}_{\ell m} \cdot \bs{\mathcal{Y}}_{\ell' m'} = \delta_{\ell \ell'} \delta_{m m'}$ (i.e.~with a factor $\sqrt{4 \pi}$ larger than the standard definition of spherical harmonics). Explicitly, we assume the GWB amplitude takes the form 
\beq
\bsm{A} = \sum_{\ell = 0}^{\ell_{\max}} \sum_{m = -\ell}^{\ell} \mathcal{A}_{\ell m} ~\bsm{Y}_{\ell m}. \label{eq:A-Ylm}
\eeq
Note that this form assumes the rather stringent prior that harmonic coefficients for $\ell > \ell_{\max}$ all vanish. 

In order for $\bsm{A}$ to represent a physical GWB, it needs to be positive everywhere. To our knowledge, there is no simple analytic expression for this physical prior. Nevertheless, we can derive an approximation as follows. We define the GWB anisotropy $\Delta \mathcal{A}(\ho) \equiv \mathcal{A}(\ho) - \mathcal{A}_{00}$. It has the following variance over directions in the sky:
\barr
\langle [\Delta \bsm{A}]^2 \rangle_{\ho} &\equiv& \int \frac{d^2 \ho}{4 \pi} [\Delta \mathcal{A}(\ho)]^2 = \Delta \bsm{A} \cdot \Delta \bsm{A} \nonumber\\
&=& \sum_{\ell = 1}^{\ell_{\max}} \sum_{m = -\ell}^{\ell} \mathcal{A}_{\ell m}^2 =  \sum_{\ell = 1}^{\ell_{\max}} (2 \ell +1) \mathcal{C}_{\ell},
\earr
where we have defined 
\beq
\mathcal{C}_{\ell} \equiv \frac1{2 \ell +1} \sum_{m = -\ell}^{\ell} \mathcal{A}_{\ell m}^2. \label{eq:Cl-def}
\eeq
A simple approximate physical prior follows: we require that the monopole amplitude $\mathcal{A}_{00}$ is larger than twice the rms anisotropy, implying
\beq
\mathcal{C}_0 \geq 4 \sum_{\ell = 1}^{\ell_{\max}} (2 \ell +1) \mathcal{C}_{\ell}.
\eeq
This implies a conservative physical prior
\beq
\mathcal{C}_{\ell} \leq \frac{\mathcal{C}_{0}}{4 (2 \ell +1)}, \ \ \ \ell \geq 1.
\eeq
This simple criterion reproduces quite accurately the numerical results of TMG15 (see inset in their Fig.~1). Our conventions map to those of TMG15 through $\mathcal{A}_{\ell m} = A_h^2 c_{\ell m}/\sqrt{4 \pi}$, $\mathcal{C}_\ell = A_h^4 C_l/4 \pi$. Note that the monopole coefficient is just the amplitude of the characteristic strain squared, $\mathcal{A}_{00} = A_h^2$, and that $\mathcal{C}_0 = \mathcal{A}_{00}^2 = A_h^4$.

\subsubsection{Forecasted sensitivity of the EPTA}

Let us start by pointing out that for a given maximum $\ell = \ell_{\max}$, the decomposition \eqref{eq:A-Ylm} includes $(\ell_{\max} + 1)^2$ independent coefficients. It thus follows that one may constrain the spherical-harmonic amplitudes from pulsar pair cross correlations only if $\ell_{\max} \leq \sqrt{N_{\rm pair}} -1$. For larger values of $\ell_{\max}$, the $\mathcal{A}_{\ell m}$ are then only constrained by the physical prior. For the full EPTA, we may therefore use $\ell_{\max} \leq 28$. For the subset of 6 EPTA pulsars, one is restricted to $\ell_{\max} \leq 2$.

We compute the covariance of the amplitudes $\mathcal{A}_{\ell m}$ as described in Section \ref{sec:known-general}. Instead of showing the sensitivity to the individual $\mathcal{A}_{\ell m}$ coefficients, we estimate the detectability of the rotationally-invariant coefficients $\mathcal{C}_{\ell}$ defined in Eq.~\eqref{eq:Cl-def}. Their mean and covariance matrix are given by 
\barr
\langle \mathcal{C}_{\ell} \rangle &=& \frac1{2 \ell +1} \sum_{m = -\ell}^{\ell} \textrm{var}( \mathcal{A}_{\ell m}), \label{eq:mean-Cl}\\
\textrm{cov}(\mathcal{C}_{\ell}, \mathcal{C}_{\ell'}) &=& 2 \sum_{m= -\ell}^{\ell} \sum_{m '= -\ell'}^{\ell'} \frac{\textrm{cov}(\mathcal{A}_{\ell m} \mathcal{A}_{\ell' m'})^2}{(2 \ell +1)(2 \ell'+1)}. \label{eq:cov-Cl}
\earr
The $\mathcal{C}_{\ell}$ are not Gaussian-distributed (especially for small $\ell$). Nevertheless, we may define an \emph{approximate} 95\% sensitivity estimate as follows:
\beq
\mathcal{C}_{\ell}^{95\%} \equiv \langle \mathcal{C}_{\ell} \rangle + 2 \sqrt{\textrm{var}(\mathcal{C}_{\ell})}. \label{eq:Cl95}
\eeq
Note that even for $\ell = 0$, this gives $\mathcal{C}_{0}^{95\%} = (1 + 2 \sqrt{2}) \textrm{var}(\mathcal{A}_{00}) \approx 3.8 ~\textrm{var}(\mathcal{A}_{00})$, which is very close to the correct 2-$\sigma$ sensitivity $\mathcal{C}_{0}^{95\%} = 4~\textrm{var}(\mathcal{A}_{00})$.

\begin{figure}
    \centering
    \includegraphics[width = \columnwidth]{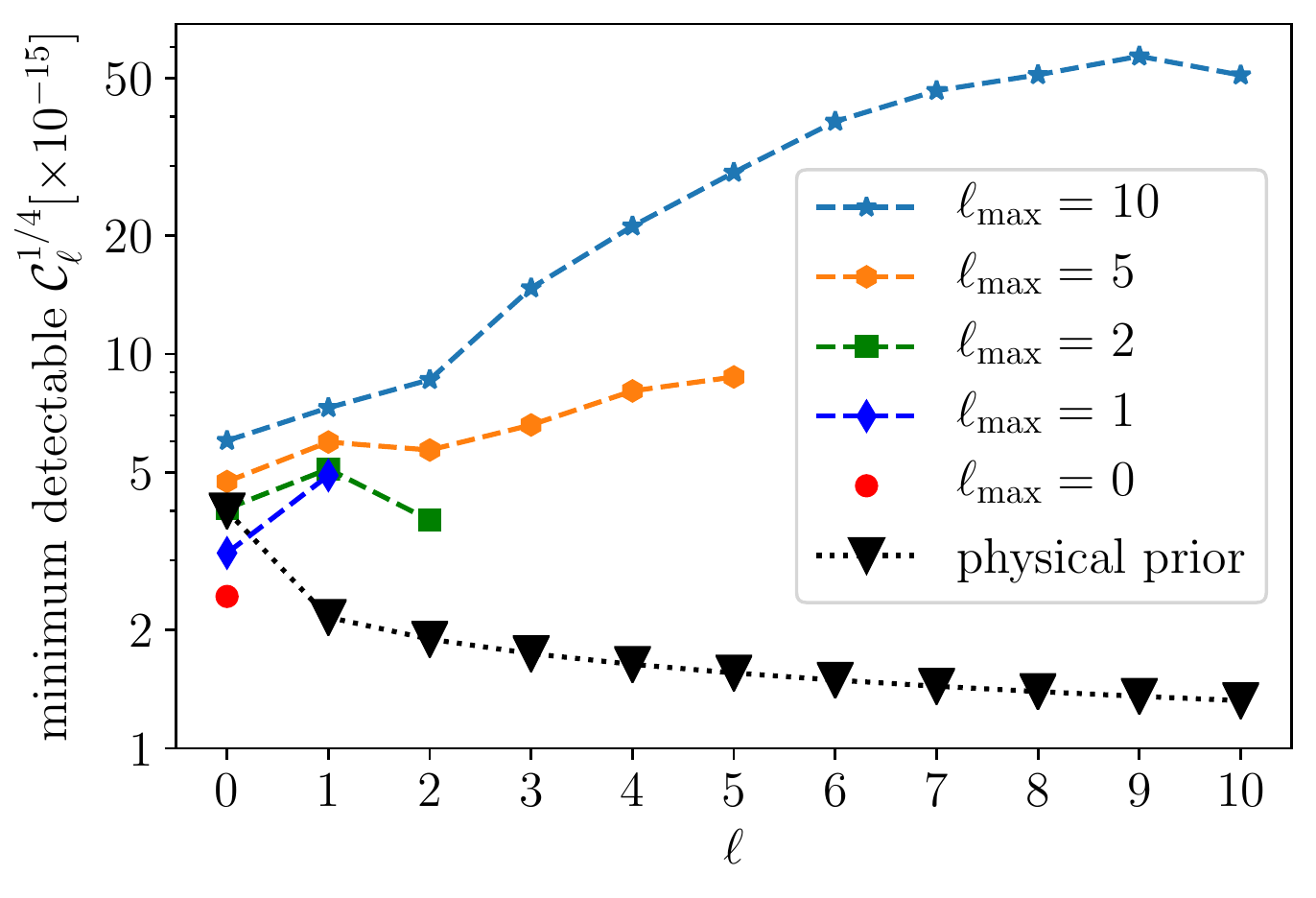}
    \caption{Approximate 95\% sensitivity of the full EPTA to the coefficients $\mathcal{C}_{\ell} \equiv \sum_m \mathcal{A}_{\ell m}^2/(2 \ell +1)$, as a function of the cutoff $\ell_{\max}$, beyond which the coefficients are assumed to strictly vanish. The physical prior is computed for a monopole upper limit $A_h \leq 4 \times 10^{-15}$ \cite{Taylor:2015udp}, and is systematically lower than the minimum detectable anisotropy amplitudes.}
    \label{fig:epta_Cl}
\end{figure}

We show $\mathcal{C}_{\ell}^{95\%}$ in Fig.~\ref{fig:epta_Cl}, for several values of $\ell_{\max}$. We see that for any given coefficient $\mathcal{C}_{\ell}$, the sensitivity systematically degrades as $\ell_{\max}$ is increased. In particular, as anticipated, the sensitivity to the monopole significantly worsens as $\ell_{\max}$ is increased. This is because the monopole is correlated with other spherical harmonics, and as one enlarges the space of functions to be searched over, the uncertainty on the monopole amplitude increases. 

Let us remark that the coefficients $\mathcal{C}_{\ell}$ are statistically correlated. We explicitly give their correlation coefficients in Tabs.~\ref{tab:corr-Cl} for $\ell_{\max} = 2$ and 5. We see that these correlations coefficients are in general not small, and depend on the chosen $\ell_{\max}$.

Finally, note that this analysis specifically estimates the minimum amplitudes necessary for a \emph{detection}, through pulsar timing cross correlations. In addition, the monopole, dipole and quadrupole can be constrained by pulsar autocorrelations, as discussed in Section \ref{sec:auto-corr}. In the limit that autocorrelations constrain the $\ell = 0, 1, 2$ harmonic coefficients much more tightly than the cross correlations, the variance of the remaining coefficients should be obtained by inverting the Fisher matrix restricted to these coefficients, leading to a lower noise for the $\mathcal{C}_{\ell}$ with $\ell \geq 3$. We have checked explicitly that this lowers the noise by no more than $\sim 10-20\%$.

In conclusion, we have demonstrated that our Fisher formalism allows to forecast the sensitivity of a PTA to the spherical-harmonic amplitudes of the GWB, for a given cutoff $\ell_{\max}$. We have applied this specifically for the EPTA, and shown that the minimum detectable amplitudes are systematically larger than allowed by the physical prior, given current upper limits on the monopole. This seems to indicate that spherical harmonics are a suboptimal choice of basis for anisotropy searches. We will specifically compare our results with those of TMG15 in Section \ref{sec:Taylor}.

\begin{table}
    \centering
    {\renewcommand{\arraystretch}{1.2}
    \begin{tabular}{|*{5}{P{0.7cm}|}} 
    \multicolumn{4}{c}{$\ell_{\max} = 2$}\\
    \hline
        & $\mathcal{C}_{0}$ & $\mathcal{C}_{1}$ & $\mathcal{C}_{2}$ \\
      \hline
     $\mathcal{C}_{0}$ & 1 & 0.28& 0.33 \\ 
$\mathcal{C}_{1}$&...& 1 & 0.12 \\ 
$\mathcal{C}_{2}$&...&...& 1  \\ 
      \hline
    \end{tabular}}
    \\[10pt]
    {\renewcommand{\arraystretch}{1.2}
    \begin{tabular}{|*{10}{P{0.7cm}|}} 
        \multicolumn{7}{c}{$\ell_{\max} = 5$}\\
    \hline
       & $\mathcal{C}_0$ & $\mathcal{C}_{1}$ & $\mathcal{C}_{2}$ & $\mathcal{C}_3$ & $\mathcal{C}_4$ & $\mathcal{C}_5$\\
      \hline
        $\mathcal{C}_0$ & 1 & 0.43& 0.45& 0.15& 0.06& 0.03 \\ 
        $\mathcal{C}_1$ &...& 1 & 0.33& 0.20& 0.07& 0.05 \\ 
        $\mathcal{C}_2$ &...&...& 1 & 0.46& 0.20& 0.17 \\ 
        $\mathcal{C}_3$ &...&...&...& 1 & 0.39& 0.27 \\ 
        $\mathcal{C}_4$ &...&...&...&...& 1 & 0.52 \\ 
        $\mathcal{C}_5$ &...&...&...&...&...& 1  \\ 
      \hline
    \end{tabular}}
 \caption{Dimensionless correlation coefficients $\textrm{cov}(\mathcal{C}_\ell,\mathcal{C}_{\ell'})/\sqrt{\textrm{var}(\mathcal{C}_{\ell}) \textrm{var}(\mathcal{C}_{\ell'})}$, for $\ell_{\max} = 2$ and $\ell_{\max} = 5$, using the EPTA pulsars. The ellipses indicate that the tables are symmetric. This illustrates that the $\mathcal{C}_{\ell}$'s are statistically correlated and that for a given pair $\ell, \ell'$, the correlation coefficient depends on the assumed $\ell_{\max}$.}
    \label{tab:corr-Cl}
\end{table}

\section{Principal maps of a PTA} \label{sec:PCA}

\subsection{Motivation and formalism}

As highlighted in the previous section, one of the challenges when using standard bases to decompose the GWB intensity is that the amplitudes of the basis maps are statistically correlated. Moreover, their covariance matrix depends on the number of maps considered. As a result, one cannot easily set model-independent limits on the map amplitudes. 

If one is completely agnostic regarding the angular dependence of the GWB, it is best to search under the lamppost, i.e.~look for the amplitudes of maps which are best-measured by a given PTA, and uncorrelated with one another. To do so, we construct the $N_{\rm pair}$ unit-norm \emph{principal maps} $\{\bsm{M}_n\}$. They are defined to extremize SNR$^2 = \bsm{M}_n \cdot \bsm{F} \cdot \bsm{M}_n$, under the normalization constraint $\bsm{M}_n \cdot \bsm{M}_n = 1$. The solutions of this constrained optimization problem are simply the eigenmaps of the Fisher matrix, i.e.~maps that satisfy
\beq
\bsm{F}\cdot \bs{\mathcal{M}}_n = \frac1{\Sigma_n^2} \bs{\mathcal{M}}_n, \label{eq:eigval}
\eeq
where we denoted the corresponding eigenvalues by $1/\Sigma_n^2$. Since $\bsm{F}$ is symmetric, its eigenmaps are orthogonal to one another and, importantly, uncorrelated with one another:
\beq
\bsm{M}_n \cdot \bsm{M}_m = \delta_{mn}, \ \ \ \ \ \ \ \bsm{M}_n \cdot \bsm{F} \cdot \bsm{M}_m = \frac{\delta_{mn}}{\Sigma_n^2}.
\eeq
We may rewrite the Fisher matrix as a direct sum of tensor products of the principal maps:
\beq
\bsm{F} = \sum_n \frac1{\Sigma_n^2} \bsm{M}_n \otimes \bsm{M}_n.
\eeq

Any GWB amplitude $\bsm{A}(\ho)$ can always be written as a linear combination of the $N_{\rm pair}$ eigenmaps and a piece orthogonal to all of them:
\barr
\bsm{A} &=& \bsm{A}_{||} + \bsm{A}_{\bot}, \\
\bsm{A}_{||} &\equiv& \sum_{n = 1}^{N_{\rm pair}} \mathcal{A}_n \bsm{M}_n, \ \ \ \ \ \ \ \ \bsm{A}_{\bot} \cdot \bsm{M}_n = 0 \ \ \ \ \forall~ n.
\earr
In that case, the total SNR$^2$ is given by 
\beq    
{\rm SNR}^2[\bsm{A}(\ho)] = \sum_{n=1}^{N_{\rm pair}} \frac{\mathcal{A}_n^2}{\Sigma_n^2}. \label{eq:SNR2-eigmaps}
\eeq
From Eq.~\eqref{eq:SNR2-eigmaps}, we see that the minimum amplitude of the eigenmap $\bsm{M}_n$ detectable with SNR = 2 is $\mathcal{A}_n = 2 \Sigma_n$.

The eigenmaps for an orthonormal basis of the $N_{\rm pair}$-dimensional space of \emph{observable} maps for a given PTA. Maps $\bsm{A}_{\bot}$ which are orthogonal to this space are completely unobservable, i.e.~have infinite noise. 

To compute the eigenmaps in practice, we first write them in the form of linear combinations of the pairwise timing response functions:
\beq
\mathcal{M}_n(\ho) = \sum_I \mathcal{M}_{n}^I ~\sqrt{\mathcal{F}_I} ~ \gamma_I(\ho).
\eeq
The eigenvalue problem \eqref{eq:eigval} is then equivalent to the finite-dimensional eigenvalue-problem
\beq
\sum_J F_{IJ} \mathcal{M}_{n}^J = \frac1{\Sigma_n^2} \mathcal{M}_n^I,  
\eeq
where the symmetric $N_{\rm pair} \times N_{\rm pair}$ matrix $F_{IJ}$ has elements
\beq
F_{IJ} \equiv \sqrt{\mathcal{F}_I \mathcal{F}_J} ~\bs{\gamma}_I \cdot \bs{\gamma}_J.
\eeq

\begin{figure*}
    \centering
    \includegraphics[width = 2\columnwidth]{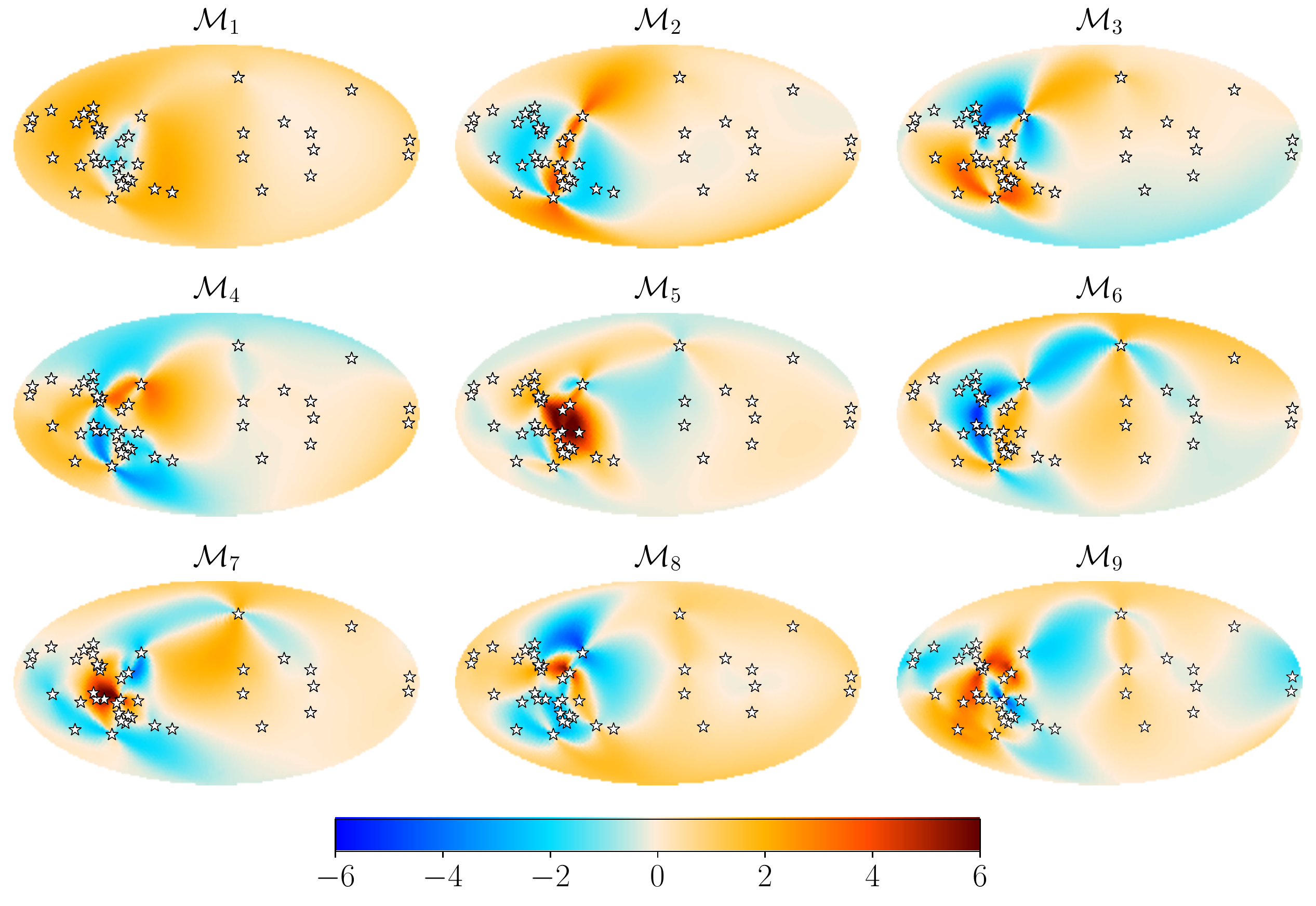}
    \caption{First 9 principal maps of the EPTA, ordered by increasing noise. The maps have unit norm and their sign was chosen so that their average value (i.e. projection on the monopole) is positive, i.e. $\bsm{M}_n \cdot \bs{1} > 0$.}
    \label{fig:epta_PM}
\end{figure*}

\begin{figure}
    \centering
    \includegraphics[width = \columnwidth]{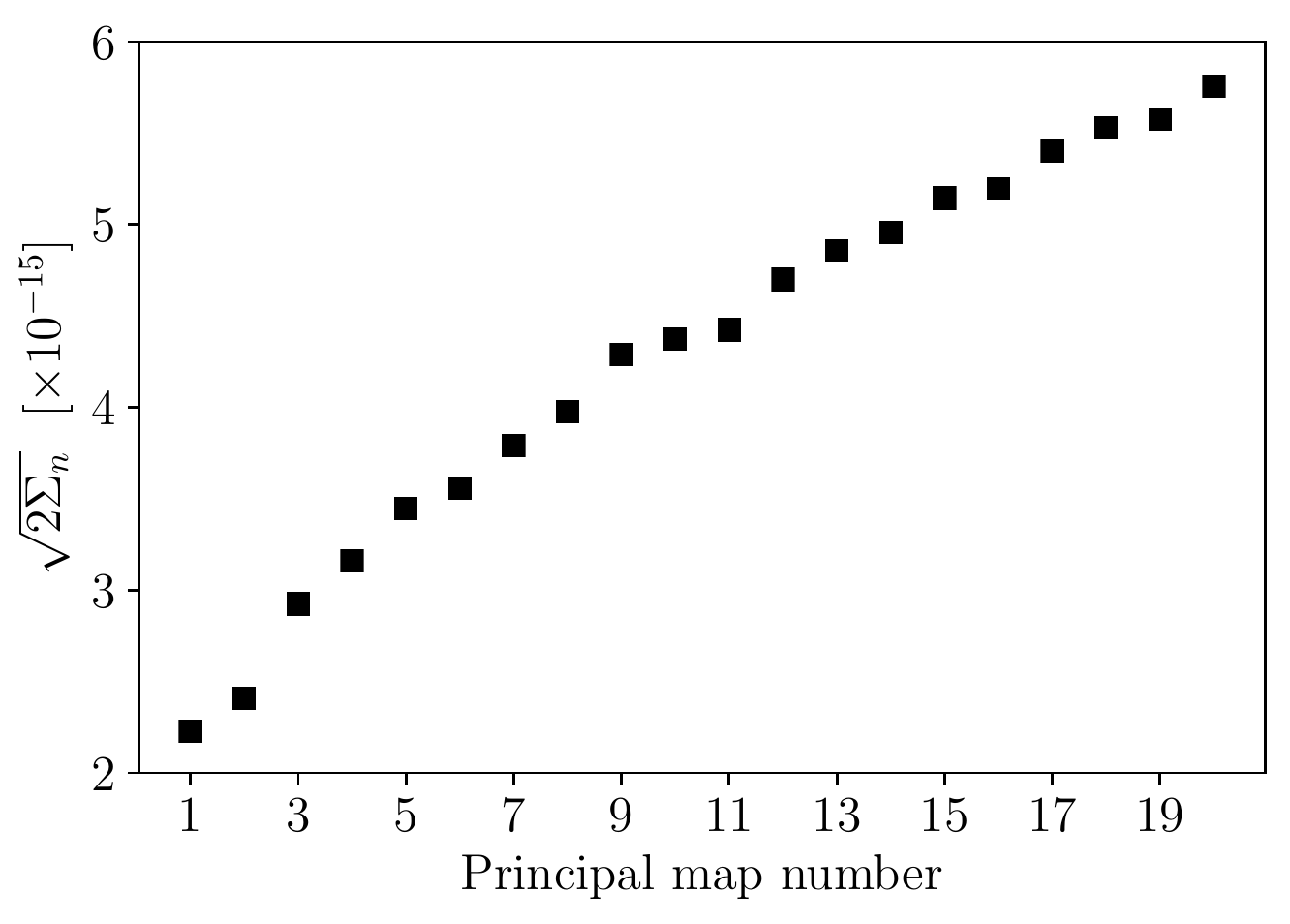}
    \caption{Noise eigenvalues of the first 20 principal maps of the EPTA. Specifically, the principal maps are normalized to unity, and here we show the minimum amplitude of the coefficients multiplying them (with dimensions of characteristic strain) required for a 95\%-confidence detection.}
    \label{fig:epta_eigval}
\end{figure}

We show the first 9 eigenmaps of the EPTA in Fig.~\ref{fig:epta_PM}. These maps look very different from any known analytic basis functions (such as spherical harmonics): indeed, they reflect the properties of a particular PTA. We show the eigenvalues (translated into a detectability threshold in characteristic strain) of the first 20 eigenmaps in Fig.~\ref{fig:epta_eigval}. Note that beyond the first $\sim 20$ eigenmaps, the noise eigenvalues $\Sigma_n$ start increasing exponentially with $n$.

One can use the principal maps as a basis to decompose the observable GWB, and follow the same formalism as in Section \ref{sec:known-dependence}. The advantage of using this basis is that the map amplitudes are statistically uncorrelated. As a consequence, upper limits obtained on any given map amplitude do not depend on the cutoff. In particular, one could start by searching for the amplitude of the first principal map alone, then search simultaneously for the amplitudes of the first two maps, etc..., and not change any of the sensitivity estimates.

\subsection{Application: map reconstruction} \label{sec:reconstruction}

Several map-making methods for the GWB have been proposed, from a potential high-frequency GWB in the LIGO-band \cite{Allen:1996gp, Cornish:2001hg,Mitra:2007mc, Thrane:2009fp, Renzini:2018}, to millihertz LISA band \citep{Cornish:2001hg,Cornish:2002bh,Kudoh:2004he,Seto:2004np,Taruya:2005yf,Kudoh:2005as,Taruya:2006kqa}, to the nanohertz PTA band \cite{Anholm:2008wy,Mingarelli:2013dsa,Taylor:2013esa,Gair:2014rwa,Hotinli_19}. There are also studies exploring methods to extract maps of continuous gravitational waves \cite{Cornish:2014rva,Romano:2015uma}. Here we propose a new reconstruction method for the GWB, relying on the principal maps. 

In principle, given $N_{\rm pair}$ estimated timing residual cross-spectra $\widehat{\mathcal{R}}_I(f)$, one can build an estimate of the underlying GWB intensity from Eqs.~\eqref{eq:hat-A}-\eqref{eq:hat-AI}. However, in the weak-signal limit, all the estimators $\widehat{\mathcal{A}}_I$ are individually noise-dominated. Collectively, however, a PTA may still measure some independent pieces of information with high signal-to-noise ratio: these are precisely the amplitudes of the lowest-noise principal maps, $\widehat{\mathcal{A}}_n\equiv \bsm{M}_n \cdot {\bsmhat{A}}$. Concretely, using this expression with Eq.~\eqref{eq:hat-AI}, we see that the $\widehat{\mathcal{A}}_n$ are linear combinations of the timing residual cross-spectra $\widehat{\mathcal{R}}_I(f)$, appropriately integrated over frequencies.

We can then use the principal maps to attempt to ``reconstruct" the GWB angular dependence. Provided some of the individual principal map amplitudes $\widehat{\mathcal{A}}_n$ are measured with sufficiently high individual SNR$_n \equiv \widehat{\mathcal{A}}_n/\Sigma_n$ (say, SNR$_n > 3$), we can define the reconstructed map 
\beq
\bsm{A}_{\rm recon} \equiv \sum_{n; \textrm{SNR}_n > 3}  \widehat{\mathcal{A}}_n ~\bsm{M}_n.
\eeq
This procedure is analogous to the production of ``dirty map" in radio interferometry, i.e.~including the contributions of observed ``visibilities", and setting the remaining (unobserved) piece to zero. Unlike interferometry, however, the ability to reconstruct a map is a strong function of its amplitude. Indeed, for a single interferometer, different visibilities typically have comparable noise, and as a consequence all contribute to the dirty map once any one of them is detected with sufficiently high SNR. In contrast, the noise of principal maps of a PTA increases steeply with the principal map number, see Fig.~\ref{fig:epta_eigval} for the principal maps of the EPTA. Note that this property is \emph{not} a result of unequal pulsar noises: even for an array of equal pulsars, densely and isotropically distributed on the sky, the eigenvalues of the Fisher matrix are steep function of principal map index (see Fig.~4 in Paper I).

\begin{figure*}
    \centering
    \includegraphics[width = 0.66\columnwidth]{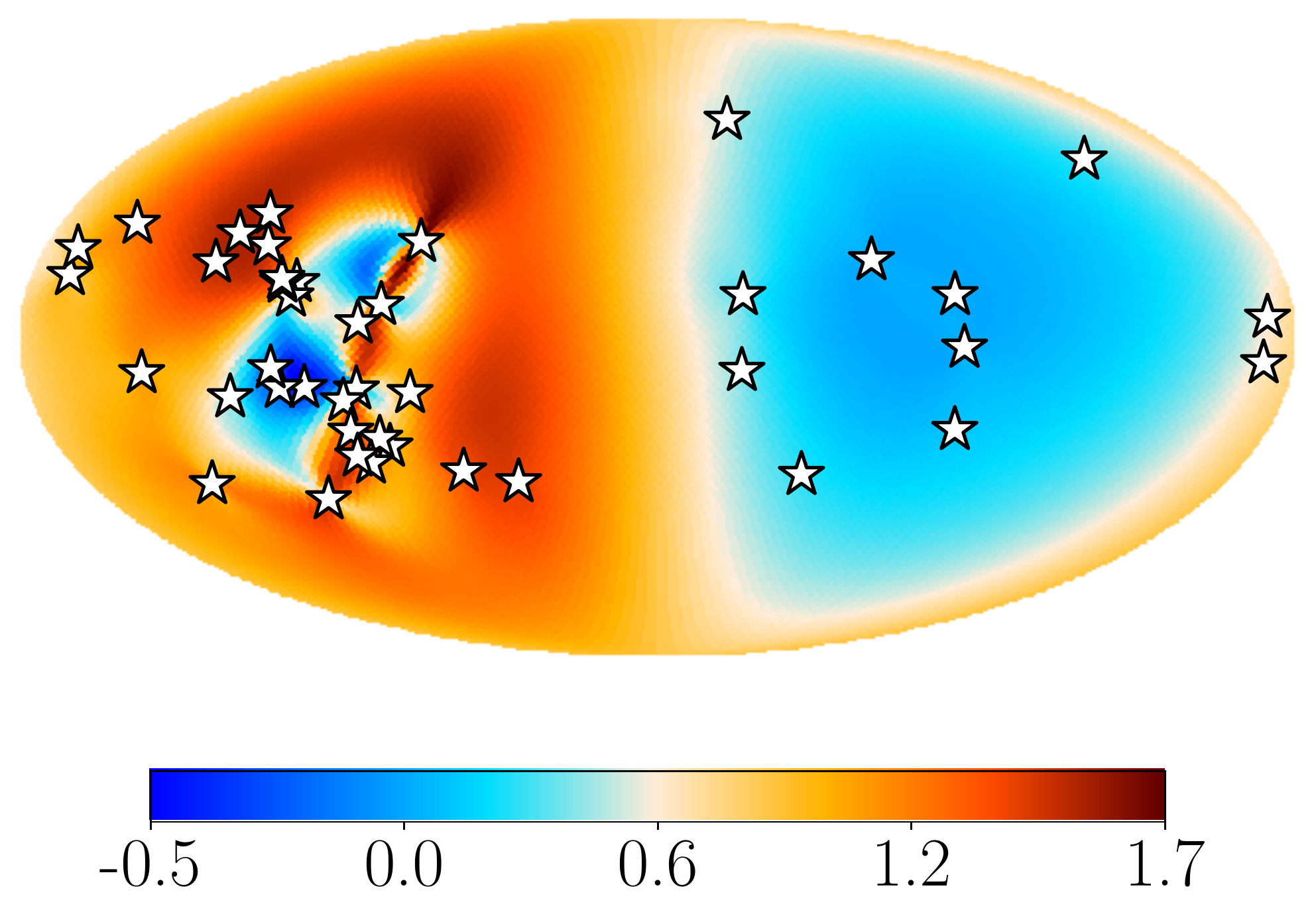}
    \includegraphics[width = 0.66\columnwidth]{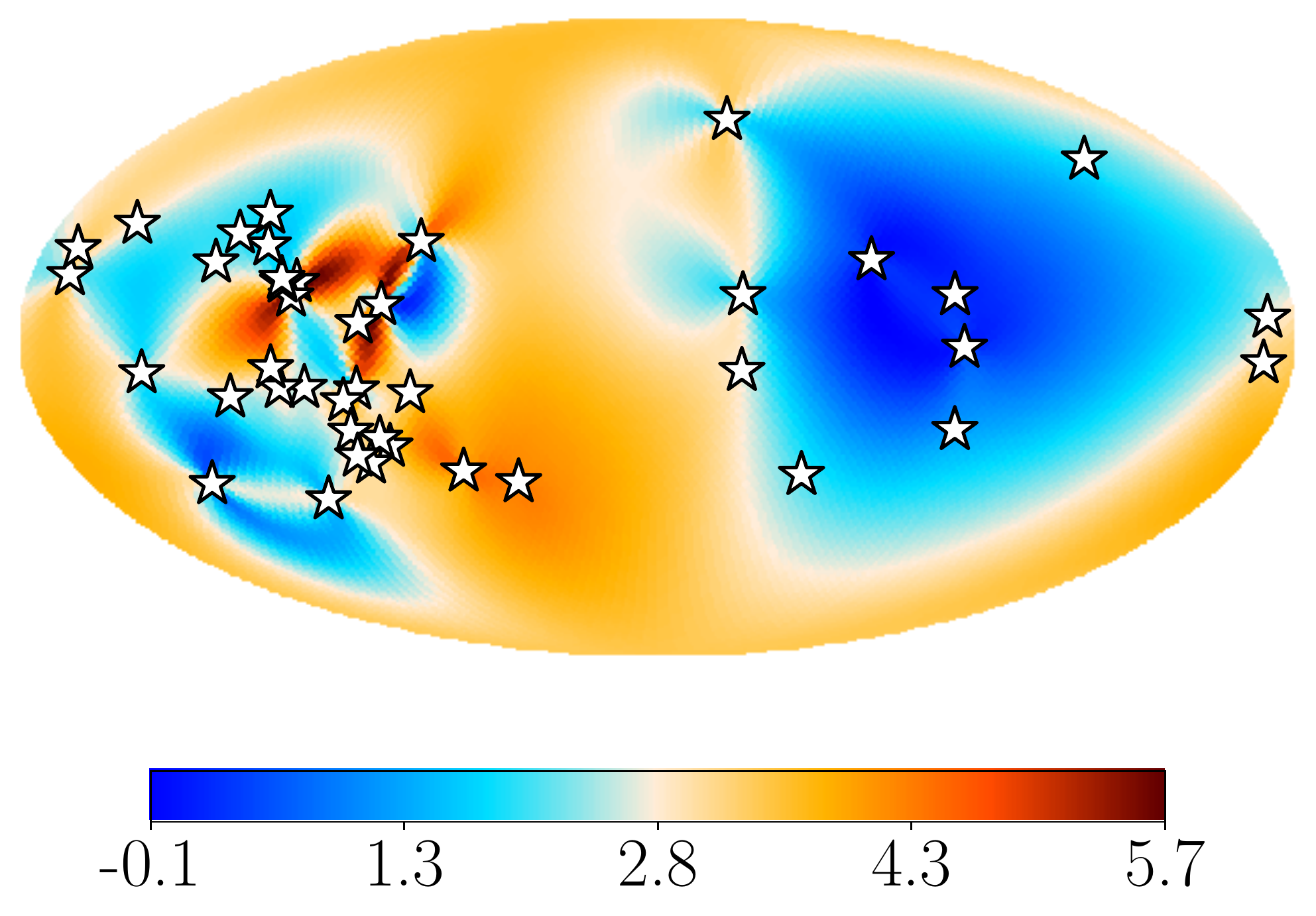}
    \includegraphics[width = 0.66\columnwidth]{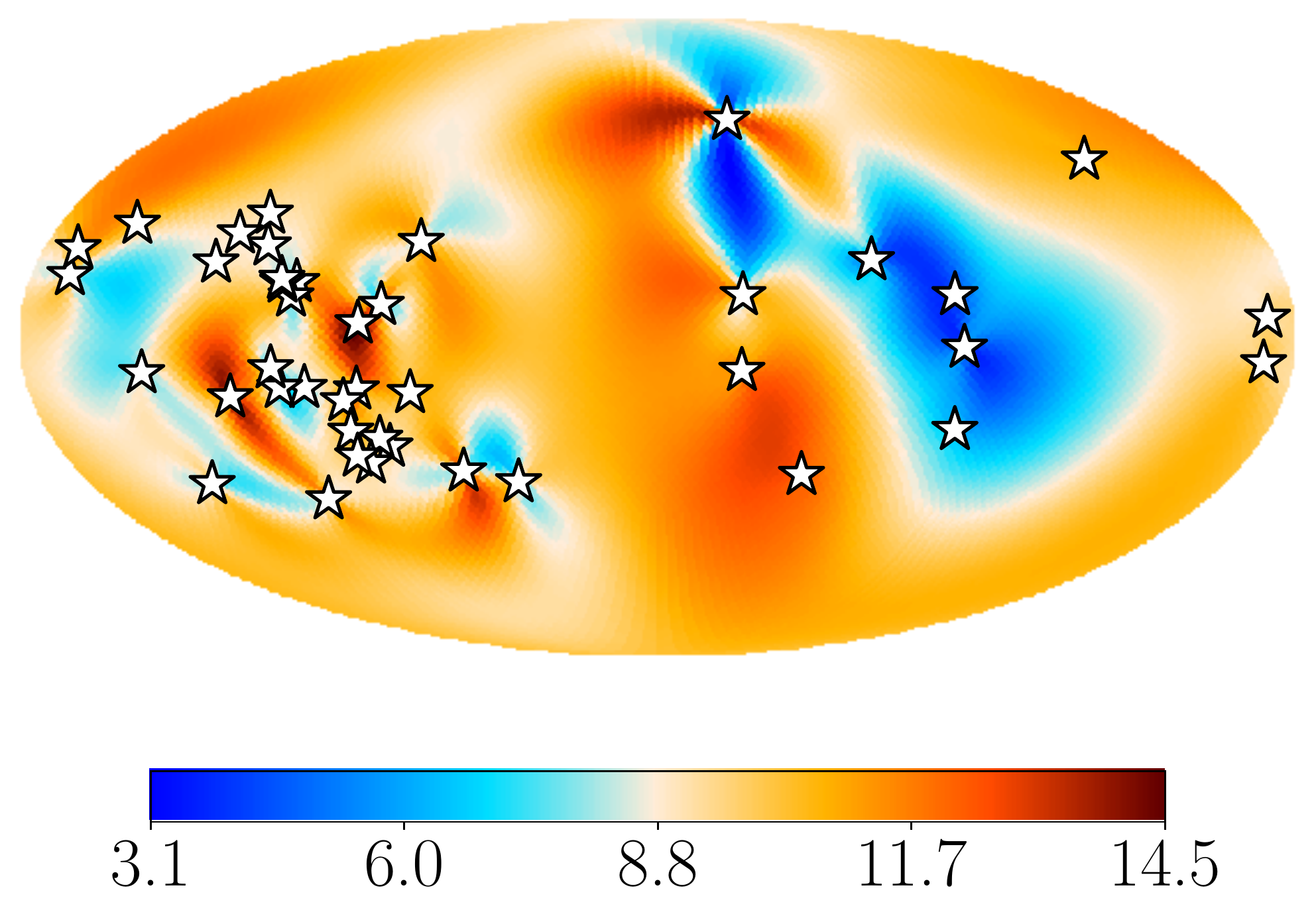}
    \includegraphics[width = 0.66\columnwidth]{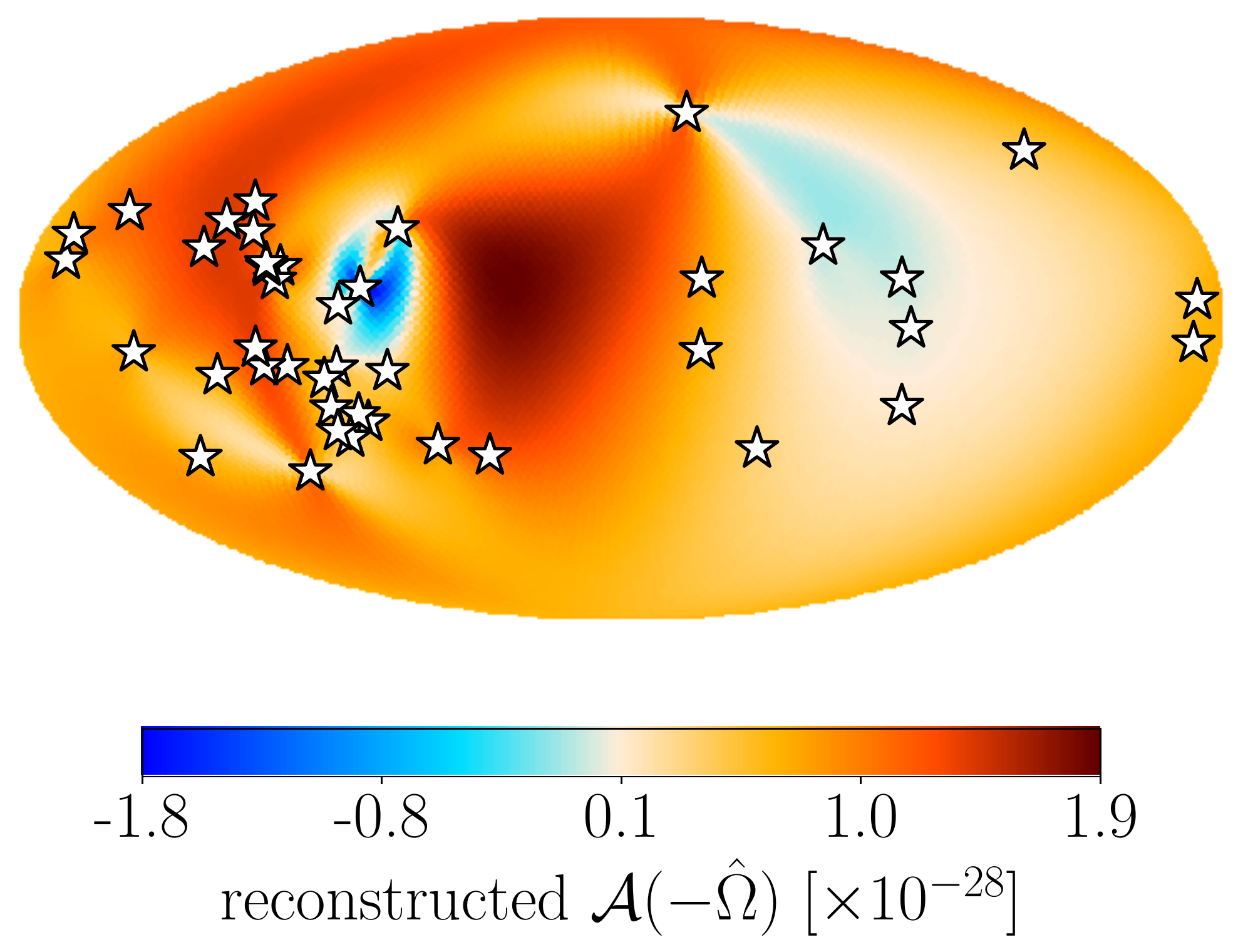}
    \includegraphics[width = 0.66\columnwidth]{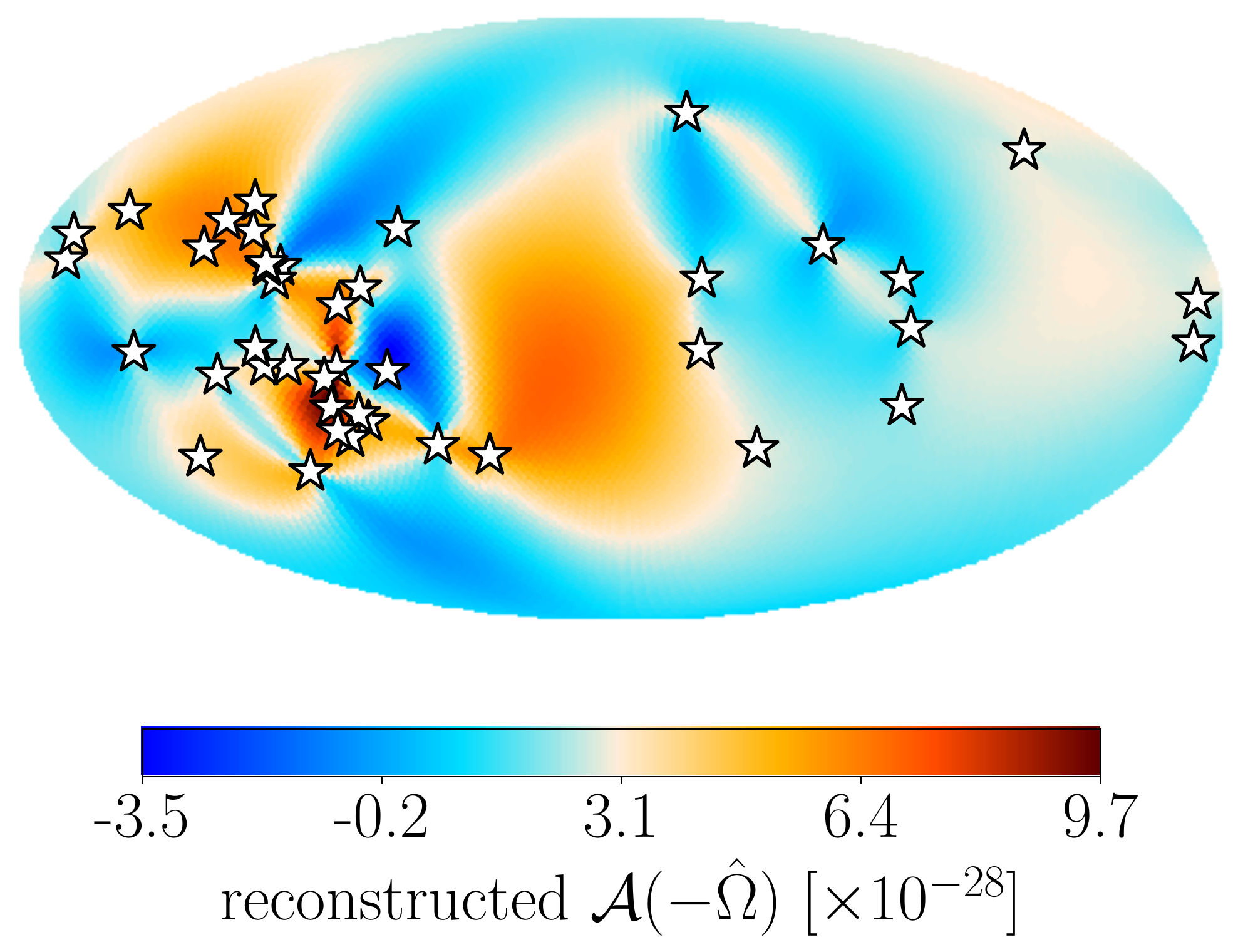}
    \includegraphics[width = 0.66\columnwidth]{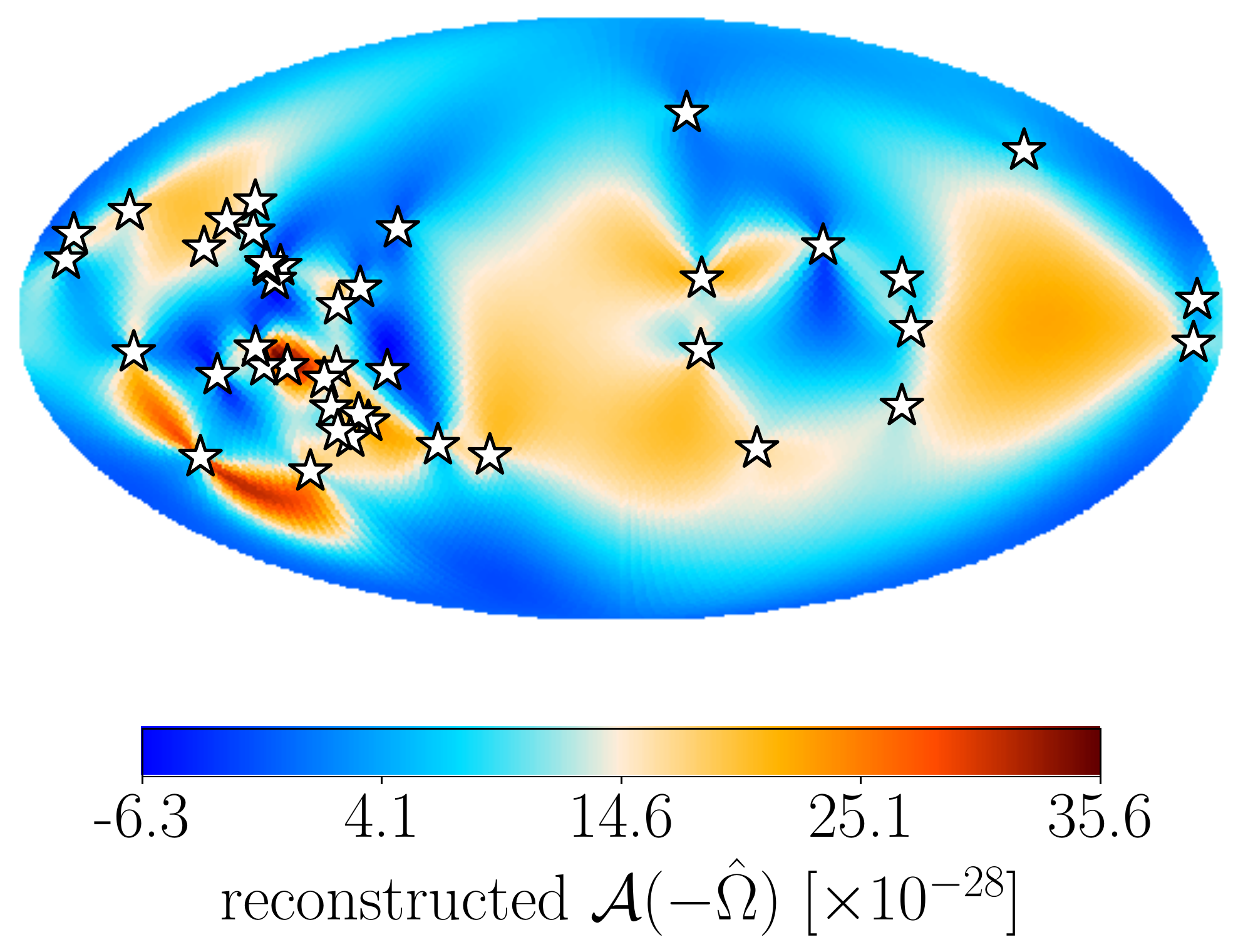}
    \caption{Reconstructions of a purely isotropic GWB (top row) and of the ``GWB" map shown in Fig.~\ref{fig:gwb} (bottom row) with the EPTA. In each column, from left to right, the input map has monopole amplitude $A_h^2 = 10^{-28}, 3\times 10^{-28}, 10^{-27}$, respectively. For the isotropic input map, these reconstructed maps are built from 3, 9 and 21 principal maps detected with individual SNR $ > 3$. For the ``GWB" map, the reconstructed maps are built from 3, 13 and 26 principal maps.} 
    \label{fig:gwb_reconstructed}
\end{figure*}

\begin{figure}
    \centering
    \includegraphics[width = \columnwidth]{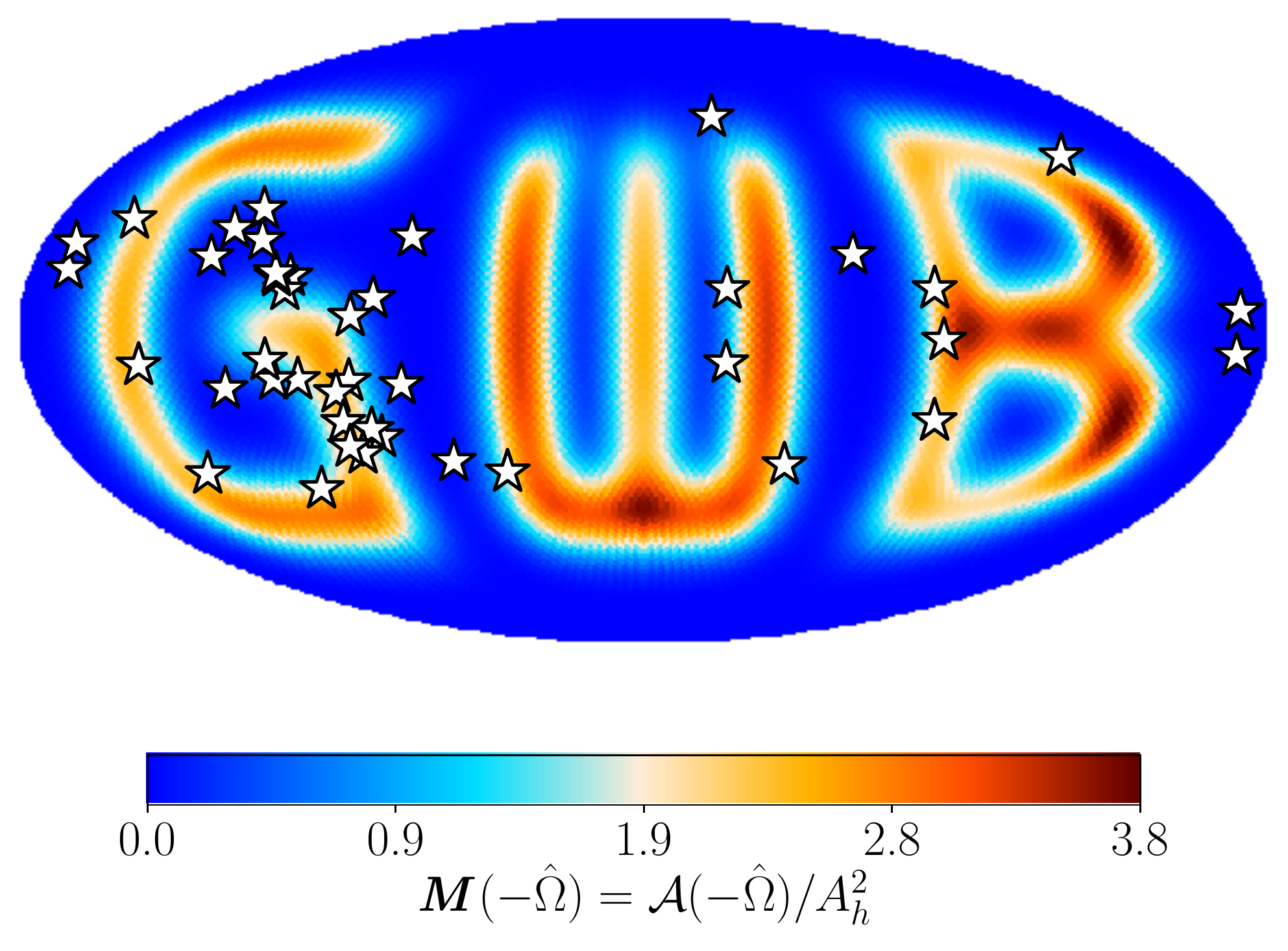}
    \caption{Dimensionless GWB intensity map used to illustrate the map reconstruction technique in Section \ref{sec:reconstruction}, as well as the frequentist approach in Sec.~\ref{sec:frequentist}. This map is normalized to a unit monopole.}
    \label{fig:gwb}
\end{figure}

We illustrate the map reconstruction technique with the EPTA in Fig.~\ref{fig:gwb_reconstructed}. In the top row, we show the reconstructed maps obtained if the underlying GWB is a pure monopole, with amplitudes $A_h = 10^{-14}, \sqrt{3}\times 10^{-14}, 3 \times 10^{-14}$, from left to right. The bottom row shows the reconstructed maps obtained if the underlying GWB is proportional to the map shown in Fig.~\ref{fig:gwb}, with the same monopole amplitude. We see that even with these very large amplitudes, the reconstructed maps have little resemblance with the underlying GWB intensity distribution. Note, also, that the overall amplitude of these maps (corresponding to total SNR of 30, 100 and 300, respectively, from left to right), is so large that they are most likely inconsistent with the weak-signal limit.

\subsection{Possible extensions of principal maps singling out the monopole}

\begin{figure}
    \centering
    \includegraphics[width = \columnwidth]{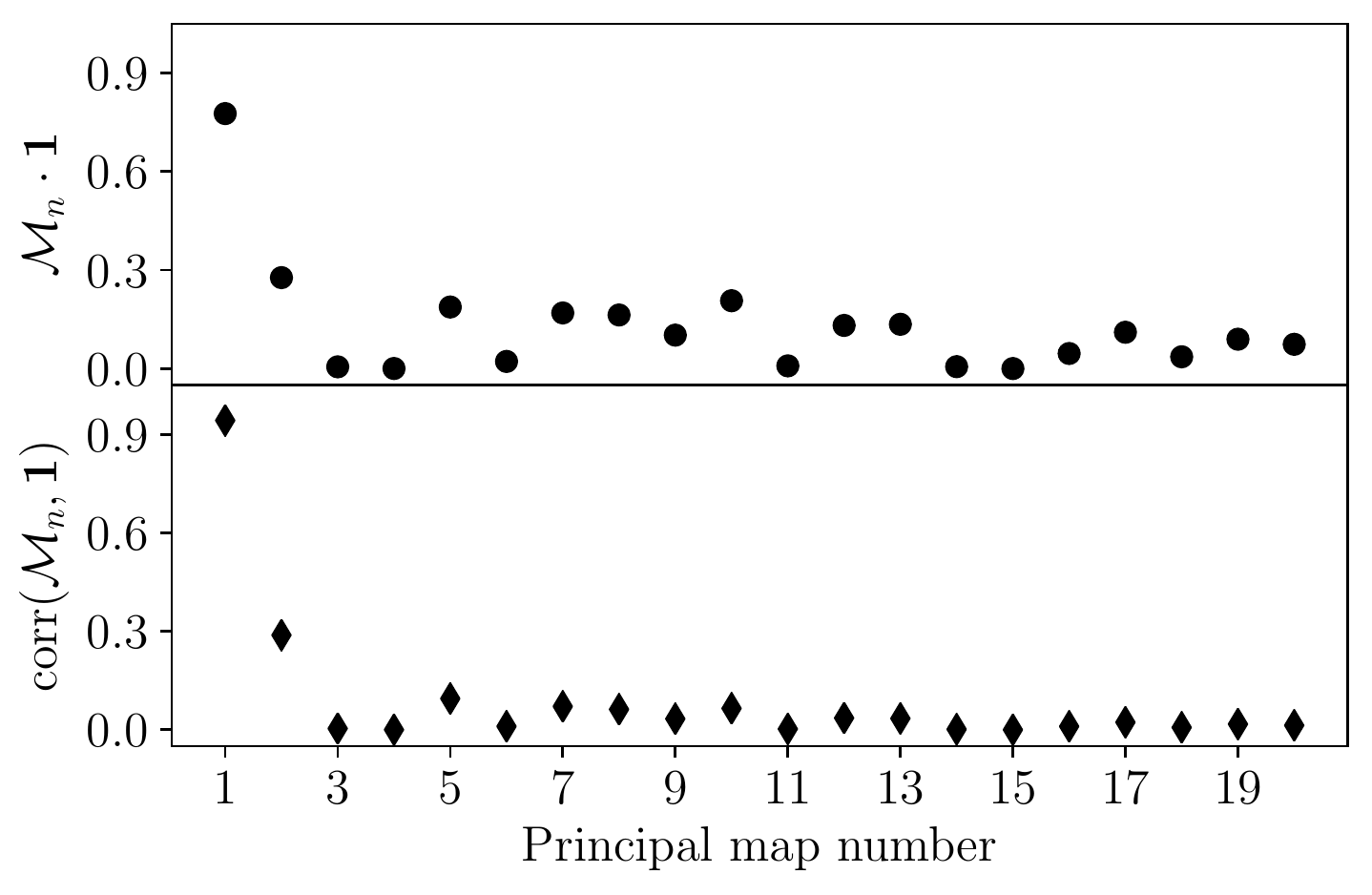}
    \caption{Monopole component (upper panel) and correlation coefficient for the first 20 EPTA principal maps. The correlation coefficient is defined as $\textrm{corr}(\bsm{M}_n, \bs{1}) \equiv \bsm{M}_n \cdot \bsm{F} \cdot \bs{1}/[(\bsm{M}_n \cdot \bsm{F} \cdot \bsm{M}_n)(\bs{1} \cdot \bsm{F} \cdot \bs{1})]^{1/2}$.}
    \label{fig:epta_eig_mono}
\end{figure}

The principal maps provide a useful basis to decompose the GWB if one is fully agnostic about its angular distribution. In practice, however, this is not the case: one does expect a significant monopole component in the GWB. In other words, physical considerations isolate a preferred map, which should always be included in GWB searches. This map is in general different from the principal maps, which are entirely based on the noise and geometric properties of the PTA, rather than external physical considerations.

Fig.~\ref{fig:epta_eig_mono} shows the monopole components of the first 20 EPTA principal maps, as well as their correlation coefficient with the monopole. Even though the monopole has the largest projection on the first principal map, we see that it still has significant projections on and correlations with some of the higher-order eigenmaps. In other words, higher-order eigenmaps are $(i)$ not anisotropic and $(ii)$ statistically correlated with the monopole. 

In order to alleviate issue $(i)$, one could try to construct a set of ``principal anisotropies", $\{\bsm{A}_n\}$, which are orthogonal to the monopole, i.e.~satisfy $\bsm{A}_n \cdot \bs{1} = 0$. One would then find the unit-norm maps extremizing $\bsm{A}_n \cdot \bsm{F} \cdot \bsm{A}_n$ under this additional constraint. This optimization problem admits $N_{\rm pair} -1$ solutions, which, in addition to the monopole, form a basis of $N_{\rm pair}$ maps which can be used to search for a monopole and anisotropies. However, the principal anisotropies derived in this fashion are in general statistically correlated with the monopole, $\bsm{A}_n \cdot \bsm{F} \cdot \bs{1} \neq 0$, thus would inflate the error bar on the monopole when included in a search. 

To alleviate issue $(ii)$, one could construct a set of ``monopole-uncorrelated maps", $\{\bsm{B}_n\}$, which are uncorrelated with the monopole, i.e.~satisfy $\bsm{B}_n \cdot \bsm{F} \cdot \bs{1} = 0$. One would then find the unit-norm maps extremizing $\bsm{B}_n \cdot \bsm{F} \cdot \bsm{B}_n$ under this additional constraint. This optimization problem also admits $N_{\rm pair} -1$ solutions, which, in addition to the monopole, form a basis of $N_{\rm pair}$ maps which can be used to search for a monopole and uncorrelated maps. However, the maps $\bsm{B}_n$ derived in this fashion in general have a non-zero projection on the monopole, $\bsm{B}_n \cdot \bs{1} \neq 0$. Thus, detecting a non-zero amplitude for these maps would not imply that one has detected anisotropies in the data.

One could try and alleviate both issues simultaneously by searching for a set of normalized maps $\bsm{C}_n$which are anisotropic \emph{and} uncorrelated with the monopole, with extremal SNR$^2$. As long as $\bsm{F} \cdot \bs{1}$ is not colinear with $\bs{1}$ (which is the case if the monopole is not an eigenmap of the Fisher matrix), the two constraints are independent. The resulting optimization problem thus admits $N_{\rm pair} - 2$ solutions. To span the $N_{\rm pair}$-dimensional set of observable maps, one must supplement these monopole-uncorrelated principal anisotropies with the monopole \emph{and} $\bsm{F} \cdot \bs{1} - (\bsm{F} \cdot \bs{1}) \bs{1}$ (properly normalized). Indeed, this additional map is orthogonal to the monopole and to all the $\bsm{C}_n$ maps (which stems from the condition that the $\bsm{C}_n$ are orthogonal to and uncorrelated with the monopole), thus linearly independent from all of them. However, this additional map is in general not statistically independent from the monopole nor from the maps $\bsm{C}_n$. Therefore, this construction still does not alleviate the issue.

To conclude, unless the monopole is an eigenmap of the Fisher matrix, there is no good strategy to \emph{agnostically} search for anisotropies within the standard Bayesian setup. In Section \ref{sec:frequentist}, we explore a frequentist approach to this problem.

\section{Frequentist approach to agnostic anisotropy searches} \label{sec:frequentist}

In the previous sections we have highlighted the difficulties in carrying a standard Bayesian search for anisotropies of pre-determined shapes alongside a monopole. Here we take a different approach, and derive a criterion to assess the presence of anisotropies in the data, regardless of their specific shape, without requiring a basis of maps on which to decompose the GWB.

\subsection{Derivation of the frequentist criterion}

Given the data translated into an estimator ${\bsmhat{A}}$ for the GWB amplitude, one may seek the monopole amplitude that minimizes the $\chi^2$, given by
\beq
\chi^2(\mathcal{A}_0) = \left(\mathcal{A}_0 \bs{1} - {\bsmhat{A}}\right) \cdot \bsm{F} \cdot \left(\mathcal{A}_0 \bs{1} - {\bsmhat{A}}\right).
\eeq
The best-fit monopole amplitude is then simply
\beq
\mathcal{A}_0^{\rm bf} = \frac{\bs{1} \cdot \bsm{F} \cdot {\bsmhat{A}}}{\bs{1} \cdot \bsm{F} \cdot \bs{1}}. \label{eq:A0_bf}
\eeq
The SNR of the best-fit monopole is then SNR$^{\rm bf} = |\mathcal{A}_0^{\rm bf}| \sqrt{\bs{1} \cdot \bsm{F} \cdot \bs{1}}$. This quantifies how well the best-fit monopole is detected relative to statistical noise. However, it does not quantify whether a pure monopole is a good fit to the data or not.

To quantify the goodness of fit, one can examine the $\chi^2$ at this best-fit value. After simplification, one obtains
\barr
\chi^2_{\rm bf} \equiv \chi^2\left(\mathcal{A}_0^{\rm bf}\right) &=& {\bsmhat{A}} \cdot \bsm{F} \cdot {\bsmhat{A}} - \frac{(\bs{1}\cdot \bsm{F} \cdot {\bsmhat{A}})^2}{\bs{1} \cdot \bsm{F} \cdot \bs{1}} \nonumber\\
&=& {\bsmhat{A}} \cdot \widetilde{\bsm{F}} \cdot {\bsmhat{A}},
\earr
where we have defined the projected Fisher matrix
\beq
\widetilde{\bsm{F}} \equiv \bsm{F} - \frac{(\bsm{F} \cdot \bs{1}) \otimes (\bs{1} \cdot \bsm{F})}{\bs{1} \cdot \bsm{F} \cdot \bs{1}},
\eeq
which satisfies $\bs{1} \cdot \widetilde{\bsm{F}} = \widetilde{\bsm{F}} \cdot \bs{1} = 0$. This implies that any monopole contribution to ${\bsmhat{A}}$ drops out of $\chi^2_{\rm bf}$. 

If the underlying GWB is purely isotropic, so that ${\bsmhat{A}}$ is a monopole plus pure statistical noise, only the latter contributes to $\chi^2_{\rm bf}$, which has a chi-square distribution with $N_{\rm pair}-1$ degrees of freedom. In particular, it has mean $N_{\rm pair} -1$ and variance $2 (N_{\rm pair} -1)$. 

In the limit that $N_{\rm pair}  \gg 1$, $\chi^2_{\rm bf}$ is approximately Gaussian-distributed. A simple criterion for the detection of a anisotropy  with 95\% confidence (i.e.~for the presence of a non-pure-statistical noise in ${\bsmhat{A}}$) is then
\beq
| \chi^2_{\rm bf} - (N_{\rm pair} -1)| > 2 \sqrt{2} \sqrt{N_{\rm pair} -1}.
\eeq
Note that this criterion does not provide any information about the nature of the anisotropy (besides the fact that it must have some non-zero projections on one of the $N_{\rm pair} -1$ eigenmaps of $\widetilde{\bsm{F}}$). It merely asserts that the data cannot be well described purely by a monopole.

Consider a map $\bsm{A} = \mathcal{A}_1 \bs{M}$, where $\bs{M} \neq \bs{1}$ is a dimensionless map. The minimum amplitude $\mathcal{A}_1$ required for an anisotropy to be confidently detected in the data (without knowledge of its precise form) is thus 
\beq
\mathcal{A}_1^{\rm min} = \left(\frac{2 \sqrt{2} \sqrt{N_{\rm pair} -1}}{\bs{M} \cdot \widetilde{\bsm{F}} \cdot \bs{M}}\right)^{1/2}. \label{eq:A1min}
\eeq
This corresponds to a best-fit monopole $\mathcal{A}_0^{\rm bf, min} = \mathcal{A}_1^{\rm min} (\bs{1} \cdot \bsm{F} \cdot \bs{M})/(\bs{1} \cdot \bsm{F} \cdot \bs{1})$, hence to a minimum SNR for the best-fit monopole
\beq
\textrm{SNR}^{\rm min} =  \frac{ \left(2 \sqrt{2} \sqrt{N_{\rm pair} -1}\right)^{1/2}|\bs{1} \cdot \bsm{F} \cdot \bs{M}|}{\sqrt{(\bs{M} \cdot \widetilde{\bsm{F}} \cdot \bs{M})(\bs{1} \cdot \bsm{F} \cdot \bs{1})}}.
\eeq

\subsection{Illustration with the EPTA}

We illustrate this frequentist approach to the map $\bs{M}$ shown in Fig~\ref{fig:gwb}, which is normalized to have a unit monopole. First, we compute the best-fit monopole amplitude for this map, $\mathcal{A}_0^{\rm bf} \equiv \mathcal{A}_1 (\bs{1} \cdot \bsm{F} \cdot \bs{M})/(\bs{1} \cdot \bsm{F} \cdot \bs{1})$. We find $\mathcal{A}_0^{\rm bf} = 0.96 \mathcal{A}_1$: the best-fit monopole amplitude turns out to be rather close to the actual value. This is because the projections of $\bs{M}$ and $\bs{1}$ on the first few principal maps happen to be similar in this case. This need not be always the case for other maps or PTAs. From Eq.~\eqref{eq:A1min}, and using the EPTA Fisher matrix, we find $\sqrt{\mathcal{A}_1^{\rm min}} \approx 1.2 \times 10^{-14}$ (this quantity has dimensions of characteristic strain). This corresponds to an SNR for the best-fit monopole amplitude $0.96 \times \mathcal{A}_1^{\rm min} \sqrt{\bs{1} \cdot \bsm{F} \cdot 1} \approx 46$. In other words, if the GWB has the angular dependence shown in Fig.~\ref{fig:gwb}, the full EPTA would only be able to establish the presence of anisotropies with 95\% confidence (without being able to specify their precise form) after it detects the best-fit monopole with SNR $\gtrsim $ 46.

\section{Reproducing TMG15's results} \label{sec:Taylor}

To our knowledge, TMG15 is the only study of anisotropies in the GWB with real PTA data. In this section, we show how to use our Fisher formalism to reproduce their results. We moreover clarify the meaning of some of their upper limits. Indeed, as discussed earlier, a PTA is sensitive to $N_{\rm pair}$ independent maps at most. This means that it is not possible to simultaneously constrain the GWB amplitude in more than $N_{\rm pair}$ pixels in the sky, or constrain more than $N_{\rm pair}$ of its spherical harmonic amplitudes. Yet, TMG15 present a 95\% upper-limit map on the GWB in 12288 pixels, using only the 15 pairs afforded by 6 EPTA pulsars. They also present upper limits on spherical harmonic coefficients with $\ell_{\max}$ as large as 10, corresponding to 121 independent coefficients. Clearly, these cannot be true upper limits. In this section we shall understand precisely what these limits are, and reproduce them approximately using the Fisher matrix and its eigenmap decomposition.

\subsection{Pixel-by-pixel upper limit maps}

Following TMG15, we define $\bs{P} \equiv \bsm{A}/A_h^2$, which satisfies $\int d^2 \ho~ P(\ho) = 4 \pi$. With our notation, the $N_{\rm pair}$ overlap reduction functions are given by $\Gamma_I \equiv \frac38 \bs{\gamma}_I \cdot \bs{P}$. We now discretize the sky into $N_{\rm pix}$ pixels centered at $\ho_i$, with equal area $\Delta \Omega$. As in TMG15, we define $c_i \equiv P(\ho_i)$. The overlap reduction functions are then 
\beq
\Gamma_I = \frac38 \int \frac{d^2 \ho}{4 \pi} \gamma_I(\ho) \bs{P}(\ho) \approx \sum_{i = 1}^{N_{\rm pix}} H_{Ii} c_i, 
\eeq
where the $N_{\rm pair} \times N_{\rm pix}$ matrix $\bs{H}$ has elements
\beq
H_{I i} = \frac38 \gamma_I(\ho_i) \frac{\Delta \Omega}{4 \pi}.
\eeq
We can thus rewrite the relationship between the overlap-reduction functions and the power distribution $\bs{P}(\ho)$ in matrix form, as $\vec{\Gamma} = \bs{H} \vec{c}$, where $\vec{\Gamma}$ is the $N_{\rm pair}$-dimensional vector with components $\Gamma_I$ and $\vec{c}$ is the the $N_{\rm pix}$-dimensional vector with components $c_i$. 

Unless $N_{\rm pix} = N_{\rm pair}$, the matrix $\bs{H}$ is not square, and a fortiori not invertible. One can however define its Moore-Penrose pseudo-inverse \cite{Moore_20, Penrose_55} $\bs{H}^+$. We may then define the $N_{\rm pix}$-dimensional vector
\beq
\vec{\widetilde{c}} \equiv \bs{H}^+ \vec{\Gamma} = (\bs{H}^+ \bs{H}) \vec{c}.
\eeq
Unless $N_{\rm pix} = N_{\rm pair}$, $\bs{H}^+ \neq \bs{H}^{-1}$ and $\vec{\widetilde{c}} \neq \vec{c}$. Instead, as we demonstrate in Appendix \ref{app:proj}, $\widetilde{c}_i = P_{||}(\ho_i)$ is the (discretized) power distribution \emph{projected} on the $N_{\rm pair}$-dimensional space spanned by the $\gamma_I$'s. In other words, $\vec{\widetilde{c}}$ corresponds to the observable component of the GWB power, i.e.~the only piece of $\bs{P}(\ho)$ that is measurable by a PTA. Importantly, it \emph{cannot} be identified with $c_i = P(\ho_i)$ as done in TMG15. As a consequence, the map shown in TMG15 is only an upper-limit map for the \emph{projected} (i.e.~observable) GWB intensity. We refer to it as the Moore-Penrose pseudo-upper-limit map.

We now explicitly construct an approximation of this pseudo-upper-limit map. Given estimators for the timing residual correlations, one may construct an estimator for the projected GWB map $\bsmhat{A}_{||} = \bsmhat{A}$ given by Eqs.~\eqref{eq:hat-A}-\eqref{eq:hat-AI}. The pixel values of this estimator map are precisely $\widehat{\mathcal{A}}_{||}(\ho_i) = A_h^2 P_{||}(\ho_i) = A_h^2 \widetilde{c}_i$. In order to compute the variance of these values, it is best to first decompose it on the basis of principal maps $\bsm{M}_n$:
\beq
\widehat{\mathcal{A}}_{||}(\ho_i) = \sum_{n = 1}^{N_{\rm pair}} \widehat{\mathcal{A}}_n \mathcal{M}_n(\ho_i),  \ \ \ \ \widehat{\mathcal{A}}_n \equiv \bsm{M}_n \cdot \bsmhat{A} \label{eq:A||Cn}
\eeq
The $N_{\rm pair}$ amplitudes $\widehat{\mathcal{A}}_n$ are uncorrelated Gaussian variables with covariance 
\beq
\textrm{cov}(\widehat{\mathcal{A}}_n, \widehat{\mathcal{A}}_m) = \delta_{nm} \Sigma_n^{2}. \label{eq:cov-An}
\eeq
We thus find
\beq
\textrm{var}[\widehat{\mathcal{A}}_{||}(\ho_i)] = \sum_n \left(\Sigma_n\mathcal{M}_n(\ho_i)\right)^2. \label{eq:sigma_i^2}
\eeq
From this we can estimate a 95\%-sensitivity to $\mathcal{A}_{||}(\ho_i)$, 
\beq
\mathcal{A}_{||}^{95\%}(\ho_i) = 2 \sqrt{\textrm{var}[\widehat{\mathcal{A}}_{||}(\ho_i)]} \equiv \left[\widetilde{A}_h^{95\%}(\ho_i)\right]^2, \label{eq:A||ul}
\eeq
where the last equality defines the characteristic strain $\widetilde{A}_h^{95\%}(\ho_i)$. 

\begin{figure}
    \centering
    \includegraphics[width = \columnwidth]{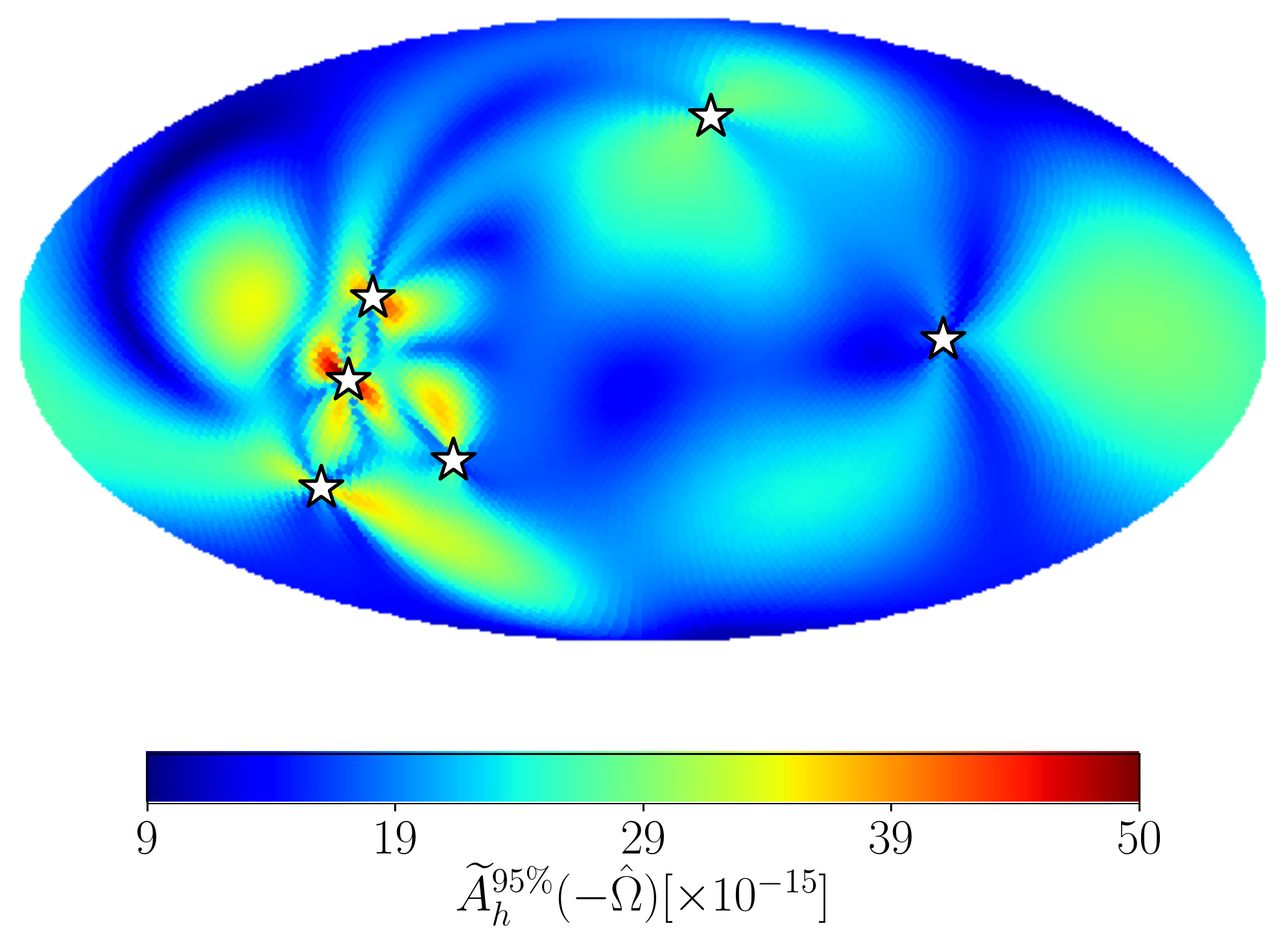}
    \includegraphics[width = \columnwidth]{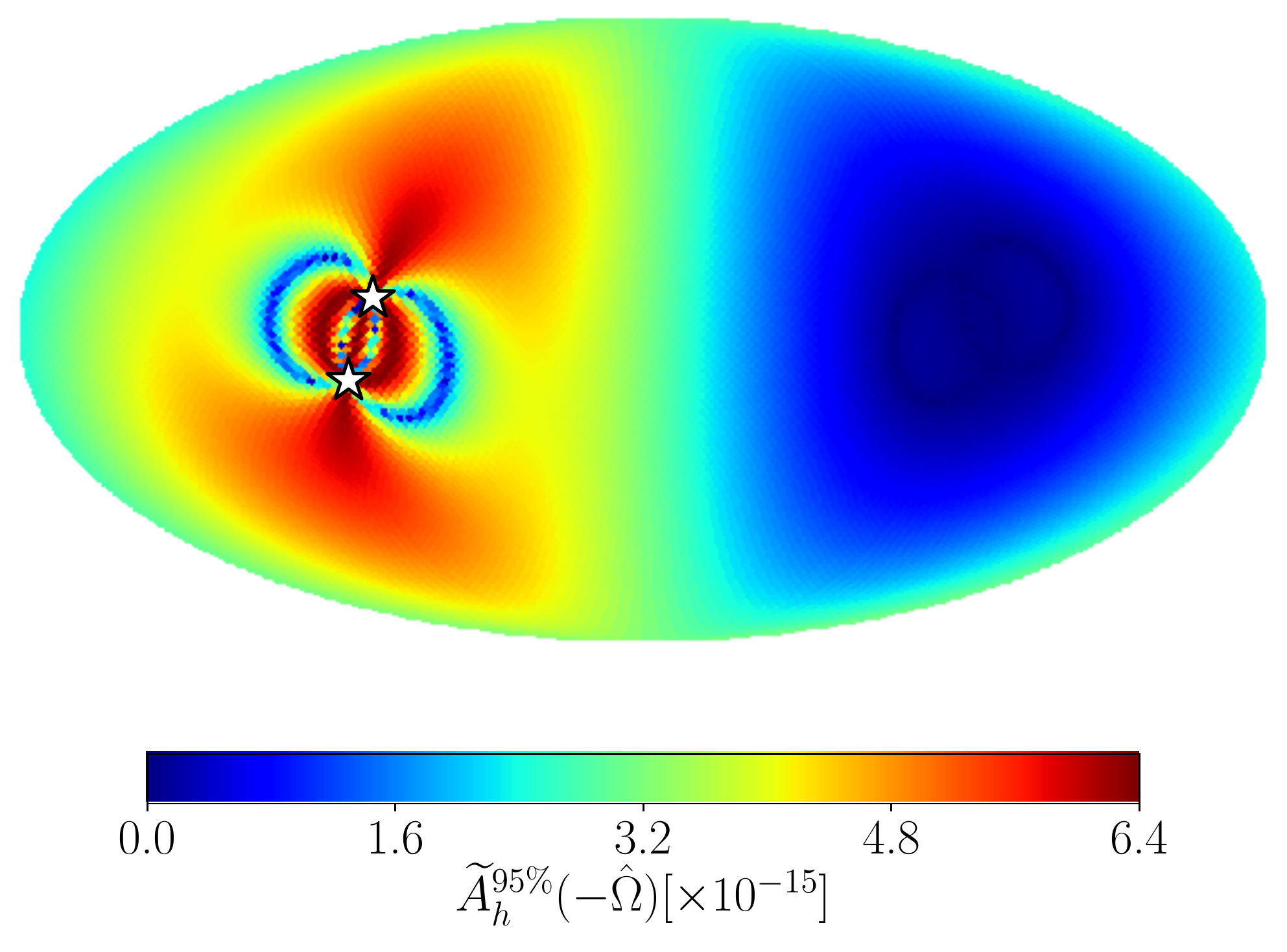}
    \caption{Moore-Penrose pseudo-sensitivity maps produced with 6 EPTA pulsars (top) and one single EPTA pulsar pair (bottom). The qualitative and quantitative similarities of the top map with Fig.~2 of TMG15 are striking. The overall lower amplitude of our map relative to that of TMG15 is consistent with our lower 95\% sensitivity to the monopole amplitude $A_h$. These maps \emph{cannot} be interpreted as actual upper limits on the GWB intensity in each pixel, as a PTA cannot constrain more than $N_{\rm pair}$ independent components simultaneoulsy. This is further demonstrated by the fact that the pseudo-sensitivity map produced with only two pulsars (hence less data) has lower amplitude than the one produced with 6 pulsars.}
    \label{fig:Taylor-map}
\end{figure}

We show $\widetilde{A}_h^{95\%}(\ho)$ in the top panel of Fig.~\ref{fig:Taylor-map}. The qualitative \emph{and} quantitative similarities with Fig.~2 of TMG15 are striking. The overall amplitude of our map is lower than that of TMG15, consistent with our lower 95\% sensitivity estimate for the monopole amplitude $A_h^{95\%}$. Note, also, that we cannot hope to reproduce precisely the same map as TMG15, as their map is constructed from real data, i.e.~a non-zero realization of the noise.

Let us point out that the pseudo-upper-limit map derived with this procedure gets \emph{worse} if one includes \emph{more} pulsars. Indeed, as can be seen from Eq.~\eqref{eq:sigma_i^2}, the variance of $\widehat{\mathcal{A}}_{||}(\ho_i)$ is dominated by the noisiest eigenmaps. As more pulsars are added, and the number of eigenmaps grows, the noisiest ones become increasingly noisy. Conversely, using only one pulsar pair results in a pseudo-upper-limit map with an overall amplitude lower by nearly one order of magnitude, as can be seen in the bottom panel of Fig.~\ref{fig:Taylor-map}. In fact, in the right hemisphere, the map amplitude nearly vanishes, as does the pairwise-timing response function of the single pulsar pair. Clearly, adding more information should only tighten true upper limits, not make them weaker. This demonstrates once again that these Moore-Penrose pseudo-upper-limit maps cannot be interpreted as true pixel-by-pixel upper limits.

\subsection{Constraints to spherical harmonic coefficients}

\begin{figure}
    \centering
    \includegraphics[width = \columnwidth]{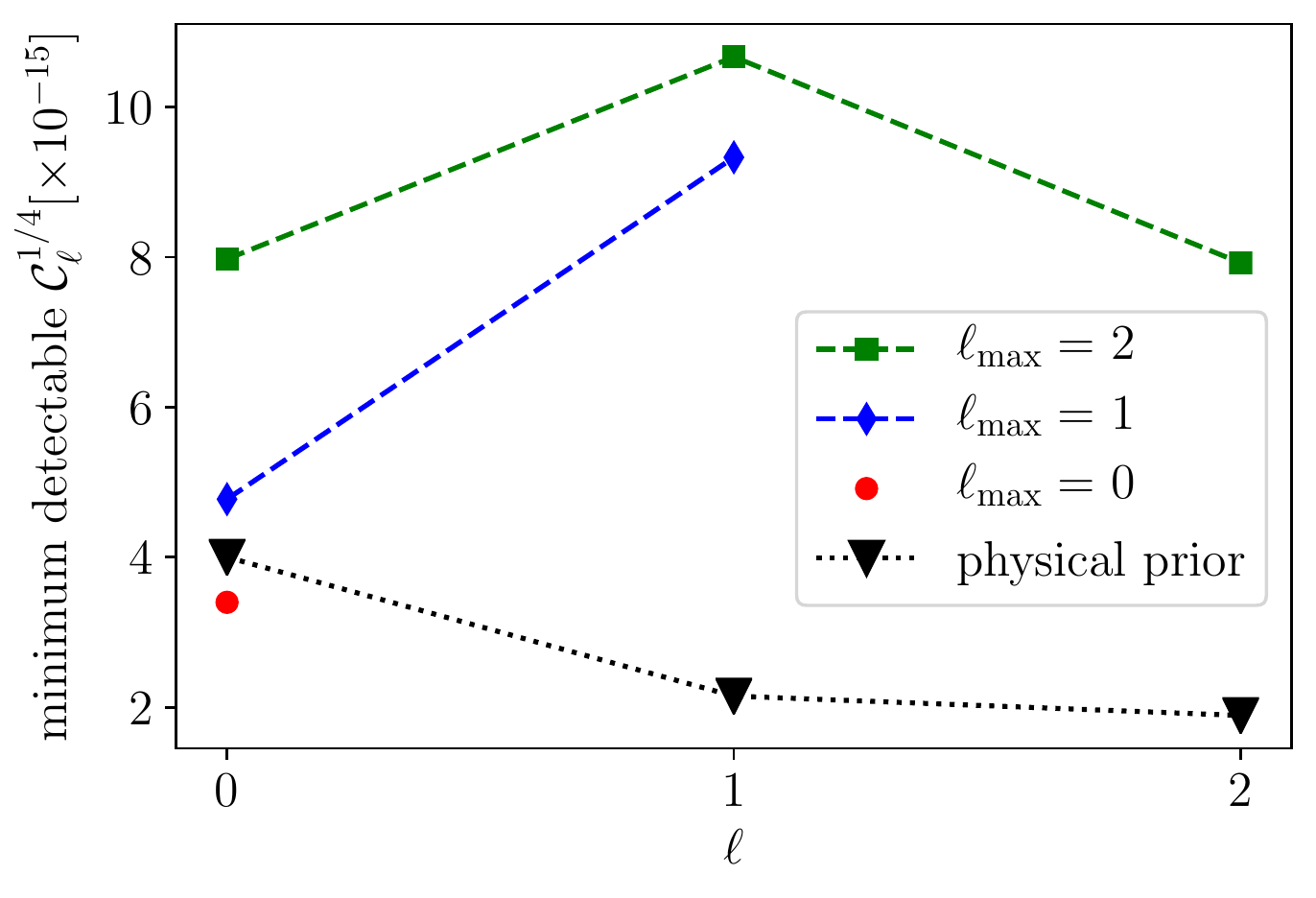}
    \caption{95\% sensitivity estimates of the 6 EPTA pulsars to spherical harmonic amplitudes. Compare with Fig.~1 of TMG15, right panel, for $\ell_{\max} \leq 2$. The physical prior assumes a monopole upper limit $A_h \leq 4 \times 10^{-15}$.}
    \label{fig:epta6_Cl}
\end{figure}

We now apply the formalism of Section \ref{sec:sph_harm} to the subset of 6 EPTA pulsars used in TMG15, and compare our results to theirs. With the 15 pairs afforded by 6 pulsars, cross-correlation data can simultaneously constrain spherical-harmonic coefficients only if $\ell_{\max} \leq 2$. For these values of $\ell_{\max}$, our 95\% sensitivity estimates on $\mathcal{C}_1$ and $\mathcal{C}_2$, shown in Fig.~\ref{fig:epta6_Cl}, are remarkably close to their 95\% upper limits in the absence of physical prior (see their Fig.~1, right panel, for $\ell_{\max} = 1, 2$). However, we find that the 95\% sensitivity to the monopole amplitude $\mathcal{C}_0$ gets significantly degraded as $\ell_{\max}$ increases from 0 to 2, while the upper limit of TMG15 is virtually unaffected. We attribute this to their use of autocorrelations, which we do not include in our analysis. We note, however, that since the GWB dipole and quadrupole affect autocorrelations, increasing $\ell_{\max}$ from 0 to 1 and then 2 should result in poorer constraints on the monopole. It is unclear why this is not what is found in TMG15.

\begin{figure}
    \centering
    \includegraphics[width = \columnwidth]{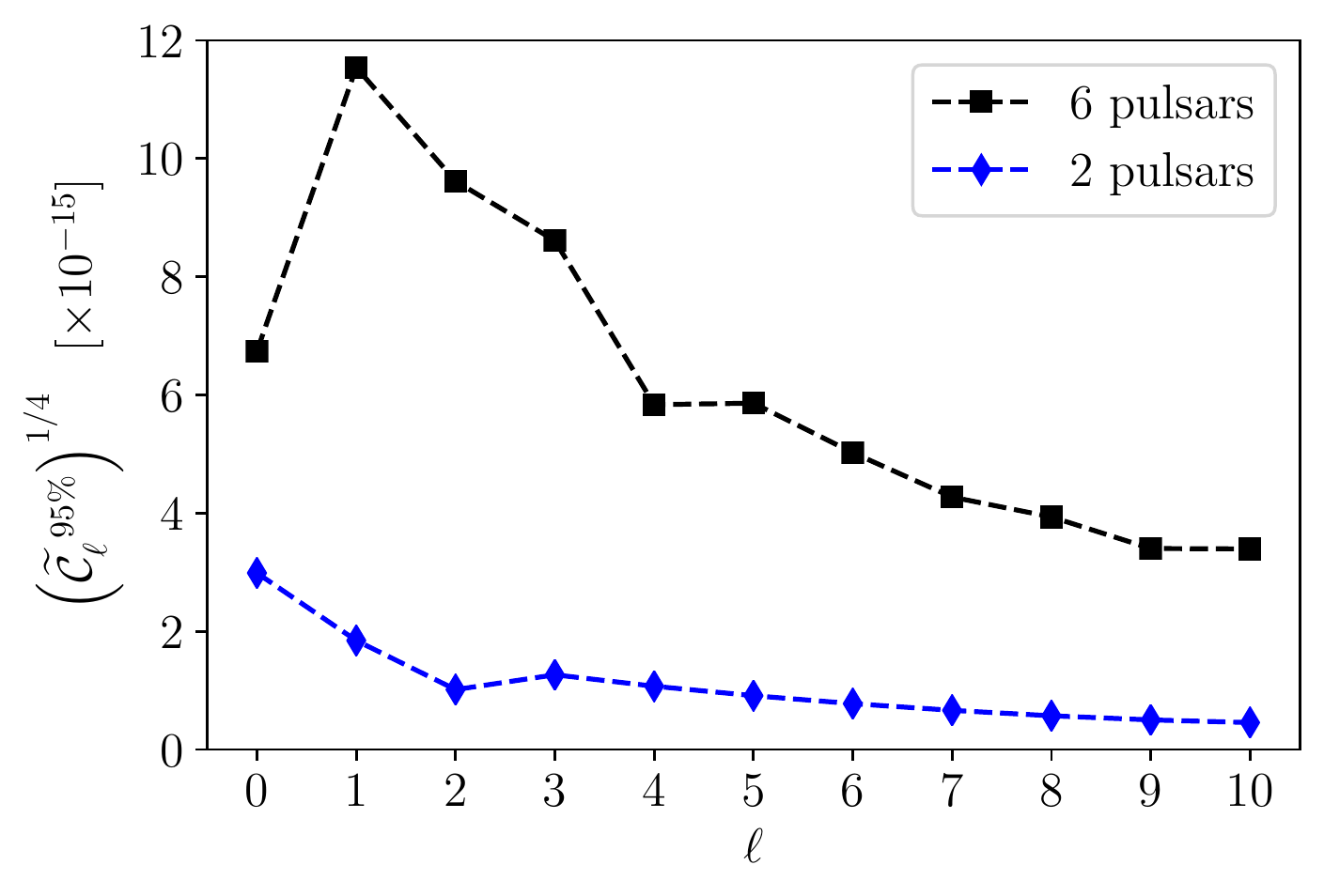}
    \caption{95\% sensitivity to the Moore-Penrose pseudo $\mathcal{C}_{\ell}$'s in the limit $\ell_{\max} \rightarrow \infty$, for 6 EPTA pulsars (upper curve) and 2 EPTA pulsars (lower curve). Compare the upper curve with the right panel of Fig.~1 of TMG15, for $\ell_{\max} = 7, 10$. The $\widetilde{\mathcal{C}}_{\ell}$ are obtained from the harmonic amplitudes of the \emph{projection} of the GWB on the $N_{\rm pair}$-dimensional space of observable maps, and therefore are \emph{not} equal to the true harmonic amplitudes of the GWB, see text for details.}
    \label{fig:pseudo-Cl}
\end{figure}

Let us now understand TMG15's upper limits on the spherical-harmonic amplitudes for $\ell_{\max} > 2$, for which there are more independent coefficients than independent data. In this case, TMG15 also define $\widetilde{c}_{\ell m}$ given the overlap reduction function $\Gamma_I$ through the Moore-Penrose pseudo inverse of the linear relation between $\Gamma_I$ and $c_{\ell m}$. Again, unless the number of harmonic coefficients is equal to $N_{\rm pair}$, $\widetilde{c}_{\ell m} \neq c_{\ell m}$. For finite $\ell_{\max}$, there is no simple intuitive interpretation of the $\tilde{c}_{\ell m}$, but for $\ell_{\max} \rightarrow \infty$, it is straightforward to show that the $\widetilde{c}_{\ell m}$ are the spherical-harmonic coefficients of the \emph{projected} GWB. Specifically, using our normalization conventions, we define $\widetilde{\mathcal{A}}_{\ell m} \equiv A_h^2 \widetilde{c}_{\ell m}/\sqrt{4 \pi} = \bsm{Y}_{\ell m}  \cdot \bsmhat{A}_{||}$. Using Eq.~\eqref{eq:A||Cn}, we thus have
\beq
\widetilde{\mathcal{A}}_{\ell m} = \sum_{n = 1}^{N_{\rm pair}} \widehat{\mathcal{A}}_n ~ \bsm{Y}_{\ell m}  \cdot \bsm{M}_n.
\eeq
Using Eq.~\eqref{eq:cov-An}, we thus obtain
\beq
\textrm{cov}(\widetilde{\mathcal{A}}_{\ell m}, \widetilde{\mathcal{A}}_{\ell' m'}) = \sum_{n = 1}^{N_{\rm pair}} \Sigma_n^2 (\bsm{Y}_{\ell m}  \cdot \bsm{M}_n) (\bsm{Y}_{\ell' m'}  \cdot \bsm{M}_n).
\eeq
We then define $\widetilde{\mathcal{C}}_{\ell} \equiv \sum_m \widetilde{A}_{\ell m}^2/(2 \ell +1)$. The mean and variance of these coefficients is given by Eqs.~\eqref{eq:mean-Cl} and \eqref{eq:cov-Cl}, with the substitution $\textrm{cov}(\mathcal{A}_{\ell m},\mathcal{A}_{\ell' m'}) \rightarrow \textrm{cov}(\widetilde{\mathcal{A}}_{\ell m}, \widetilde{\mathcal{A}}_{\ell' m'})$, from which we can infer a 95\% sensitivity $\widetilde{C}_{\ell}^{95\%}$ as in Eq.~\eqref{eq:Cl95}. We show $\widetilde{C}_{\ell}^{95\%}$ in Fig.~\ref{fig:pseudo-Cl}. This can be compared to the right panel of Fig.~1 of TMG15. The similarity of their $\ell_{\max} = 7, 10$ results with ours (which holds in the limit $\ell_{\max} \rightarrow \infty$) is striking. Note again that we cannot expect to reproduce exactly TMG15's results, which are based on actual data, i.e.~a particular realization of the noise. Here again, a sharp illustration of the fact that these Moore-Penrose pseudo-upper-limits cannot be interpreted as upper limits on the $\mathcal{C}_{\ell}$ is the fact that $\widetilde{C}_{\ell}^{95\%}$ decreases if one only includes 2 pulsars instead of 6, as can be seen in Fig.~\ref{fig:pseudo-Cl}. 

\section{Conclusions} \label{sec:discussion}

In this paper, we have built on the Fisher formalism we introduced in Paper I, and applied it to EPTA data, since this PTA is currently the only one with published limits on GWB anisotropy (TMG15). We showed through multiple examples how this framework can serve to understand the sensitivity of a PTA to a general, anisotropic GWB, and can be used to guide and optimize data analysis. We also showed how to recover existing EPTA results, and clarified their meaning, in the light of our improved understanding of the sensitivity of PTAs to GWB anisotropies. In particular, we pointed out that one cannot simultaneously detect -- or constrain beyond physical priors -- more than $N_{\rm pair}$ components of the GWB, be they pixel values or spherical harmonic amplitudes.

The fundamental tool of our formalism is the Fisher ``matrix" for the direction-dependent GWB intensity. This matrix condenses the essential noise characteristics of a PTA in a compact form. It can be computed given the positions of the pulsars and their characteristic noise strains. The latter requires analyzing the timing data of individual pulsars, but not conducting any cross-correlation analysis with real data.

We started by using our formalism to find the pulsar pairs which are the most sensitive to an isotropic GWB with a power-law frequency dependence $h_c(f) \propto f^{-2/3}$. We identified the 44 best EPTA pairs (out of the 861 available) that provide 90\% of the SNR$^2$. With the Fisher formalism, finding these best pairs for any PTA is a trivial exercise, and can be used to dramatically speed up real data analyses, at very little cost in sensitivity.

In the remainder of the paper we focused our attention to anisotropies in the GWB. We first considered the case where the GWB is assumed to be a linear combination of a finite number of known maps (no more than $N_{\rm pair}$), such as spherical harmonics, coarse pixels, or hot spots in pre-determined directions. We showed that the sensitivity of a PTA to the component of the GWB on any of the basis maps systematically and quickly degrades as the number of basis functions is increased. This stems from the fact that the noise properties of standard basis maps are correlated; therefore, adding more functions to the search changes the overall covariance structure of the entire set of amplitudes. In particular, the ability to \emph{detect} the monopole through pulsar cross correlations is systematically degraded as one searches simultaneously for anisotropies. We did not quantify the ability of a PTA to set upper limits through pulsar autocorrelations (or ``common red noise" analyses), but pointed out that these are sensitive to the GWB monopole, dipole and quadrupole. Hence, upper limits on the monopole should also degrade as one accounts for anisotropies.

Having shown that standard basis maps are poorly adapted for agnostic GWB searches, we then constructed the ``principal maps" of a PTA, which are the eigenmaps of the Fisher matrix. These maps form an orthonormal \emph{and uncorrelated} basis of the $N_{\rm pair}$-dimensional space of \emph{observable} maps. In other words, they represent the $N_{\rm pair}$ statistically independent pieces of information accessible to a PTA, in the space of GWB maps. We showed how one can in principle ``reconstruct" the observable part of the GWB as a linear combination of principal maps detected with sufficiently high significance, a procedure similar to the making of ``dirty maps" in radio interferometry. However, in contrast with a radio interferometer, for which different ``visibilities" have similar noise levels, principal maps have highly unequal noise properties. As a consequence, the ability to reconstruct the GWB is strongly dependent on its overall amplitude. For the EPTA, we found that even detecting a small number of principal maps would typically require a GWB amplitude largely exceeding current bounds.

Principal maps are a useful basis if one is fully agnostic about the angular dependence of the GWB: they allow for an optimal search ``under the lamppost" of a given PTA. However, the GWB is expected to be predominantly isotropic; physical considerations therefore mandate including the monopole as part of the search, and being able to identify anisotropic components of the GWB. We showed that the principal maps are in general correlated with the monopole and have a non-vanishing isotropic component, and are thus not adapted for such searches. We argued that one cannot construct a complete set of anisotropies which are uncorrelated among themselves and with the monopole, because the latter is not an eigenmap of the Fisher matrix. 

In summary, there are two options when searching for a general, direction-dependent GWB within a standard Bayesian framework. On one hand, one can include the monopole in GWB searches, alongside anisotropic basis functions, but then suffer from the loss of sensitivity resulting from correlations with the monopole. On the other hand, one may search under the lamppost of PTAs with principal maps, being fully agnostic about the angular dependence of the GWB, but then give up the special status of an isotropic GWB. To circumvent these limitations, we proposed a frequentist criterion to determine whether the GWB is \emph{consistent with isotropy}, even if one cannot determine the specific type of anisotropy. This criterion simply relies on computing the chi-squared statistics of the best-fit monopole amplitude, and comparing it with its statistical noise. We illustrated this with a fictitious, highly anisotropic GWB map, and showed that the EPTA would have to first detect its best-fit monopole with a SNR of 46 before being able to confidently assert that the data is inconsistent with a purely isotropic GWB.

The prospects for current PTAs to detect fully unknown anisotropies in the GWB appear to be somewhat limited, but the situation may improve dramatically with the several hundreds (or even thousands) of millisecond pulsars that SKA is expected to detect~\cite{Keane_15}. The Fisher formalism presented here provides a useful set of tools to address this question, under realistic assumptions for the properties of a future PTA built with SKA pulsars, and without requiring to run expensive simulations. We defer such forecasting analyses to future studies.
In addition, there may be hope to detect anisotropy in the GWB through its \emph{statistical properties}, for instance through its correlations with tracers of large-scale structure \cite{Alonso_20}. We will explore these avenues in future work. 

In closing, let us note that our Fisher formalism has far broader applications than what we considered in this work. First, we assumed an overall known frequency dependence, identical for the monopole and anisotropies. It would be straightforward to generalize our work to arbitrary frequency dependence, possibly dependent on the angular structure --  something which would be difficult to do with a full-on data analysis. One could also include other sources of correlated noise, such as global clock or ephemeris errors~\cite{Tiburzi16, Vallisneri20}, and go beyond current studies which ascribe them the same frequency dependence as the GWB. In addition, we could generalize our Fisher matrix beyond the weak-signal limit, and consider, for instance, the weak-anisotropy limit. Lastly, one could easily extend the formalism to include polarized GWBs. Overall, this formalism is a valuable tool that can be used to guide and cross-validate any full-blown data analysis. It will be especially important when SKA harvests hundreds of new millisecond pulsars, out of which PTAs with tens of thousands of independent pulsar pairs can be constructed. Once the GWB monopole is confidently detected by a PTA, the Fisher formalism will be extremely valuable in determining the next steps forward. 

\section*{Acknowledgments}

We thank Jeffrey Hazboun, Stephen Taylor, and Joseph Romano for useful discussions. This work used the Strelka Computing Cluster, which is run by Swarthmore College. YAH acknowledges support from the National Science Foundation through grant number 1820861. TLS acknowledges support from NASA (though grant number 80NSSC18K0728) and the Research Corporation. The Flatiron Institute is supported by the Simons Foundation. 

\appendix

\section{Proof that $\widetilde{c}_i = P_{||}(\ho_i)$} \label{app:proj}

In this appendix, completing Section \ref{sec:Taylor}, we prove that the vector $\vec{\widetilde{c}} \equiv \bs{H}^+ \vec{\Gamma} = (\bs{H}^+ \bs{H}) \vec{c}$ is the GWB angular distribution \emph{projected} on the $N_{\rm pair}$-dimensional space observable by a PTA. Thus, in general, $\vec{\widetilde{c}}\neq \vec{c}$.

To see this, recall that the Moore-Penrose pseudo-inverse matrix is defined as $\bs{H}^+ = \bs{H}^{\rm T} (\bs{H} \bs{H}^{\rm T})^{-1}$. As a consequence, $\bs{H}^+ \bs{H}$ is clearly a projector, since $[\bs{H}^+ \bs{H}]^2 = \bs{H}^+ \bs{H}$. Since $H_{Ii} \propto \gamma_I(\ho_i)$, $(\bs{H}^+\bs{H}) \bs{M}_{\bot} = 0$ for any map $\bs{M}_{\bot}$ orthogonal to all the $\gamma_I$ (in the limit of large $N_{\rm pix}$, and replacing sums by integrals). Moreover, we have
\barr
(\bs{H} \bs{H}^{\rm T})_{IJ} &=& \sum_i H_{Ij} H_{Ji} \nonumber\\
&=& \left(\frac38 \frac{\Delta \Omega}{4 \pi}\right)^2 \sum_i \gamma_I(\ho_i)\gamma_J(\ho_i) \nonumber\\
&\approx& \frac{(3/8)^2 \Delta \Omega}{4 \pi}  \bs{\gamma}_I \cdot \bs{\gamma}_J.
\earr
For a given pair index $I$, we define the $N_{\rm pix}$-dimensional vector $\vec{\gamma}_I$ with elements $\gamma_I(\ho_i)$. The $J$-th component of the $N_{\rm pair}$-dimensional vector $\bs{H} \vec{\gamma}_I$ is
\barr
\left[\bs{H} \vec{\gamma}_I\right]_J &=& \sum_i H_{Ji} \gamma_I(\ho_i) = \frac38 \frac{\Delta \Omega}{4 \pi} \sum_i \gamma_J(\ho_i) \gamma_I(\ho_i) \nonumber\\
&\approx& \frac38 \bs{\gamma}_J \cdot \bs{\gamma}_I = \left(\frac38 \frac{\Delta \Omega}{4 \pi}\right)^{-1} (\bs{H} \bs{H}^{\rm T})_{IJ}.
\earr
Therefore, the $K$-th component of the $N_{\rm pair}$-dimensional vector $(\bs{H} \bs{H}^{\rm T})^{-1} \bs{H} \vec{\gamma}_I$ is 
\beq
\left[(\bs{H} \bs{H}^{\rm T})^{-1} \bs{H} \vec{\gamma}_I\right]_K = \left(\frac38 \frac{\Delta \Omega}{4 \pi}\right)^{-1} \delta_{KI}.
\eeq
From this, we conclude that 
\beq
(\bs{H}^+ \bs{H})\vec{\gamma}_I = \bs{H}^{\rm T} (\bs{H} \bs{H}^{\rm T})^{-1} \bs{H} \vec{\gamma}_I = \vec{\gamma}_I.
\eeq
This achieves the proof that $\bs{H}^+\bs{H}$ is the projector on the space spanned by the $N_{\rm pair}$ maps $\bs{\gamma}_I$. Therefore, as announced, $\widetilde{c}_i = P_{||}(\ho_i)$ is the GWB angular power projected onto that $N_{\rm pair}$-dimensional space. 

\bibliography{PTA_map.bib}

\begin{thebibliography}{71}%
\makeatletter
\providecommand \@ifxundefined [1]{%
 \@ifx{#1\undefined}
}%
\providecommand \@ifnum [1]{%
 \ifnum #1\expandafter \@firstoftwo
 \else \expandafter \@secondoftwo
 \fi
}%
\providecommand \@ifx [1]{%
 \ifx #1\expandafter \@firstoftwo
 \else \expandafter \@secondoftwo
 \fi
}%
\providecommand \natexlab [1]{#1}%
\providecommand \enquote  [1]{``#1''}%
\providecommand \bibnamefont  [1]{#1}%
\providecommand \bibfnamefont [1]{#1}%
\providecommand \citenamefont [1]{#1}%
\providecommand \href@noop [0]{\@secondoftwo}%
\providecommand \href [0]{\begingroup \@sanitize@url \@href}%
\providecommand \@href[1]{\@@startlink{#1}\@@href}%
\providecommand \@@href[1]{\endgroup#1\@@endlink}%
\providecommand \@sanitize@url [0]{\catcode `\\12\catcode `\$12\catcode
  `\&12\catcode `\#12\catcode `\^12\catcode `\_12\catcode `\%12\relax}%
\providecommand \@@startlink[1]{}%
\providecommand \@@endlink[0]{}%
\providecommand \url  [0]{\begingroup\@sanitize@url \@url }%
\providecommand \@url [1]{\endgroup\@href {#1}{\urlprefix }}%
\providecommand \urlprefix  [0]{URL }%
\providecommand \Eprint [0]{\href }%
\providecommand \doibase [0]{http://dx.doi.org/}%
\providecommand \selectlanguage [0]{\@gobble}%
\providecommand \bibinfo  [0]{\@secondoftwo}%
\providecommand \bibfield  [0]{\@secondoftwo}%
\providecommand \translation [1]{[#1]}%
\providecommand \BibitemOpen [0]{}%
\providecommand \bibitemStop [0]{}%
\providecommand \bibitemNoStop [0]{.\EOS\space}%
\providecommand \EOS [0]{\spacefactor3000\relax}%
\providecommand \BibitemShut  [1]{\csname bibitem#1\endcsname}%
\let\auto@bib@innerbib\@empty
\bibitem [{\citenamefont {{Burke-Spolaor}}\ \emph {et~al.}(2019)\citenamefont
  {{Burke-Spolaor}} \emph {et~al.}}]{Burke-Spolaor:2018bvk}%
  \BibitemOpen
  \bibfield  {author} {\bibinfo {author} {\bibfnamefont {S.}~\bibnamefont
  {{Burke-Spolaor}}} \emph {et~al.},\ }\href {\doibase
  10.1007/s00159-019-0115-7} {\bibfield  {journal} {\bibinfo  {journal}
  {\aapr}\ }\textbf {\bibinfo {volume} {27}},\ \bibinfo {eid} {5} (\bibinfo
  {year} {2019})},\ \Eprint {http://arxiv.org/abs/1811.08826}
  {arXiv:1811.08826} \BibitemShut {NoStop}%
\bibitem [{\citenamefont {{Mingarelli}}(2019)}]{Mingarelli2019}%
  \BibitemOpen
  \bibfield  {author} {\bibinfo {author} {\bibfnamefont {C.~M.~F.}\
  \bibnamefont {{Mingarelli}}},\ }\href {\doibase 10.1038/s41550-018-0666-y}
  {\bibfield  {journal} {\bibinfo  {journal} {Nature Astronomy}\ }\textbf
  {\bibinfo {volume} {3}},\ \bibinfo {pages} {8} (\bibinfo {year} {2019})},\
  \Eprint {http://arxiv.org/abs/1901.06785} {arXiv:1901.06785} \BibitemShut
  {NoStop}%
\bibitem [{\citenamefont {{Taylor}}\ \emph {et~al.}(2016)\citenamefont
  {{Taylor}}, \citenamefont {{Vallisneri}}, \citenamefont {{Ellis}},
  \citenamefont {{Mingarelli}}, \citenamefont {{Lazio}},\ and\ \citenamefont
  {{van Haasteren}}}]{Taylor+:2016}%
  \BibitemOpen
  \bibfield  {author} {\bibinfo {author} {\bibfnamefont {S.~R.}\ \bibnamefont
  {{Taylor}}}, \bibinfo {author} {\bibfnamefont {M.}~\bibnamefont
  {{Vallisneri}}}, \bibinfo {author} {\bibfnamefont {J.~A.}\ \bibnamefont
  {{Ellis}}}, \bibinfo {author} {\bibfnamefont {C.~M.~F.}\ \bibnamefont
  {{Mingarelli}}}, \bibinfo {author} {\bibfnamefont {T.~J.~W.}\ \bibnamefont
  {{Lazio}}}, \ and\ \bibinfo {author} {\bibfnamefont {R.}~\bibnamefont {{van
  Haasteren}}},\ }\href {\doibase 10.3847/2041-8205/819/1/L6} {\bibfield
  {journal} {\bibinfo  {journal} {\apjl}\ }\textbf {\bibinfo {volume} {819}},\
  \bibinfo {eid} {L6} (\bibinfo {year} {2016})},\ \Eprint
  {http://arxiv.org/abs/1511.05564} {arXiv:1511.05564} \BibitemShut {NoStop}%
\bibitem [{\citenamefont {Kelley}\ \emph {et~al.}(2017)\citenamefont {Kelley},
  \citenamefont {Blecha}, \citenamefont {Hernquist}, \citenamefont {Sesana},\
  and\ \citenamefont {Taylor}}]{Kelley:2017lek}%
  \BibitemOpen
  \bibfield  {author} {\bibinfo {author} {\bibfnamefont {L.~Z.}\ \bibnamefont
  {Kelley}}, \bibinfo {author} {\bibfnamefont {L.}~\bibnamefont {Blecha}},
  \bibinfo {author} {\bibfnamefont {L.}~\bibnamefont {Hernquist}}, \bibinfo
  {author} {\bibfnamefont {A.}~\bibnamefont {Sesana}}, \ and\ \bibinfo {author}
  {\bibfnamefont {S.~R.}\ \bibnamefont {Taylor}},\ }\href {\doibase
  10.1093/mnras/stx1638} {\bibfield  {journal} {\bibinfo  {journal} {Mon. Not.
  Roy. Astron. Soc.}\ }\textbf {\bibinfo {volume} {471}},\ \bibinfo {pages}
  {4508} (\bibinfo {year} {2017})},\ \Eprint {http://arxiv.org/abs/1702.02180}
  {arXiv:1702.02180} \BibitemShut {NoStop}%
\bibitem [{\citenamefont {{Sesana}}(2013)}]{Sesana13}%
  \BibitemOpen
  \bibfield  {author} {\bibinfo {author} {\bibfnamefont {A.}~\bibnamefont
  {{Sesana}}},\ }\href {\doibase 10.1088/0264-9381/30/22/224014} {\bibfield
  {journal} {\bibinfo  {journal} {Classical and Quantum Gravity}\ }\textbf
  {\bibinfo {volume} {30}},\ \bibinfo {eid} {224014} (\bibinfo {year}
  {2013})},\ \Eprint {http://arxiv.org/abs/1307.2600} {arXiv:1307.2600
  [astro-ph.CO]} \BibitemShut {NoStop}%
\bibitem [{\citenamefont {Simon}\ and\ \citenamefont
  {Burke-Spolaor}(2016)}]{Simon:2016ibt}%
  \BibitemOpen
  \bibfield  {author} {\bibinfo {author} {\bibfnamefont {J.}~\bibnamefont
  {Simon}}\ and\ \bibinfo {author} {\bibfnamefont {S.}~\bibnamefont
  {Burke-Spolaor}},\ }\href {\doibase 10.3847/0004-637X/826/1/11} {\bibfield
  {journal} {\bibinfo  {journal} {Astrophys. J.}\ }\textbf {\bibinfo {volume}
  {826}},\ \bibinfo {pages} {11} (\bibinfo {year} {2016})},\ \Eprint
  {http://arxiv.org/abs/1603.06577} {arXiv:1603.06577} \BibitemShut {NoStop}%
\bibitem [{\citenamefont {{Mingarelli}}\ \emph {et~al.}(2017)\citenamefont
  {{Mingarelli}}, \citenamefont {{Lazio}}, \citenamefont {{Sesana}},
  \citenamefont {{Greene}}, \citenamefont {{Ellis}}, \citenamefont {{Ma}},
  \citenamefont {{Croft}}, \citenamefont {{Burke-Spolaor}},\ and\ \citenamefont
  {{Taylor}}}]{2017NatAs...1..886M}%
  \BibitemOpen
  \bibfield  {author} {\bibinfo {author} {\bibfnamefont {C.~M.~F.}\
  \bibnamefont {{Mingarelli}}}, \bibinfo {author} {\bibfnamefont {T.~J.~W.}\
  \bibnamefont {{Lazio}}}, \bibinfo {author} {\bibfnamefont {A.}~\bibnamefont
  {{Sesana}}}, \bibinfo {author} {\bibfnamefont {J.~E.}\ \bibnamefont
  {{Greene}}}, \bibinfo {author} {\bibfnamefont {J.~A.}\ \bibnamefont
  {{Ellis}}}, \bibinfo {author} {\bibfnamefont {C.-P.}\ \bibnamefont {{Ma}}},
  \bibinfo {author} {\bibfnamefont {S.}~\bibnamefont {{Croft}}}, \bibinfo
  {author} {\bibfnamefont {S.}~\bibnamefont {{Burke-Spolaor}}}, \ and\ \bibinfo
  {author} {\bibfnamefont {S.~R.}\ \bibnamefont {{Taylor}}},\ }\href {\doibase
  10.1038/s41550-017-0299-6} {\bibfield  {journal} {\bibinfo  {journal} {Nature
  Astronomy}\ }\textbf {\bibinfo {volume} {1}},\ \bibinfo {pages} {886}
  (\bibinfo {year} {2017})},\ \Eprint {http://arxiv.org/abs/1708.03491}
  {arXiv:1708.03491} \BibitemShut {NoStop}%
\bibitem [{\citenamefont {{Siemens}}\ \emph {et~al.}(2019)\citenamefont
  {{Siemens}}, \citenamefont {{Hazboun}}, \citenamefont {{Baker}},
  \citenamefont {{Burke-Spolaor}}, \citenamefont {{Madison}}, \citenamefont
  {{Mingarelli}}, \citenamefont {{Simon}},\ and\ \citenamefont
  {{Smith}}}]{siemensetal2019}%
  \BibitemOpen
  \bibfield  {author} {\bibinfo {author} {\bibfnamefont {X.}~\bibnamefont
  {{Siemens}}}, \bibinfo {author} {\bibfnamefont {J.~S.}\ \bibnamefont
  {{Hazboun}}}, \bibinfo {author} {\bibfnamefont {P.~T.}\ \bibnamefont
  {{Baker}}}, \bibinfo {author} {\bibfnamefont {S.}~\bibnamefont
  {{Burke-Spolaor}}}, \bibinfo {author} {\bibfnamefont {D.}~\bibnamefont
  {{Madison}}}, \bibinfo {author} {\bibfnamefont {C.}~\bibnamefont
  {{Mingarelli}}}, \bibinfo {author} {\bibfnamefont {J.}~\bibnamefont
  {{Simon}}}, \ and\ \bibinfo {author} {\bibfnamefont {T.}~\bibnamefont
  {{Smith}}},\ }\href@noop {} {\  (\bibinfo {year} {2019})},\ \Eprint
  {http://arxiv.org/abs/1907.04960} {arXiv:1907.04960} \BibitemShut {NoStop}%
\bibitem [{\citenamefont {{Lasky}}\ \emph {et~al.}(2016)\citenamefont {{Lasky}}
  \emph {et~al.}}]{lms+16}%
  \BibitemOpen
  \bibfield  {author} {\bibinfo {author} {\bibfnamefont {P.~D.}\ \bibnamefont
  {{Lasky}}} \emph {et~al.},\ }\href {\doibase 10.1103/PhysRevX.6.011035}
  {\bibfield  {journal} {\bibinfo  {journal} {Physical Review X}\ }\textbf
  {\bibinfo {volume} {6}},\ \bibinfo {eid} {011035} (\bibinfo {year} {2016})},\
  \Eprint {http://arxiv.org/abs/1511.05994} {arXiv:1511.05994} \BibitemShut
  {NoStop}%
\bibitem [{Note1()}]{Note1}%
  \BibitemOpen
  \bibinfo {note} {\protect \url {http://nanograv.org}}\BibitemShut {NoStop}%
\bibitem [{Note2()}]{Note2}%
  \BibitemOpen
  \bibinfo {note} {\protect \url {http://www.epta.eu.org}}\BibitemShut
  {NoStop}%
\bibitem [{Note3()}]{Note3}%
  \BibitemOpen
  \bibinfo {note} {\protect \url
  {https://www.atnf.csiro.au/research/pulsar/ppta/}}\BibitemShut {NoStop}%
\bibitem [{Note4()}]{Note4}%
  \BibitemOpen
  \bibinfo {note} {\protect \url {http://ipta4gw.org}}\BibitemShut {NoStop}%
\bibitem [{Note5()}]{Note5}%
  \BibitemOpen
  \bibinfo {note} {\protect \url {https://www.skatelescope.org}}\BibitemShut
  {NoStop}%
\bibitem [{\citenamefont {{Keane}}\ \emph {et~al.}(2015)\citenamefont {{Keane}}
  \emph {et~al.}}]{Keane_15}%
  \BibitemOpen
  \bibfield  {author} {\bibinfo {author} {\bibfnamefont {E.}~\bibnamefont
  {{Keane}}} \emph {et~al.},\ }in\ \href@noop {} {\emph {\bibinfo {booktitle}
  {Advancing Astrophysics with the Square Kilometre Array (AASKA14)}}}\
  (\bibinfo {year} {2015})\ p.~\bibinfo {pages} {40},\ \Eprint
  {http://arxiv.org/abs/1501.00056} {arXiv:1501.00056} \BibitemShut {NoStop}%
\bibitem [{\citenamefont {{Janssen}}\ \emph {et~al.}(2015)\citenamefont
  {{Janssen}} \emph {et~al.}}]{Janssen_15}%
  \BibitemOpen
  \bibfield  {author} {\bibinfo {author} {\bibfnamefont {G.}~\bibnamefont
  {{Janssen}}} \emph {et~al.},\ }in\ \href@noop {} {\emph {\bibinfo {booktitle}
  {Advancing Astrophysics with the Square Kilometre Array (AASKA14)}}}\
  (\bibinfo {year} {2015})\ p.~\bibinfo {pages} {37},\ \Eprint
  {http://arxiv.org/abs/1501.00127} {arXiv:1501.00127} \BibitemShut {NoStop}%
\bibitem [{\citenamefont {Ali-Ha\"\i{}moud}\ \emph {et~al.}(2020)\citenamefont
  {Ali-Ha\"\i{}moud}, \citenamefont {Smith},\ and\ \citenamefont
  {Mingarelli}}]{Ali-Haimoud:2020ozu}%
  \BibitemOpen
  \bibfield  {author} {\bibinfo {author} {\bibfnamefont {Y.}~\bibnamefont
  {Ali-Ha\"\i{}moud}}, \bibinfo {author} {\bibfnamefont {T.~L.}\ \bibnamefont
  {Smith}}, \ and\ \bibinfo {author} {\bibfnamefont {C.~M.}\ \bibnamefont
  {Mingarelli}},\ }\href@noop {} {\  (\bibinfo {year} {2020})},\ \Eprint
  {http://arxiv.org/abs/2006.14570} {arXiv:2006.14570} \BibitemShut {NoStop}%
\bibitem [{\citenamefont {Arzoumanian}\ \emph {et~al.}(2020)\citenamefont
  {Arzoumanian} \emph {et~al.}}]{Arzoumanian:2020vkk}%
  \BibitemOpen
  \bibfield  {author} {\bibinfo {author} {\bibfnamefont {Z.}~\bibnamefont
  {Arzoumanian}} \emph {et~al.} (\bibinfo {collaboration} {NANOGrav}),\
  }\href@noop {} {\  (\bibinfo {year} {2020})},\ \Eprint
  {http://arxiv.org/abs/2009.04496} {arXiv:2009.04496} \BibitemShut {NoStop}%
\bibitem [{\citenamefont {{Arzoumanian}}\ \emph {et~al.}(2018)\citenamefont
  {{Arzoumanian}}, \citenamefont {others},\ and\ \citenamefont {{NANOGrav
  Collaboration}}}]{Arzoumanian_18}%
  \BibitemOpen
  \bibfield  {author} {\bibinfo {author} {\bibfnamefont {Z.}~\bibnamefont
  {{Arzoumanian}}}, \bibinfo {author} {\bibnamefont {others}}, \ and\ \bibinfo
  {author} {\bibnamefont {{NANOGrav Collaboration}}},\ }\href {\doibase
  10.3847/1538-4357/aabd3b} {\bibfield  {journal} {\bibinfo  {journal} {\apj}\
  }\textbf {\bibinfo {volume} {859}},\ \bibinfo {eid} {47} (\bibinfo {year}
  {2018})},\ \Eprint {http://arxiv.org/abs/1801.02617} {arXiv:1801.02617}
  \BibitemShut {NoStop}%
\bibitem [{\citenamefont {{Lentati}}\ \emph {et~al.}(2015)\citenamefont
  {{Lentati}} \emph {et~al.}}]{Lentati_15}%
  \BibitemOpen
  \bibfield  {author} {\bibinfo {author} {\bibfnamefont {L.}~\bibnamefont
  {{Lentati}}} \emph {et~al.} (\bibinfo {collaboration} {EPTA collaboration}),\
  }\href {\doibase 10.1093/mnras/stv1538} {\bibfield  {journal} {\bibinfo
  {journal} {\mnras}\ }\textbf {\bibinfo {volume} {453}},\ \bibinfo {pages}
  {2576} (\bibinfo {year} {2015})},\ \Eprint {http://arxiv.org/abs/1504.03692}
  {arXiv:1504.03692} \BibitemShut {NoStop}%
\bibitem [{\citenamefont {{Shannon}}\ \emph {et~al.}(2015)\citenamefont
  {{Shannon}} \emph {et~al.}}]{Shannon_15}%
  \BibitemOpen
  \bibfield  {author} {\bibinfo {author} {\bibfnamefont {R.~M.}\ \bibnamefont
  {{Shannon}}} \emph {et~al.},\ }\href {\doibase 10.1126/science.aab1910}
  {\bibfield  {journal} {\bibinfo  {journal} {Science}\ }\textbf {\bibinfo
  {volume} {349}},\ \bibinfo {pages} {1522} (\bibinfo {year} {2015})},\ \Eprint
  {http://arxiv.org/abs/1509.07320} {arXiv:1509.07320} \BibitemShut {NoStop}%
\bibitem [{\citenamefont {{Verbiest}}\ \emph {et~al.}(2016)\citenamefont
  {{Verbiest}},  \emph {et~al.}}]{Verbiest_16}%
  \BibitemOpen
  \bibfield  {author} {\bibinfo {author} {\bibfnamefont {J.~P.~W.}\
  \bibnamefont {{Verbiest}}}, ,  \emph {et~al.} (\bibinfo {collaboration} {IPTA
  collaboration}),\ }\href {\doibase 10.1093/mnras/stw347} {\bibfield
  {journal} {\bibinfo  {journal} {\mnras}\ }\textbf {\bibinfo {volume} {458}},\
  \bibinfo {pages} {1267} (\bibinfo {year} {2016})},\ \Eprint
  {http://arxiv.org/abs/1602.03640} {arXiv:1602.03640} \BibitemShut {NoStop}%
\bibitem [{\citenamefont {Mingarelli}\ \emph {et~al.}(2013)\citenamefont
  {Mingarelli}, \citenamefont {Sidery}, \citenamefont {Mandel},\ and\
  \citenamefont {Vecchio}}]{Mingarelli:2013dsa}%
  \BibitemOpen
  \bibfield  {author} {\bibinfo {author} {\bibfnamefont {C.~M.}\ \bibnamefont
  {Mingarelli}}, \bibinfo {author} {\bibfnamefont {T.}~\bibnamefont {Sidery}},
  \bibinfo {author} {\bibfnamefont {I.}~\bibnamefont {Mandel}}, \ and\ \bibinfo
  {author} {\bibfnamefont {A.}~\bibnamefont {Vecchio}},\ }\href {\doibase
  10.1103/PhysRevD.88.062005} {\bibfield  {journal} {\bibinfo  {journal} {Phys.
  Rev. D}\ }\textbf {\bibinfo {volume} {88}},\ \bibinfo {pages} {062005}
  (\bibinfo {year} {2013})},\ \Eprint {http://arxiv.org/abs/1306.5394}
  {arXiv:1306.5394} \BibitemShut {NoStop}%
\bibitem [{\citenamefont {Blanco-Pillado}\ and\ \citenamefont
  {Olum}(2017)}]{Blanco-Pillado:2017oxo}%
  \BibitemOpen
  \bibfield  {author} {\bibinfo {author} {\bibfnamefont {J.~J.}\ \bibnamefont
  {Blanco-Pillado}}\ and\ \bibinfo {author} {\bibfnamefont {K.~D.}\
  \bibnamefont {Olum}},\ }\href {\doibase 10.1103/PhysRevD.96.104046}
  {\bibfield  {journal} {\bibinfo  {journal} {Phys. Rev.}\ }\textbf {\bibinfo
  {volume} {D96}},\ \bibinfo {pages} {104046} (\bibinfo {year} {2017})},\
  \Eprint {http://arxiv.org/abs/1709.02693} {arXiv:1709.02693} \BibitemShut
  {NoStop}%
\bibitem [{\citenamefont {Olmez}\ \emph {et~al.}(2012)\citenamefont {Olmez},
  \citenamefont {Mandic},\ and\ \citenamefont {Siemens}}]{Olmez:2011cg}%
  \BibitemOpen
  \bibfield  {author} {\bibinfo {author} {\bibfnamefont {S.}~\bibnamefont
  {Olmez}}, \bibinfo {author} {\bibfnamefont {V.}~\bibnamefont {Mandic}}, \
  and\ \bibinfo {author} {\bibfnamefont {X.}~\bibnamefont {Siemens}},\ }\href
  {\doibase 10.1088/1475-7516/2012/07/009} {\bibfield  {journal} {\bibinfo
  {journal} {JCAP}\ }\textbf {\bibinfo {volume} {1207}},\ \bibinfo {pages}
  {009} (\bibinfo {year} {2012})},\ \Eprint {http://arxiv.org/abs/1106.5555}
  {arXiv:1106.5555} \BibitemShut {NoStop}%
\bibitem [{\citenamefont {{Jenkins}}\ and\ \citenamefont
  {{Sakellariadou}}(2018)}]{Jenkins_18}%
  \BibitemOpen
  \bibfield  {author} {\bibinfo {author} {\bibfnamefont {A.~C.}\ \bibnamefont
  {{Jenkins}}}\ and\ \bibinfo {author} {\bibfnamefont {M.}~\bibnamefont
  {{Sakellariadou}}},\ }\href {\doibase 10.1103/PhysRevD.98.063509} {\bibfield
  {journal} {\bibinfo  {journal} {\prd}\ }\textbf {\bibinfo {volume} {98}},\
  \bibinfo {eid} {063509} (\bibinfo {year} {2018})},\ \Eprint
  {http://arxiv.org/abs/1802.06046} {arXiv:1802.06046} \BibitemShut {NoStop}%
\bibitem [{\citenamefont {{Liu}}\ \emph {et~al.}(2020)\citenamefont {{Liu}},
  \citenamefont {{Cai}},\ and\ \citenamefont {{Guo}}}]{Liu_20}%
  \BibitemOpen
  \bibfield  {author} {\bibinfo {author} {\bibfnamefont {J.}~\bibnamefont
  {{Liu}}}, \bibinfo {author} {\bibfnamefont {R.-G.}\ \bibnamefont {{Cai}}}, \
  and\ \bibinfo {author} {\bibfnamefont {Z.-K.}\ \bibnamefont {{Guo}}},\
  }\href@noop {} {\bibfield  {journal} {\bibinfo  {journal} {arXiv e-prints}\
  ,\ \bibinfo {eid} {arXiv:2010.03225}} (\bibinfo {year} {2020})},\ \Eprint
  {http://arxiv.org/abs/2010.03225} {arXiv:2010.03225} \BibitemShut {NoStop}%
\bibitem [{\citenamefont {Gair}\ \emph {et~al.}(2014)\citenamefont {Gair},
  \citenamefont {Romano}, \citenamefont {Taylor},\ and\ \citenamefont
  {Mingarelli}}]{Gair:2014rwa}%
  \BibitemOpen
  \bibfield  {author} {\bibinfo {author} {\bibfnamefont {J.}~\bibnamefont
  {Gair}}, \bibinfo {author} {\bibfnamefont {J.~D.}\ \bibnamefont {Romano}},
  \bibinfo {author} {\bibfnamefont {S.}~\bibnamefont {Taylor}}, \ and\ \bibinfo
  {author} {\bibfnamefont {C.~M.~F.}\ \bibnamefont {Mingarelli}},\ }\href
  {\doibase 10.1103/PhysRevD.90.082001} {\bibfield  {journal} {\bibinfo
  {journal} {Phys. Rev. D}\ }\textbf {\bibinfo {volume} {90}},\ \bibinfo
  {pages} {082001} (\bibinfo {year} {2014})},\ \Eprint
  {http://arxiv.org/abs/1406.4664} {arXiv:1406.4664} \BibitemShut {NoStop}%
\bibitem [{\citenamefont {{Hotinli}}\ \emph {et~al.}(2019)\citenamefont
  {{Hotinli}}, \citenamefont {{Kamionkowski}},\ and\ \citenamefont
  {{Jaffe}}}]{Hotinli_19}%
  \BibitemOpen
  \bibfield  {author} {\bibinfo {author} {\bibfnamefont {S.~C.}\ \bibnamefont
  {{Hotinli}}}, \bibinfo {author} {\bibfnamefont {M.}~\bibnamefont
  {{Kamionkowski}}}, \ and\ \bibinfo {author} {\bibfnamefont {A.~H.}\
  \bibnamefont {{Jaffe}}},\ }\href@noop {} {\bibfield  {journal} {\bibinfo
  {journal} {arXiv e-prints}\ ,\ \bibinfo {eid} {arXiv:1904.05348}} (\bibinfo
  {year} {2019})},\ \Eprint {http://arxiv.org/abs/1904.05348}
  {arXiv:1904.05348} \BibitemShut {NoStop}%
\bibitem [{\citenamefont {Taylor}\ and\ \citenamefont
  {Gair}(2013)}]{Taylor:2013esa}%
  \BibitemOpen
  \bibfield  {author} {\bibinfo {author} {\bibfnamefont {S.~R.}\ \bibnamefont
  {Taylor}}\ and\ \bibinfo {author} {\bibfnamefont {J.~R.}\ \bibnamefont
  {Gair}},\ }\href {\doibase 10.1103/PhysRevD.88.084001} {\bibfield  {journal}
  {\bibinfo  {journal} {Phys. Rev. D}\ }\textbf {\bibinfo {volume} {88}},\
  \bibinfo {pages} {084001} (\bibinfo {year} {2013})},\ \Eprint
  {http://arxiv.org/abs/1306.5395} {arXiv:1306.5395} \BibitemShut {NoStop}%
\bibitem [{\citenamefont {{Taylor}}\ \emph {et~al.}(2020)\citenamefont
  {{Taylor}}, \citenamefont {{van Haasteren}},\ and\ \citenamefont
  {{Sesana}}}]{Taylor_20}%
  \BibitemOpen
  \bibfield  {author} {\bibinfo {author} {\bibfnamefont {S.~R.}\ \bibnamefont
  {{Taylor}}}, \bibinfo {author} {\bibfnamefont {R.}~\bibnamefont {{van
  Haasteren}}}, \ and\ \bibinfo {author} {\bibfnamefont {A.}~\bibnamefont
  {{Sesana}}},\ }\href@noop {} {\bibfield  {journal} {\bibinfo  {journal}
  {arXiv e-prints}\ ,\ \bibinfo {eid} {arXiv:2006.04810}} (\bibinfo {year}
  {2020})},\ \Eprint {http://arxiv.org/abs/2006.04810} {arXiv:2006.04810}
  \BibitemShut {NoStop}%
\bibitem [{\citenamefont {Taylor}\ \emph {et~al.}(2015)\citenamefont {Taylor}
  \emph {et~al.}}]{Taylor:2015udp}%
  \BibitemOpen
  \bibfield  {author} {\bibinfo {author} {\bibfnamefont {S.~R.}\ \bibnamefont
  {Taylor}} \emph {et~al.},\ }\href {\doibase 10.1103/PhysRevLett.115.041101}
  {\bibfield  {journal} {\bibinfo  {journal} {Phys. Rev. Lett.}\ }\textbf
  {\bibinfo {volume} {115}},\ \bibinfo {pages} {041101} (\bibinfo {year}
  {2015})},\ \Eprint {http://arxiv.org/abs/1506.08817} {arXiv:1506.08817}
  \BibitemShut {NoStop}%
\bibitem [{\citenamefont {Babak}\ \emph {et~al.}(2016)\citenamefont {Babak}
  \emph {et~al.}}]{Babak:2015lua}%
  \BibitemOpen
  \bibfield  {author} {\bibinfo {author} {\bibfnamefont {S.}~\bibnamefont
  {Babak}} \emph {et~al.},\ }\href {\doibase 10.1093/mnras/stv2092} {\bibfield
  {journal} {\bibinfo  {journal} {Mon. Not. Roy. Astron. Soc.}\ }\textbf
  {\bibinfo {volume} {455}},\ \bibinfo {pages} {1665} (\bibinfo {year}
  {2016})},\ \Eprint {http://arxiv.org/abs/1509.02165} {arXiv:1509.02165}
  \BibitemShut {NoStop}%
\bibitem [{\citenamefont {Hazboun}\ \emph
  {et~al.}(2019{\natexlab{a}})\citenamefont {Hazboun}, \citenamefont {Romano},\
  and\ \citenamefont {Smith}}]{Hazboun:2019nqt}%
  \BibitemOpen
  \bibfield  {author} {\bibinfo {author} {\bibfnamefont {J.}~\bibnamefont
  {Hazboun}}, \bibinfo {author} {\bibfnamefont {J.}~\bibnamefont {Romano}}, \
  and\ \bibinfo {author} {\bibfnamefont {T.}~\bibnamefont {Smith}},\ }\href
  {\doibase 10.21105/joss.01775} {\bibfield  {journal} {\bibinfo  {journal} {J.
  Open Source Softw.}\ }\textbf {\bibinfo {volume} {4}},\ \bibinfo {pages}
  {1775} (\bibinfo {year} {2019}{\natexlab{a}})}\BibitemShut {NoStop}%
\bibitem [{\citenamefont {Hazboun}\ \emph
  {et~al.}(2019{\natexlab{b}})\citenamefont {Hazboun}, \citenamefont {Romano},\
  and\ \citenamefont {Smith}}]{Hazboun:2019vhv}%
  \BibitemOpen
  \bibfield  {author} {\bibinfo {author} {\bibfnamefont {J.~S.}\ \bibnamefont
  {Hazboun}}, \bibinfo {author} {\bibfnamefont {J.~D.}\ \bibnamefont {Romano}},
  \ and\ \bibinfo {author} {\bibfnamefont {T.~L.}\ \bibnamefont {Smith}},\
  }\href {\doibase 10.1103/PhysRevD.100.104028} {\bibfield  {journal} {\bibinfo
   {journal} {Phys. Rev. D}\ }\textbf {\bibinfo {volume} {100}},\ \bibinfo
  {pages} {104028} (\bibinfo {year} {2019}{\natexlab{b}})},\ \Eprint
  {http://arxiv.org/abs/1907.04341} {arXiv:1907.04341} \BibitemShut {NoStop}%
\bibitem [{\citenamefont {{Tiburzi}}\ \emph {et~al.}(2016)\citenamefont
  {{Tiburzi}}, \citenamefont {{Hobbs}}, \citenamefont {{Kerr}}, \citenamefont
  {{Coles}}, \citenamefont {{Dai}}, \citenamefont {{Manchester}}, \citenamefont
  {{Possenti}}, \citenamefont {{Shannon}},\ and\ \citenamefont
  {{You}}}]{Tiburzi16}%
  \BibitemOpen
  \bibfield  {author} {\bibinfo {author} {\bibfnamefont {C.}~\bibnamefont
  {{Tiburzi}}}, \bibinfo {author} {\bibfnamefont {G.}~\bibnamefont {{Hobbs}}},
  \bibinfo {author} {\bibfnamefont {M.}~\bibnamefont {{Kerr}}}, \bibinfo
  {author} {\bibfnamefont {W.~A.}\ \bibnamefont {{Coles}}}, \bibinfo {author}
  {\bibfnamefont {S.}~\bibnamefont {{Dai}}}, \bibinfo {author} {\bibfnamefont
  {R.~N.}\ \bibnamefont {{Manchester}}}, \bibinfo {author} {\bibfnamefont
  {A.}~\bibnamefont {{Possenti}}}, \bibinfo {author} {\bibfnamefont {R.~M.}\
  \bibnamefont {{Shannon}}}, \ and\ \bibinfo {author} {\bibfnamefont {X.~P.}\
  \bibnamefont {{You}}},\ }\href {\doibase 10.1093/mnras/stv2143} {\bibfield
  {journal} {\bibinfo  {journal} {\mnras}\ }\textbf {\bibinfo {volume} {455}},\
  \bibinfo {pages} {4339} (\bibinfo {year} {2016})},\ \Eprint
  {http://arxiv.org/abs/1510.02363} {arXiv:1510.02363 [astro-ph.IM]}
  \BibitemShut {NoStop}%
\bibitem [{\citenamefont {{Vallisneri}}\ \emph {et~al.}(2020)\citenamefont
  {{Vallisneri}} \emph {et~al.}}]{Vallisneri20}%
  \BibitemOpen
  \bibfield  {author} {\bibinfo {author} {\bibfnamefont {M.}~\bibnamefont
  {{Vallisneri}}} \emph {et~al.},\ }\href {\doibase 10.3847/1538-4357/ab7b67}
  {\bibfield  {journal} {\bibinfo  {journal} {\apj}\ }\textbf {\bibinfo
  {volume} {893}},\ \bibinfo {eid} {112} (\bibinfo {year} {2020})},\ \Eprint
  {http://arxiv.org/abs/2001.00595} {arXiv:2001.00595} \BibitemShut {NoStop}%
\bibitem [{\citenamefont {Hellings}\ and\ \citenamefont
  {Downs}(1983)}]{Hellings:1983fr}%
  \BibitemOpen
  \bibfield  {author} {\bibinfo {author} {\bibfnamefont {R.~w.}\ \bibnamefont
  {Hellings}}\ and\ \bibinfo {author} {\bibfnamefont {G.~s.}\ \bibnamefont
  {Downs}},\ }\href {\doibase 10.1086/183954} {\bibfield  {journal} {\bibinfo
  {journal} {Astrophys. J.}\ }\textbf {\bibinfo {volume} {265}},\ \bibinfo
  {pages} {L39} (\bibinfo {year} {1983})}\BibitemShut {NoStop}%
\bibitem [{\citenamefont {{Hazboun}}\ \emph {et~al.}(2019)\citenamefont
  {{Hazboun}}, \citenamefont {{Romano}},\ and\ \citenamefont
  {{Smith}}}]{Hazboun_19}%
  \BibitemOpen
  \bibfield  {author} {\bibinfo {author} {\bibfnamefont {J.~S.}\ \bibnamefont
  {{Hazboun}}}, \bibinfo {author} {\bibfnamefont {J.~D.}\ \bibnamefont
  {{Romano}}}, \ and\ \bibinfo {author} {\bibfnamefont {T.~L.}\ \bibnamefont
  {{Smith}}},\ }\href@noop {} {\bibfield  {journal} {\bibinfo  {journal} {arXiv
  e-prints}\ ,\ \bibinfo {eid} {arXiv:1907.04341}} (\bibinfo {year} {2019})},\
  \Eprint {http://arxiv.org/abs/1907.04341} {arXiv:1907.04341} \BibitemShut
  {NoStop}%
\bibitem [{\citenamefont {Phinney}(2001)}]{Phinney:2001di}%
  \BibitemOpen
  \bibfield  {author} {\bibinfo {author} {\bibfnamefont {E.~S.}\ \bibnamefont
  {Phinney}},\ }\href@noop {} {\  (\bibinfo {year} {2001})},\ \Eprint
  {http://arxiv.org/abs/astro-ph/0108028} {arXiv:astro-ph/0108028} \BibitemShut
  {NoStop}%
\bibitem [{\citenamefont {{Chen}}\ \emph {et~al.}(2017)\citenamefont {{Chen}},
  \citenamefont {{Sesana}},\ and\ \citenamefont {{Del Pozzo}}}]{Chen_17}%
  \BibitemOpen
  \bibfield  {author} {\bibinfo {author} {\bibfnamefont {S.}~\bibnamefont
  {{Chen}}}, \bibinfo {author} {\bibfnamefont {A.}~\bibnamefont {{Sesana}}}, \
  and\ \bibinfo {author} {\bibfnamefont {W.}~\bibnamefont {{Del Pozzo}}},\
  }\href {\doibase 10.1093/mnras/stx1093} {\bibfield  {journal} {\bibinfo
  {journal} {\mnras}\ }\textbf {\bibinfo {volume} {470}},\ \bibinfo {pages}
  {1738} (\bibinfo {year} {2017})},\ \Eprint {http://arxiv.org/abs/1612.00455}
  {arXiv:1612.00455} \BibitemShut {NoStop}%
\bibitem [{\citenamefont {Anholm}\ \emph {et~al.}(2009)\citenamefont {Anholm},
  \citenamefont {Ballmer}, \citenamefont {Creighton}, \citenamefont {Price},\
  and\ \citenamefont {Siemens}}]{Anholm:2008wy}%
  \BibitemOpen
  \bibfield  {author} {\bibinfo {author} {\bibfnamefont {M.}~\bibnamefont
  {Anholm}}, \bibinfo {author} {\bibfnamefont {S.}~\bibnamefont {Ballmer}},
  \bibinfo {author} {\bibfnamefont {J.~D.~E.}\ \bibnamefont {Creighton}},
  \bibinfo {author} {\bibfnamefont {L.~R.}\ \bibnamefont {Price}}, \ and\
  \bibinfo {author} {\bibfnamefont {X.}~\bibnamefont {Siemens}},\ }\href
  {\doibase 10.1103/PhysRevD.79.084030} {\bibfield  {journal} {\bibinfo
  {journal} {Phys. Rev.}\ }\textbf {\bibinfo {volume} {D79}},\ \bibinfo {pages}
  {084030} (\bibinfo {year} {2009})},\ \Eprint {http://arxiv.org/abs/0809.0701}
  {arXiv:0809.0701} \BibitemShut {NoStop}%
\bibitem [{\citenamefont {Siemens}\ \emph {et~al.}(2013)\citenamefont
  {Siemens}, \citenamefont {Ellis}, \citenamefont {Jenet},\ and\ \citenamefont
  {Romano}}]{Siemens:2013zla}%
  \BibitemOpen
  \bibfield  {author} {\bibinfo {author} {\bibfnamefont {X.}~\bibnamefont
  {Siemens}}, \bibinfo {author} {\bibfnamefont {J.}~\bibnamefont {Ellis}},
  \bibinfo {author} {\bibfnamefont {F.}~\bibnamefont {Jenet}}, \ and\ \bibinfo
  {author} {\bibfnamefont {J.~D.}\ \bibnamefont {Romano}},\ }\href {\doibase
  10.1088/0264-9381/30/22/224015} {\bibfield  {journal} {\bibinfo  {journal}
  {Class. Quant. Grav.}\ }\textbf {\bibinfo {volume} {30}},\ \bibinfo {pages}
  {224015} (\bibinfo {year} {2013})},\ \Eprint {http://arxiv.org/abs/1305.3196}
  {arXiv:1305.3196} \BibitemShut {NoStop}%
\bibitem [{\citenamefont {{Desvignes}}\ \emph {et~al.}(2016)\citenamefont
  {{Desvignes}} \emph {et~al.}}]{Desvignes_16}%
  \BibitemOpen
  \bibfield  {author} {\bibinfo {author} {\bibfnamefont {G.}~\bibnamefont
  {{Desvignes}}} \emph {et~al.} (\bibinfo {collaboration} {EPTA
  collaboration}),\ }\href {\doibase 10.1093/mnras/stw483} {\bibfield
  {journal} {\bibinfo  {journal} {\mnras}\ }\textbf {\bibinfo {volume} {458}},\
  \bibinfo {pages} {3341} (\bibinfo {year} {2016})},\ \Eprint
  {http://arxiv.org/abs/1602.08511} {arXiv:1602.08511} \BibitemShut {NoStop}%
\bibitem [{\citenamefont {{Arzoumanian}}\ \emph {et~al.}(2015)\citenamefont
  {{Arzoumanian}} \emph {et~al.}}]{2015ApJ...813...65N}%
  \BibitemOpen
  \bibfield  {author} {\bibinfo {author} {\bibfnamefont {Z.}~\bibnamefont
  {{Arzoumanian}}} \emph {et~al.} (\bibinfo {collaboration} {NANOGrav
  Collaboration}),\ }\href {\doibase 10.1088/0004-637X/813/1/65} {\bibfield
  {journal} {\bibinfo  {journal} {\apj}\ }\textbf {\bibinfo {volume} {813}},\
  \bibinfo {eid} {65} (\bibinfo {year} {2015})},\ \Eprint
  {http://arxiv.org/abs/1505.07540} {arXiv:1505.07540} \BibitemShut {NoStop}%
\bibitem [{\citenamefont {{Reich}}\ and\ \citenamefont
  {{Reich}}(1986)}]{Reich_86}%
  \BibitemOpen
  \bibfield  {author} {\bibinfo {author} {\bibfnamefont {P.}~\bibnamefont
  {{Reich}}}\ and\ \bibinfo {author} {\bibfnamefont {W.}~\bibnamefont
  {{Reich}}},\ }\href@noop {} {\bibfield  {journal} {\bibinfo  {journal}
  {\aaps}\ }\textbf {\bibinfo {volume} {63}},\ \bibinfo {pages} {205} (\bibinfo
  {year} {1986})}\BibitemShut {NoStop}%
\bibitem [{\citenamefont {{Reich}}\ \emph {et~al.}(2001)\citenamefont
  {{Reich}}, \citenamefont {{Testori}},\ and\ \citenamefont
  {{Reich}}}]{Reich_01}%
  \BibitemOpen
  \bibfield  {author} {\bibinfo {author} {\bibfnamefont {P.}~\bibnamefont
  {{Reich}}}, \bibinfo {author} {\bibfnamefont {J.~C.}\ \bibnamefont
  {{Testori}}}, \ and\ \bibinfo {author} {\bibfnamefont {W.}~\bibnamefont
  {{Reich}}},\ }\href {\doibase 10.1051/0004-6361:20011000} {\bibfield
  {journal} {\bibinfo  {journal} {\aap}\ }\textbf {\bibinfo {volume} {376}},\
  \bibinfo {pages} {861} (\bibinfo {year} {2001})}\BibitemShut {NoStop}%
\bibitem [{\citenamefont {van Haasteren}\ and\ \citenamefont
  {Levin}(2013)}]{vanHaasteren:2012hj}%
  \BibitemOpen
  \bibfield  {author} {\bibinfo {author} {\bibfnamefont {R.}~\bibnamefont {van
  Haasteren}}\ and\ \bibinfo {author} {\bibfnamefont {Y.}~\bibnamefont
  {Levin}},\ }\href {\doibase 10.1093/mnras/sts097} {\bibfield  {journal}
  {\bibinfo  {journal} {Mon. Not. Roy. Astron. Soc.}\ }\textbf {\bibinfo
  {volume} {428}},\ \bibinfo {pages} {1147} (\bibinfo {year} {2013})},\ \Eprint
  {http://arxiv.org/abs/1202.5932} {arXiv:1202.5932} \BibitemShut {NoStop}%
\bibitem [{\citenamefont {Ellis}\ and\ \citenamefont {van
  Haasteren}()}]{PTMCMCSampler}%
  \BibitemOpen
  \bibfield  {author} {\bibinfo {author} {\bibfnamefont {J.~A.}\ \bibnamefont
  {Ellis}}\ and\ \bibinfo {author} {\bibfnamefont {R.}~\bibnamefont {van
  Haasteren}},\ }\href {\doibase 10.5281/zenodo.1037579} {\
  10.5281/zenodo.1037579}\BibitemShut {NoStop}%
\bibitem [{\citenamefont {Ellis}\ \emph {et~al.}()\citenamefont {Ellis},
  \citenamefont {Vallisneri}, \citenamefont {Taylor},\ and\ \citenamefont
  {Baker}}]{enterprise}%
  \BibitemOpen
  \bibfield  {author} {\bibinfo {author} {\bibfnamefont {J.~A.}\ \bibnamefont
  {Ellis}}, \bibinfo {author} {\bibfnamefont {M.}~\bibnamefont {Vallisneri}},
  \bibinfo {author} {\bibfnamefont {S.~R.}\ \bibnamefont {Taylor}}, \ and\
  \bibinfo {author} {\bibfnamefont {P.~T.}\ \bibnamefont {Baker}},\ }\href@noop
  {} {\bibinfo  {journal} {\url{http://ascl.net/1912.015}}\ }\BibitemShut
  {NoStop}%
\bibitem [{\citenamefont {Desvignes}\ \emph {et~al.}(2016)\citenamefont
  {Desvignes} \emph {et~al.}}]{Desvignes:2016yex}%
  \BibitemOpen
\bibfield  {journal} {  }\bibfield  {author} {\bibinfo {author} {\bibfnamefont
  {G.}~\bibnamefont {Desvignes}} \emph {et~al.},\ }\href {\doibase
  10.1093/mnras/stw483} {\bibfield  {journal} {\bibinfo  {journal} {Mon. Not.
  Roy. Astron. Soc.}\ }\textbf {\bibinfo {volume} {458}},\ \bibinfo {pages}
  {3341} (\bibinfo {year} {2016})},\ \Eprint {http://arxiv.org/abs/1602.08511}
  {arXiv:1602.08511} \BibitemShut {NoStop}%
\bibitem [{Note6()}]{Note6}%
  \BibitemOpen
  \bibinfo {note} {\protect \url {http://www.epta.eu.org/aom.html}}\BibitemShut
  {NoStop}%
\bibitem [{\citenamefont {Caballero}\ \emph {et~al.}(2016)\citenamefont
  {Caballero} \emph {et~al.}}]{Caballero:2015srj}%
  \BibitemOpen
  \bibfield  {author} {\bibinfo {author} {\bibfnamefont {R.}~\bibnamefont
  {Caballero}} \emph {et~al.},\ }\href {\doibase 10.1093/mnras/stw179}
  {\bibfield  {journal} {\bibinfo  {journal} {Mon. Not. Roy. Astron. Soc.}\
  }\textbf {\bibinfo {volume} {457}},\ \bibinfo {pages} {4421} (\bibinfo {year}
  {2016})},\ \Eprint {http://arxiv.org/abs/1510.09194} {arXiv:1510.09194}
  \BibitemShut {NoStop}%
\bibitem [{\citenamefont {{Yardley}}\ \emph {et~al.}(2011)\citenamefont
  {{Yardley}} \emph {et~al.}}]{Yardley_11}%
  \BibitemOpen
  \bibfield  {author} {\bibinfo {author} {\bibfnamefont {D.~R.~B.}\
  \bibnamefont {{Yardley}}} \emph {et~al.},\ }\href {\doibase
  10.1111/j.1365-2966.2011.18517.x} {\bibfield  {journal} {\bibinfo  {journal}
  {\mnras}\ }\textbf {\bibinfo {volume} {414}},\ \bibinfo {pages} {1777}
  (\bibinfo {year} {2011})},\ \Eprint {http://arxiv.org/abs/1102.2230}
  {arXiv:1102.2230} \BibitemShut {NoStop}%
\bibitem [{\citenamefont {{G{\'o}rski}}\ \emph {et~al.}(2005)\citenamefont
  {{G{\'o}rski}}, \citenamefont {{Hivon}}, \citenamefont {{Banday}},
  \citenamefont {{Wand elt}}, \citenamefont {{Hansen}}, \citenamefont
  {{Reinecke}},\ and\ \citenamefont {{Bartelmann}}}]{Healpix}%
  \BibitemOpen
  \bibfield  {author} {\bibinfo {author} {\bibfnamefont {K.~M.}\ \bibnamefont
  {{G{\'o}rski}}}, \bibinfo {author} {\bibfnamefont {E.}~\bibnamefont
  {{Hivon}}}, \bibinfo {author} {\bibfnamefont {A.~J.}\ \bibnamefont
  {{Banday}}}, \bibinfo {author} {\bibfnamefont {B.~D.}\ \bibnamefont {{Wand
  elt}}}, \bibinfo {author} {\bibfnamefont {F.~K.}\ \bibnamefont {{Hansen}}},
  \bibinfo {author} {\bibfnamefont {M.}~\bibnamefont {{Reinecke}}}, \ and\
  \bibinfo {author} {\bibfnamefont {M.}~\bibnamefont {{Bartelmann}}},\ }\href
  {\doibase 10.1086/427976} {\bibfield  {journal} {\bibinfo  {journal} {\apj}\
  }\textbf {\bibinfo {volume} {622}},\ \bibinfo {pages} {759} (\bibinfo {year}
  {2005})},\ \Eprint {http://arxiv.org/abs/astro-ph/0409513}
  {arXiv:astro-ph/0409513} \BibitemShut {NoStop}%
\bibitem [{\citenamefont {Allen}\ and\ \citenamefont
  {Ottewill}(1997)}]{Allen:1996gp}%
  \BibitemOpen
  \bibfield  {author} {\bibinfo {author} {\bibfnamefont {B.}~\bibnamefont
  {Allen}}\ and\ \bibinfo {author} {\bibfnamefont {A.~C.}\ \bibnamefont
  {Ottewill}},\ }\href {\doibase 10.1103/PhysRevD.56.545} {\bibfield  {journal}
  {\bibinfo  {journal} {Phys. Rev. D}\ }\textbf {\bibinfo {volume} {56}},\
  \bibinfo {pages} {545} (\bibinfo {year} {1997})},\ \Eprint
  {http://arxiv.org/abs/gr-qc/9607068} {arXiv:gr-qc/9607068} \BibitemShut
  {NoStop}%
\bibitem [{\citenamefont {Cornish}(2001)}]{Cornish:2001hg}%
  \BibitemOpen
  \bibfield  {author} {\bibinfo {author} {\bibfnamefont {N.~J.}\ \bibnamefont
  {Cornish}},\ }\href {\doibase 10.1088/0264-9381/18/20/307} {\bibfield
  {journal} {\bibinfo  {journal} {Class. Quant. Grav.}\ }\textbf {\bibinfo
  {volume} {18}},\ \bibinfo {pages} {4277} (\bibinfo {year} {2001})},\ \Eprint
  {http://arxiv.org/abs/astro-ph/0105374} {arXiv:astro-ph/0105374} \BibitemShut
  {NoStop}%
\bibitem [{\citenamefont {Mitra}\ \emph {et~al.}(2008)\citenamefont {Mitra},
  \citenamefont {Dhurandhar}, \citenamefont {Souradeep}, \citenamefont
  {Lazzarini}, \citenamefont {Mandic}, \citenamefont {Bose},\ and\
  \citenamefont {Ballmer}}]{Mitra:2007mc}%
  \BibitemOpen
  \bibfield  {author} {\bibinfo {author} {\bibfnamefont {S.}~\bibnamefont
  {Mitra}}, \bibinfo {author} {\bibfnamefont {S.}~\bibnamefont {Dhurandhar}},
  \bibinfo {author} {\bibfnamefont {T.}~\bibnamefont {Souradeep}}, \bibinfo
  {author} {\bibfnamefont {A.}~\bibnamefont {Lazzarini}}, \bibinfo {author}
  {\bibfnamefont {V.}~\bibnamefont {Mandic}}, \bibinfo {author} {\bibfnamefont
  {S.}~\bibnamefont {Bose}}, \ and\ \bibinfo {author} {\bibfnamefont
  {S.}~\bibnamefont {Ballmer}},\ }\href {\doibase 10.1103/PhysRevD.77.042002}
  {\bibfield  {journal} {\bibinfo  {journal} {Phys. Rev. D}\ }\textbf {\bibinfo
  {volume} {77}},\ \bibinfo {pages} {042002} (\bibinfo {year} {2008})},\
  \Eprint {http://arxiv.org/abs/0708.2728} {arXiv:0708.2728} \BibitemShut
  {NoStop}%
\bibitem [{\citenamefont {Thrane}\ \emph {et~al.}(2009)\citenamefont {Thrane},
  \citenamefont {Ballmer}, \citenamefont {Romano}, \citenamefont {Mitra},
  \citenamefont {Talukder}, \citenamefont {Bose},\ and\ \citenamefont
  {Mandic}}]{Thrane:2009fp}%
  \BibitemOpen
  \bibfield  {author} {\bibinfo {author} {\bibfnamefont {E.}~\bibnamefont
  {Thrane}}, \bibinfo {author} {\bibfnamefont {S.}~\bibnamefont {Ballmer}},
  \bibinfo {author} {\bibfnamefont {J.~D.}\ \bibnamefont {Romano}}, \bibinfo
  {author} {\bibfnamefont {S.}~\bibnamefont {Mitra}}, \bibinfo {author}
  {\bibfnamefont {D.}~\bibnamefont {Talukder}}, \bibinfo {author}
  {\bibfnamefont {S.}~\bibnamefont {Bose}}, \ and\ \bibinfo {author}
  {\bibfnamefont {V.}~\bibnamefont {Mandic}},\ }\href {\doibase
  10.1103/PhysRevD.80.122002} {\bibfield  {journal} {\bibinfo  {journal} {Phys.
  Rev. D}\ }\textbf {\bibinfo {volume} {80}},\ \bibinfo {pages} {122002}
  (\bibinfo {year} {2009})},\ \Eprint {http://arxiv.org/abs/0910.0858}
  {arXiv:0910.0858} \BibitemShut {NoStop}%
\bibitem [{\citenamefont {{Renzini}}\ and\ \citenamefont
  {{Contaldi}}(2018)}]{Renzini:2018}%
  \BibitemOpen
  \bibfield  {author} {\bibinfo {author} {\bibfnamefont {A.~I.}\ \bibnamefont
  {{Renzini}}}\ and\ \bibinfo {author} {\bibfnamefont {C.~R.}\ \bibnamefont
  {{Contaldi}}},\ }\href {\doibase 10.1093/mnras/sty2546} {\bibfield  {journal}
  {\bibinfo  {journal} {\mnras}\ }\textbf {\bibinfo {volume} {481}},\ \bibinfo
  {pages} {4650} (\bibinfo {year} {2018})},\ \Eprint
  {http://arxiv.org/abs/1806.11360} {arXiv:1806.11360} \BibitemShut {NoStop}%
\bibitem [{\citenamefont {Cornish}(2002)}]{Cornish:2002bh}%
  \BibitemOpen
  \bibfield  {author} {\bibinfo {author} {\bibfnamefont {N.}~\bibnamefont
  {Cornish}},\ }\href {\doibase 10.1088/0264-9381/19/7/306} {\bibfield
  {journal} {\bibinfo  {journal} {Class. Quant. Grav.}\ }\textbf {\bibinfo
  {volume} {19}},\ \bibinfo {pages} {1279} (\bibinfo {year}
  {2002})}\BibitemShut {NoStop}%
\bibitem [{\citenamefont {Kudoh}\ and\ \citenamefont
  {Taruya}(2005)}]{Kudoh:2004he}%
  \BibitemOpen
  \bibfield  {author} {\bibinfo {author} {\bibfnamefont {H.}~\bibnamefont
  {Kudoh}}\ and\ \bibinfo {author} {\bibfnamefont {A.}~\bibnamefont {Taruya}},\
  }\href {\doibase 10.1103/PhysRevD.71.024025} {\bibfield  {journal} {\bibinfo
  {journal} {Phys. Rev. D}\ }\textbf {\bibinfo {volume} {71}},\ \bibinfo
  {pages} {024025} (\bibinfo {year} {2005})},\ \Eprint
  {http://arxiv.org/abs/gr-qc/0411017} {arXiv:gr-qc/0411017} \BibitemShut
  {NoStop}%
\bibitem [{\citenamefont {Seto}\ and\ \citenamefont
  {Cooray}(2004)}]{Seto:2004np}%
  \BibitemOpen
  \bibfield  {author} {\bibinfo {author} {\bibfnamefont {N.}~\bibnamefont
  {Seto}}\ and\ \bibinfo {author} {\bibfnamefont {A.}~\bibnamefont {Cooray}},\
  }\href {\doibase 10.1103/PhysRevD.70.123005} {\bibfield  {journal} {\bibinfo
  {journal} {Phys. Rev. D}\ }\textbf {\bibinfo {volume} {70}},\ \bibinfo
  {pages} {123005} (\bibinfo {year} {2004})},\ \Eprint
  {http://arxiv.org/abs/astro-ph/0403259} {arXiv:astro-ph/0403259} \BibitemShut
  {NoStop}%
\bibitem [{\citenamefont {Taruya}\ and\ \citenamefont
  {Kudoh}(2005)}]{Taruya:2005yf}%
  \BibitemOpen
  \bibfield  {author} {\bibinfo {author} {\bibfnamefont {A.}~\bibnamefont
  {Taruya}}\ and\ \bibinfo {author} {\bibfnamefont {H.}~\bibnamefont {Kudoh}},\
  }\href {\doibase 10.1103/PhysRevD.72.104015} {\bibfield  {journal} {\bibinfo
  {journal} {Phys. Rev. D}\ }\textbf {\bibinfo {volume} {72}},\ \bibinfo
  {pages} {104015} (\bibinfo {year} {2005})},\ \Eprint
  {http://arxiv.org/abs/gr-qc/0507114} {arXiv:gr-qc/0507114} \BibitemShut
  {NoStop}%
\bibitem [{\citenamefont {Kudoh}\ \emph {et~al.}(2006)\citenamefont {Kudoh},
  \citenamefont {Taruya}, \citenamefont {Hiramatsu},\ and\ \citenamefont
  {Himemoto}}]{Kudoh:2005as}%
  \BibitemOpen
  \bibfield  {author} {\bibinfo {author} {\bibfnamefont {H.}~\bibnamefont
  {Kudoh}}, \bibinfo {author} {\bibfnamefont {A.}~\bibnamefont {Taruya}},
  \bibinfo {author} {\bibfnamefont {T.}~\bibnamefont {Hiramatsu}}, \ and\
  \bibinfo {author} {\bibfnamefont {Y.}~\bibnamefont {Himemoto}},\ }\href
  {\doibase 10.1103/PhysRevD.73.064006} {\bibfield  {journal} {\bibinfo
  {journal} {Phys. Rev. D}\ }\textbf {\bibinfo {volume} {73}},\ \bibinfo
  {pages} {064006} (\bibinfo {year} {2006})},\ \Eprint
  {http://arxiv.org/abs/gr-qc/0511145} {arXiv:gr-qc/0511145} \BibitemShut
  {NoStop}%
\bibitem [{\citenamefont {Taruya}(2006)}]{Taruya:2006kqa}%
  \BibitemOpen
  \bibfield  {author} {\bibinfo {author} {\bibfnamefont {A.}~\bibnamefont
  {Taruya}},\ }\href {\doibase 10.1103/PhysRevD.74.104022} {\bibfield
  {journal} {\bibinfo  {journal} {Phys. Rev. D}\ }\textbf {\bibinfo {volume}
  {74}},\ \bibinfo {pages} {104022} (\bibinfo {year} {2006})},\ \Eprint
  {http://arxiv.org/abs/gr-qc/0607080} {arXiv:gr-qc/0607080} \BibitemShut
  {NoStop}%
\bibitem [{\citenamefont {Cornish}\ and\ \citenamefont {van
  Haasteren}(2014)}]{Cornish:2014rva}%
  \BibitemOpen
  \bibfield  {author} {\bibinfo {author} {\bibfnamefont {N.~J.}\ \bibnamefont
  {Cornish}}\ and\ \bibinfo {author} {\bibfnamefont {R.}~\bibnamefont {van
  Haasteren}},\ }\href@noop {} {\  (\bibinfo {year} {2014})},\ \Eprint
  {http://arxiv.org/abs/1406.4511} {arXiv:1406.4511} \BibitemShut {NoStop}%
\bibitem [{\citenamefont {Romano}\ \emph {et~al.}(2015)\citenamefont {Romano},
  \citenamefont {Taylor}, \citenamefont {Cornish}, \citenamefont {Gair},
  \citenamefont {Mingarelli},\ and\ \citenamefont {van
  Haasteren}}]{Romano:2015uma}%
  \BibitemOpen
  \bibfield  {author} {\bibinfo {author} {\bibfnamefont {J.~D.}\ \bibnamefont
  {Romano}}, \bibinfo {author} {\bibfnamefont {S.~R.}\ \bibnamefont {Taylor}},
  \bibinfo {author} {\bibfnamefont {N.~J.}\ \bibnamefont {Cornish}}, \bibinfo
  {author} {\bibfnamefont {J.}~\bibnamefont {Gair}}, \bibinfo {author}
  {\bibfnamefont {C.~M.}\ \bibnamefont {Mingarelli}}, \ and\ \bibinfo {author}
  {\bibfnamefont {R.}~\bibnamefont {van Haasteren}},\ }\href {\doibase
  10.1103/PhysRevD.92.042003} {\bibfield  {journal} {\bibinfo  {journal} {Phys.
  Rev. D}\ }\textbf {\bibinfo {volume} {92}},\ \bibinfo {pages} {042003}
  (\bibinfo {year} {2015})},\ \Eprint {http://arxiv.org/abs/1505.07179}
  {arXiv:1505.07179} \BibitemShut {NoStop}%
\bibitem [{\citenamefont {{Moore}}(1920)}]{Moore_20}%
  \BibitemOpen
  \bibfield  {author} {\bibinfo {author} {\bibfnamefont {E.~H.}\ \bibnamefont
  {{Moore}}},\ }\href@noop {} {\bibfield  {journal} {\bibinfo  {journal}
  {Bulletin of the American Mathematical Society}\ }\textbf {\bibinfo {volume}
  {26}},\ \bibinfo {pages} {394–95} (\bibinfo {year} {1920})}\BibitemShut
  {NoStop}%
\bibitem [{\citenamefont {{Penrose}}(1955)}]{Penrose_55}%
  \BibitemOpen
  \bibfield  {author} {\bibinfo {author} {\bibfnamefont {R.}~\bibnamefont
  {{Penrose}}},\ }\href@noop {} {\bibfield  {journal} {\bibinfo  {journal}
  {Proceedings of the Cambridge Philosophical Society}\ }\textbf {\bibinfo
  {volume} {51}},\ \bibinfo {pages} {406–13} (\bibinfo {year}
  {1955})}\BibitemShut {NoStop}%
\bibitem [{\citenamefont {{Alonso}}\ \emph {et~al.}(2020)\citenamefont
  {{Alonso}}, \citenamefont {{Cusin}}, \citenamefont {{Ferreira}},\ and\
  \citenamefont {{Pitrou}}}]{Alonso_20}%
  \BibitemOpen
  \bibfield  {author} {\bibinfo {author} {\bibfnamefont {D.}~\bibnamefont
  {{Alonso}}}, \bibinfo {author} {\bibfnamefont {G.}~\bibnamefont {{Cusin}}},
  \bibinfo {author} {\bibfnamefont {P.~G.}\ \bibnamefont {{Ferreira}}}, \ and\
  \bibinfo {author} {\bibfnamefont {C.}~\bibnamefont {{Pitrou}}},\ }\href
  {\doibase 10.1103/PhysRevD.102.023002} {\bibfield  {journal} {\bibinfo
  {journal} {\prd}\ }\textbf {\bibinfo {volume} {102}},\ \bibinfo {eid}
  {023002} (\bibinfo {year} {2020})},\ \Eprint
  {http://arxiv.org/abs/2002.02888} {arXiv:2002.02888} \BibitemShut {NoStop}%
\end{thebibliography}%

\end{document}